\documentclass{LMCS}

\def\dOi{11(3:3)2015}
\lmcsheading%
{\dOi}
{1--37}
{}
{}
{Feb.~27, 2014}
{Aug.~13, 2015}
{}

\ACMCCS{[{\bf Theory of computation}]: Logic --- Constraint and logic programming; Semantics and reasoning --- Program reasoning --- Program specifications; [\textbf{Information systems}]:  World Wide Web --- Web services --- Service discovery and interfaces}

\usepackage{microtype}
\usepackage[english]{babel}
\usepackage{engord}

\usepackage{style/document-structure}
\usepackage{style/lists}
\usepackage{style/dashes}
\usepackage{style/acronyms}

\usepackage{macros/todo}
\usepackage{macros/align}
\usepackage{macros/delimiters}
\usepackage{macros/arrows}
\usepackage{macros/relop}
\usepackage{macros/sets}
\usepackage{macros/categories}
\usepackage{macros/comma-categories}
\usepackage{macros/indexed-categories}
\usepackage{macros/orchestrations}
\usepackage{macros/orchestrations/PE}
\usepackage{macros/orchestrations/ARN}
\usepackage{tikz}
\usepackage{macros/tikz/orchestrations}
\usepackage{macros/tikz/automata}
\usepackage{macros/institutions}
\usepackage{macros/institutions/FOL}
\usepackage{macros/institutions/POA}
\usepackage{macros/institutions/LTL}
\usepackage{macros/institutions/ALTL}
\usepackage{macros/substitutions}
\usepackage{macros/substitutions/FOL}
\usepackage{macros/substitutions/OrcScheme}
\usepackage{macros/quantified-sentences}
\usepackage{macros/LP}
\usepackage{macros/modularization}
\usepackage{macros/SL}

\usepackage[all, cmtip]{xy}
\usepackage{rotating}
\usepackage{array}
\usepackage{booktabs}

\usepackage{hyperref}

\begin{document}

\title[Service-Oriented Logic Programming]{Service-Oriented Logic Programming\rsuper*}

\author[I.\ \c{T}u\c{t}u]{Ionu\c{t} \c{T}u\c{t}u\rsuper a}
\address{{\lsuper a}Department of Computer Science, Royal Holloway University of London \newline
  Institute of Mathematics of the Romanian Academy, Research group of the project ID-3-0439}
\email{ittutu@gmail.com}

\author[J.\,L.\ Fiadeiro]{Jos\'{e} Luiz Fiadeiro\rsuper b}
\address{{\lsuper b}Department of Computer Science, Royal Holloway University of London}
\email{jose.fiadeiro@rhul.ac.uk}

\keywords{Logic programming, Institution theory, Service-oriented computing, Orchestration schemes, Service discovery and binding}

\titlecomment{{\lsuper*}A preliminary version of this work was presented at CALCO 2013~\cite{Tutu-Fiadeiro:A-logic-programming-semantics-of-services-2013}.}

\begin{abstract}
  \noindent We develop formal foundations for notions and mechanisms needed to support service-oriented computing.
  Our work builds on recent theoretical advancements in the algebraic structures that capture the way services are orchestrated and in the processes that formalize the discovery and binding of services to given client applications by means of logical representations of required and provided services.
  We show how the denotational and the operational semantics specific to conventional logic programming can be generalized using the theory of institutions to address both static and dynamic aspects of service-oriented computing.
  Our results rely upon a strong analogy between the discovery of a service that can be bound to an application and the search for a clause that can be used for computing an answer to a query; they explore the manner in which requests for external services can be described as service queries, and explain how the computation of their answers can be performed through service-oriented derivatives of unification and resolution, which characterize the binding of services and the reconfiguration of applications.
\end{abstract}

\maketitle

\section{Introduction}

\begin{minisection}{Service-Oriented Computing.}%
  Service-oriented computing is a modern computational paradigm that deals with the execution of programs over distributed information-processing infrastructures in which software applications can discover and bind dynamically, at run time, to services offered by providers.
  Whereas the paradigm has been effectively in use for a more than a decade in the form of Web services~\cite{Alonso-Casati-Kuno-Machiraju:Web-Services-2004} or Grid computing~\cite{Foster-Kesselman:The-Grid-2-2004}, research into its formal foundations has lagged somewhat behind, partly because of our lack of understanding of (or agreement on) what is really new about the paradigm, especially in relation to distributed computing in general (see, for example,~\cite{Vogels:Web-services-are-not-distributed-objects-2003}).

  It is fair to say that significant advances have been made towards formalizing new forms of distributed computation that have arisen around the notion of service (e.g.\ choreography~\cite{Su-Bultan-Fu-Zhao:Web-service-choreographies-2007}), notably through several variants of the \(\pi\)\nb-calculus.  However, service-oriented computing raises more profound challenges at the level of the structure of  systems due to their ability to discover and bind dynamically, in a non-programmed manner, to other systems.  The structure of the systems that we are now creating in the virtual space of computational networks is intrinsically dynamic, a phenomenon hitherto unknown. Formalisms such as  the \(\pi\)\nb-calculus do not address these structural properties of systems.
  This prevents us from fully controlling and developing trust in the systems that are now operating in cyberspace, and also from exploiting the power of the paradigm beyond the way it is currently deployed.

  Towards that end, we have investigated algebraic structures that account for modularity (e.g.~\cite{Fiadeiro-Lopes-Bocchi:Algebraic-semantics-of-service-component-modules-2007,Fiadeiro-Schmitt:Structured-cospans-2007}) -- referring to the way services are orchestrated as composite structures of components and how binding is performed through interaction protocols -- and the mechanisms through which discovery can be formalized in terms of logical specifications of required/provided services and constraint optimisation for service-level agreements (e.g.~\cite{Fiadeiro-Lopes-Bocchi:An-abstract-model-for-service-discovery-and-binding-2011,Fiadeiro-Lopes:Dynamic-reconfiguration-in-service-oriented-architectures-2013}).
  In the present paper, we take further this research to address the operational aspects behind \emph{dynamic} discovery and binding, i.e.\ the mechanisms through which applications discover and bind, at run time, to services.  Our aim is to develop an abstract, foundational setting -- independent of the specific technologies that are currently deployed, such as \newacronym{SOAP}{Simple Object Access protocol} for message-exchange protocols and \newacronym{UDDI}{Universal Description, Discovery and Integration} for description, discovery, and integration -- that combines both the \emph{denotational} and the \emph{operational} semantics of services. 
  The challenge here is to define an integrated algebraic framework that accounts for
  \begin{inlinenum}

  \item logical specifications of services,
    
  \item the way models of those specifications capture orchestrations of components that may depend on externally provided services  to be discovered, and
    
  \item the way the discovery of services and the binding of their orchestrations to client applications can be expressed in logical/algebraic terms.
    
  \end{inlinenum}
\end{minisection}

\begin{minisection}{Logic Programming.}%
  The approach that we propose to develop to meet this challenge builds on the relational variant of (Horn-clause) logic programming -- the paradigm that epitomizes the integration of declarative and operational aspects of logic.  In conventional logic programming, clauses have a declarative semantics as universally quantified implications that express relationships over a domain (the Herbrand universe), and an operational semantics that derives from resolution and term unification: definite clauses (provided by a given logic program) are used to resolve logic-programming queries (expressed as existentially quantified conjunctions) by generating new queries and, through term unification, computing partial answers as substitutions for the variables of the original query. 

  In a nutshell, the analogy between service-oriented computing and conventional logic programming that we propose to systematically examine in this paper unfolds as follows:
  \begin{itemize}

  \item The Herbrand universe consists of those service orchestrations that have no dependencies on external services -- what we refer to as ground orchestrations.
    
  \item Variables and terms correspond to dependencies on external services that need to be discovered and to the actual services that are made available by orchestrations.

  \item Service clauses express conditional properties of services required or provided by orchestrations, thus capturing the notion of service module described in~\cite{Fiadeiro-Lopes-Bocchi:An-abstract-model-for-service-discovery-and-binding-2011}.  Their declarative semantics is that, when bound to the orchestrations of other service clauses that ensure the required properties, they deliver, through their orchestration, services that satisfy the specified properties.

  \item Service queries express properties of orchestrations of services that an application requires in order to fulfil its goal -- what we describe in~\cite{Fiadeiro-Lopes-Bocchi:An-abstract-model-for-service-discovery-and-binding-2011} as activity modules.

  \item Logic programs define service repositories as collections of service modules.

  \item Resolution and term unification account for service discovery by matching required properties with provided ones and the binding of required with provided services.

  \end{itemize}
\end{minisection}

\begin{minisection}{The structure of the paper.}%
  Our research into the logic-programming semantics of service-oriented computing is organized in two parts.
  In Section~\ref{section:orchestration-schemes} we present a new categorical model of service orchestrations, called \emph{orchestration scheme}, that enables us to treat orchestrations as fully abstract entities required to satisfy only a few elementary properties.
  This framework is flexible enough to accommodate, for example, orchestrations in the form of program expressions, as considered in~\cite{Fiadeiro:The-many-faces-of-complexity-in-software-design-2012}, or as asynchronous relational networks similar to those defined in~\cite{Fiadeiro-Lopes:An-interface-theory-for-service-oriented-design-2013}.
  In our study, such schemes play an essential role in managing the inherent complexity of orchestrations whilst making available, at the same time, the fundamental building blocks of service-oriented logic programming.
  In Section~\ref{section:service-discovery-and-binding}, we define a logical system of orchestration schemes over which we can express properties that can be further used to guide the interconnection of orchestrations.
  We recall from~\cite{Tutu-Fiadeiro:Institution-independent-logic-programming-2015} the algebraic structures that underlie institution-independent logic programming, in particular the substitution systems that are characteristic of relational logic programming, and prove that the resulting logic of orchestration schemes constitutes a \emph{generalized substitution system}.
  This result is central to our work, not only because it provides the declarative semantics of our approach to service-oriented computing, but also because it gives a definite mathematical foundation to the analogy between service-oriented computing and conventional logic programming outlined above.
  Building on these results, we show how clauses, queries, unification and resolution can be defined over the generalized substitution system of orchestration schemes, providing in this way the corresponding operational semantics of service-oriented computing.
\end{minisection}

The work presented herein continues our investigation on logic-independent foundations of logic programming reported in~\cite{Tutu-Fiadeiro:Institution-independent-logic-programming-2015}. As such, it is based on the theory of institutions of Goguen and Burstall~\cite{Goguen-Burstall:Institutions-1992}; although familiarity with the institution-independent presentation of logic programming is not essential, some knowledge of basic notions of institution theory such as institution, (co)morphism of institutions, and also of the description of institutions as functors into the category of rooms~\cite{Diaconescu:Institution-Independent-Model-Theory-2008,Sannella-Tarlecki:Foundations-of-Algebraic-Specification-2011} is presumed.

\section{Orchestration Schemes}
\label{section:orchestration-schemes}

The first step in the development of the particular variant of logic programming that we consider in this paper consists in determining appropriate categorical abstractions of the structures that support service-oriented computing.  These will ultimately allow us to describe the process of service discovery and binding in a way that is independent of any particular formalism (such as various forms of automata, transition systems or process algebras).

Our approach is grounded on two observations: first, that orchestrations can be organized as a category whose arrows, or more precisely, cospans of arrows, can be used to model the composition of service components (as defined, for example, in~\cite{Fiadeiro-Lopes-Bocchi:Algebraic-semantics-of-service-component-modules-2007,Fiadeiro-Lopes-Bocchi:An-abstract-model-for-service-discovery-and-binding-2011,Fiadeiro-Lopes:Dynamic-reconfiguration-in-service-oriented-architectures-2013}); second, that the discovery of a service to be bound to a given client application can be formalized in terms of logical specifications of required and provided properties, ensuring that the specification of the properties offered by the service provider refines the specification of the properties requested by the client application. To this end, we explore the model-theoretic notion of refinement advanced in~\cite{Sannella-Tarlecki:Formal-development-of-programs-from-algebraic-specifications-1988}, except  that, in the present setting, the structures over which specifications are evaluated are morphisms into ground orchestrations, i.e.\ into orchestrations that have no dependencies on external services. The motivation for this choice is that, in general, the semantics of non-ground orchestrations is open: the (observable) behaviour exhibited by non-ground orchestrations varies according to the external services that they may procure at run time.
With these remarks in mind, we arrive at the following concept of orchestration scheme.

\begin{defi}[Orchestration scheme]
  An \emph{orchestration scheme} is a quadruple \(\abr{\Orc, \SSpec,\linebreak[0] \Grc, \Prop}\) consisting of
  \begin{itemize}
    
  \item a category \(\Orc\) of \emph{orchestrations} and \emph{orchestration morphisms},

  \item a functor \(\SSpec \colon \Orc \to \Set\) that defines a set \(\SSpec\rbr{\orc}\) of \emph{service specifications} over \(\orc\) for every orchestration \(\orc\),

  \item a full subcategory \(\Grc \subseteq \Orc\) of \emph{ground orchestrations}, and

  \item a functor \(\Prop \colon \Grc \to \Set\) that defines a natural subset \(\Prop\rbr{\grc} \subseteq \SSpec\rbr{\grc}\)\footnote{By describing the set \(\Prop\rbr{\grc}\) as a \emph{natural subset} of \(\SSpec\rbr{\grc}\) we mean that the family of inclusions \(\rbr{\Prop\rbr{\grc} \subseteq \SSpec\rbr{\grc}}_{\grc \in \obj{\Grc}}\) defines a natural transformation from \(\Prop\) to \(\rbr{\Grc \subseteq \Orc} \comp \SSpec\).} of \emph{properties} of \(\grc\) (specifications that are guaranteed to hold when evaluated over \(\grc\)) for every ground orchestration \(\grc\).
    
  \end{itemize}
\end{defi}

\noindent To illustrate our categorical approach to orchestrations, we consider two main running examples: program expressions as discussed in~\cite{Fiadeiro:The-many-faces-of-complexity-in-software-design-2012} (see also~\cite{Morgan:Programming-from-Specifications-1994}), which provide a way of constructing structured (sequential) programs through design-time discovery and binding, and the theory of asynchronous relational networks put forward in~\cite{Fiadeiro-Lopes:An-interface-theory-for-service-oriented-design-2013}, which emphasizes the role of services as an interface mechanism for software components that can be composed through run-time discovery and binding.

\subsection{Program Expressions}
\label{subsection:program-expressions}

\begin{sidewaysfigure}
  \centering

  \begin{tikzpicture}
    % the application
    
    \node [pvar] (app) {\(\ptm\)};
    
    \node (app-orc) [fit=(app)] {};

    \node [requires-point] (app-rspec) [right=.5em of app-orc.east] {%
      \(\osp_{1}\!\!\)% 
    };
    \draw (app) |- (app-rspec.west);
    
    \coordinate (app-north) at ($(app-rspec.north) + (0em, 1ex)$);
    \coordinate (app-south) at ($(app-rspec.south) - (0em, 1ex)$);
    \coordinate (app-west)  at ($(app-orc.west)    - (.25em, 0ex)$);
    \coordinate (app-east)  at ($(app-rspec.east)  - (1.5em, 0ex)$);

    \node (app-rspecr) [right=0em of app-rspec] {\hspace{.6em}\(\sqsubseteq\)};
    
    % sequence1

    \node [provides-point] (seq1-pspec) [below=25ex of app-orc.south west, anchor=north west] {%
      \hspace{.25em}\(\vphantom{\ssbr{}}\osp_{2}\)% 
    };
    \draw [dotted, thick] (app-rspecr)
    -- ($(app-rspecr.east) + (.5em, 0ex)$)
    -- ($(app-rspecr.east) + (.5em, -7.3ex)$)
    -| ($(seq1-pspec.west) - (.5em, -20ex)$)
    -- ($(seq1-pspec.west) - (.5em, 0ex)$)
    -- (seq1-pspec.west);
    
    \path ($(app-rspec.east) + (.5em, 0ex)$) -- (seq1-pspec.west);

    \node [pvar] (seq1-pgm1) [right=.5em of seq1-pspec.east] {\(\ptm_{1}\)};
    \node        (seq1-comp) [right=0em of seq1-pgm1]        {\(\comp\)};
    \node [pvar] (seq1-pgm2) [right=0em of seq1-comp]        {\(\ptm_{2}\)};
    
    \coordinate (seq1-onorth) at ($(seq1-comp.north) + (0em,3.5ex)$);
    \coordinate (seq1-osouth) at ($(seq1-comp.south) - (0em,3.5ex)$);
    \coordinate (seq1-owest)  at (seq1-pgm1.west);
    \coordinate (seq1-oeast)  at (seq1-pgm2.east);

    \node (seq1-orc) [fit=(seq1-onorth) (seq1-osouth) (seq1-owest) (seq1-oeast)] {};

    \node [requires-point] (seq1-rspec1) [above right=8ex and .125em of seq1-orc.east] {%
      \(\vphantom{\ssbr{}}\osp_{3}\!\!\)%
    };
    \draw (seq1-pgm1) |- (seq1-rspec1.west);
    \node [requires-point] (seq1-rspec2) [below right=8ex and .125em of seq1-orc.east] {%
      \(\vphantom{\ssbr{}}\osp_{4}\!\!\)%
    };
    \draw (seq1-pgm2) |- (seq1-rspec2.west);

    \coordinate (seq1-north) at ($(seq1-rspec1.north) + (0em, 1ex)$);
    \coordinate (seq1-south) at ($(seq1-rspec2.south) - (0em, 1ex)$);
    \coordinate (seq1-west)  at ($(seq1-pspec.west)   + (1.5em, 0ex)$);
    \coordinate (seq1-east)  at ($(seq1-rspec1.east)  - (1.5em, 0ex)$);

    % sequence2

    \node [provides-point] (seq2-pspec) [right=2em of seq1-rspec1.east] {%
      \hspace{.25em}\(\vphantom{\ssbr{}}\osp_{5}\)%
    };
    \path (seq1-rspec1) -- node {\hspace{.5em}\(\sqsubseteq\)} (seq2-pspec);

    \node [pvar] (seq2-pgm1) [right=.5em of seq2-pspec.east] {\(\ptm_{3}\)};
    \node        (seq2-comp) [right=0em of seq2-pgm1]        {\(\comp\)};
    \node [pvar] (seq2-pgm2) [right=0em of seq2-comp]        {\(\ptm_{4}\)};
    
    \coordinate (seq2-onorth) at ($(seq2-comp.north) + (0em,3.5ex)$);
    \coordinate (seq2-osouth) at ($(seq2-comp.south) - (0em,3.5ex)$);
    \coordinate (seq2-owest)  at (seq2-pgm1.west);
    \coordinate (seq2-oeast)  at (seq2-pgm2.east);

    \node (seq2-orc) [fit=(seq2-onorth) (seq2-osouth) (seq2-owest) (seq2-oeast)] {};

    \node [requires-point] (seq2-rspec1) [above right=3ex and .125em of seq2-orc.east] {%
      \(\vphantom{\ssbr{}}\osp_{6}\!\!\)%
    };
    \draw (seq2-pgm1) |- (seq2-rspec1.west);
    \node [requires-point] (seq2-rspec2) [below right=3ex and .125em of seq2-orc.east] {%
      \(\vphantom{\ssbr{}}\osp_{7}\!\!\)%
    };
    \draw (seq2-pgm2) |- (seq2-rspec2.west);

    \coordinate (seq2-north) at ($(seq2-rspec1.north) + (0em, 1ex)$);
    \coordinate (seq2-south) at ($(seq2-rspec2.south) - (0em, 1ex)$);
    \coordinate (seq2-west)  at ($(seq2-pspec.west)   + (1.5em, 0ex)$);
    \coordinate (seq2-east)  at ($(seq2-rspec1.east)  - (1.5em, 0ex)$);

    % assignment1

    \node [provides-point] (assign1-pspec) [right=2em of seq2-rspec1.east] {%
      \hspace{.25em}\(\vphantom{\ssbr{}}\osp_{8}\)%
    };
    \path (seq2-rspec1) -- node {\hspace{.5em}\(\sqsubseteq\)} (assign1-pspec);

    \node (assign1) [right=.5em of assign1-pspec.east] {\(q \coloneqq 0\)};
    
    \node (assign1-orc) [fit=(assign1)] {};

    \coordinate (assign1-north) at ($(assign1-pspec.north) + (0em, 1ex)$);
    \coordinate (assign1-south) at ($(assign1-pspec.south) - (0em, 1ex)$);
    \coordinate (assign1-west)  at ($(assign1-pspec.west)  + (1.5em, 0ex)$);
    \coordinate (assign1-east)  at ($(assign1-orc.east)    + (.125em, 0ex)$);

    % assignment2

    \node [provides-point] (assign2-pspec) [right=2em of seq2-rspec2.east] {%
      \hspace{.25em}\(\vphantom{\ssbr{}}\osp_{9}\)%
    };
    \path (seq2-rspec2) -- node {\hspace{.5em}\(\sqsubseteq\)} (assign2-pspec);

    \node (assign2) [right=.5em of assign2-pspec.east] {\(r \coloneqq x\)};
    
    \node (assign2-orc) [fit=(assign2)] {};

    \coordinate (assign2-north) at ($(assign2-pspec.north) + (0em, 1ex)$);
    \coordinate (assign2-south) at ($(assign2-pspec.south) - (0em, 1ex)$);
    \coordinate (assign2-west)  at ($(assign2-pspec.west)  + (1.5em, 0ex)$);
    \coordinate (assign2-east)  at ($(assign2-orc.east)    + (.125em, 0ex)$);

    % iteration

    \node [provides-point] (iter-pspec) [right=2em of seq1-rspec2.east] {%
      \hspace{.25em}\(\vphantom{\ssbr{}}\osp_{10}\)%
    };
    \path (seq1-rspec2) -- node {\hspace{.5em}\(\sqsubseteq\)} (iter-pspec);

    \node [pvar] (iter-pgm)   [right=1.5em of iter-pspec.east]                                   {\(\ptm_{5}\)};
    \node        (iter-while) [above left=0ex and 1em of iter-pgm.north west, anchor=south west] {\(\opname{while}\, y \leq r\, \opname{do}\)};
    \node        (iter-done)  [below left=0ex and 1em of iter-pgm.south west, anchor=north west] {\(\opname{done}\)};

    \coordinate (iter-onorth) at (iter-while.north);
    \coordinate (iter-osouth) at (iter-done.south);
    \coordinate (iter-owest)  at (iter-while.west);
    \coordinate (iter-oeast)  at (iter-while.east);

    \node (iter-orc) [fit=(iter-onorth) (iter-osouth) (iter-owest) (iter-oeast)] {};
    
    \node [requires-point] (iter-rspec) [right=.125em of iter-orc.east] {%
      \(\vphantom{\ssbr{}}\osp_{11}\!\!\)%
    };
    \draw (iter-pgm) -| (iter-rspec.west);

    \coordinate (iter-north) at (iter-rspec.north);
    \coordinate (iter-south) at (iter-rspec.south);
    \coordinate (iter-west)  at ($(iter-pspec.west) + (1.5em, 0ex)$);
    \coordinate (iter-east)  at ($(iter-rspec.east) - (1.5em, 0ex)$);

    % sequence3

    \node [provides-point] (seq3-pspec) [right=2em of iter-rspec.east] {%
      \hspace{.25em}\(\vphantom{\ssbr{}}\osp_{12}\)%
    };
    \path (iter-rspec) -- node {\hspace{.5em}\(\sqsubseteq\)} (seq3-pspec);

    \node [pvar] (seq3-pgm1) [right=.5em of seq3-pspec.east] {\(\ptm_{6}\)};
    \node        (seq3-comp) [right=0em of seq3-pgm1]        {\(\comp\)};
    \node [pvar] (seq3-pgm2) [right=0em of seq3-comp]        {\(\ptm_{7}\)};
    
    \coordinate (seq3-onorth) at ($(seq3-comp.north) + (0em,3.5ex)$);
    \coordinate (seq3-osouth) at ($(seq3-comp.south) - (0em,3.5ex)$);
    \coordinate (seq3-owest)  at (seq3-pgm1.west);
    \coordinate (seq3-oeast)  at (seq3-pgm2.east);

    \node (seq3-orc) [fit=(seq3-onorth) (seq3-osouth) (seq3-owest) (seq3-oeast)] {};

    \node [requires-point] (seq3-rspec1) [above right=3ex and .125em of seq3-orc.east] {%
      \(\vphantom{\ssbr{}}\osp_{13}\!\!\)%
    };
    \draw (seq3-pgm1) |- (seq3-rspec1.west);
    \node [requires-point] (seq3-rspec2) [below right=3ex and .125em of seq3-orc.east] {%
      \(\vphantom{\ssbr{}}\osp_{14}\!\!\)%
    };
    \draw (seq3-pgm2) |- (seq3-rspec2.west);

    \coordinate (seq3-north) at ($(seq3-rspec1.north) + (0em, 1ex)$);
    \coordinate (seq3-south) at ($(seq3-rspec2.south) - (0em, 1ex)$);
    \coordinate (seq3-west)  at ($(seq3-pspec.west)   + (1.5em, 0ex)$);
    \coordinate (seq3-east)  at ($(seq3-rspec1.east)  - (1.5em, 0ex)$);
    
    % assignment3

    \node [provides-point] (assign3-pspec) [right=2em of seq3-rspec1.east] {%
      \hspace{.25em}\(\vphantom{\ssbr{}}\osp_{15}\)%
    };
    \path (seq3-rspec1) -- node {\hspace{.5em}\(\sqsubseteq\)} (assign3-pspec);

    \node (assign3) [right=.5em of assign3-pspec.east] {\(q \coloneqq q + 1\)};
    
    \node (assign3-orc) [fit=(assign3)] {};

    \coordinate (assign3-north) at ($(assign3-pspec.north) + (0em, 1ex)$);
    \coordinate (assign3-south) at ($(assign3-pspec.south) - (0em, 1ex)$);
    \coordinate (assign3-west)  at ($(assign3-pspec.west)  + (1.5em, 0ex)$);
    \coordinate (assign3-east)  at ($(assign3-orc.east)    + (.125em, 0ex)$);
    
    % assignment4

    \node [provides-point] (assign4-pspec) [right=2em of seq3-rspec2.east] {%
      \hspace{.25em}\(\vphantom{\ssbr{}}\osp_{16}\)%
    };
    \path (seq3-rspec2) -- node {\hspace{.5em}\(\sqsubseteq\)} (assign4-pspec);

    \node (assign4) [right=.5em of assign4-pspec.east] {\(r \coloneqq r - y\)};
    
    \node (assign4-orc) [fit=(assign4)] {};

    \coordinate (assign4-north) at ($(assign4-pspec.north) + (0em, 1ex)$);
    \coordinate (assign4-south) at ($(assign4-pspec.south) - (0em, 1ex)$);
    \coordinate (assign4-west)  at ($(assign4-pspec.west)  + (1.5em, 0ex)$);
    \coordinate (assign4-east)  at ($(assign4-orc.east)    + (.125em, 0ex)$);
    
    \begin{pgfonlayer}{background} 
      \node [orchestration] [fit=(app-orc) (app-north) (app-south) (app-west) (app-east)] {};
      \node [orchestration] [fit=(seq1-orc) (seq1-north) (seq1-south) (seq1-west) (seq1-east)] {};
      \node [orchestration] [fit=(seq2-orc) (seq2-north) (seq2-south) (seq2-west) (seq2-east)] {};
      \node [orchestration] [fit=(assign1-orc) (assign1-north) (assign1-south) (assign1-west) (assign1-east)] {};
      \node [orchestration] [fit=(assign2-orc) (assign2-north) (assign2-south) (assign2-west) (assign2-east)] {};
      \node [orchestration] [fit=(iter-orc) (iter-north) (iter-south) (iter-west) (iter-east)] {};
      \node [orchestration] [fit=(seq3-orc) (seq3-north) (seq3-south) (seq3-west) (seq3-east)] {};
      \node [orchestration] [fit=(assign3-orc) (assign3-north) (assign3-south) (assign3-west) (assign3-east)] {};
      \node [orchestration] [fit=(assign4-orc) (assign4-north) (assign4-south) (assign4-west) (assign4-east)] {};
    \end{pgfonlayer}
  \end{tikzpicture}
  
  \caption{The derivation of a program that computes the quotient \(q\) and the remainder \(r\) obtained from the division of \(x\) by \(y\)}
  \label{figure:program-derivation}

  \begin{alignat*}{4}
    & \osp_{1}  && {} \colon \pltrue, \ssbr{x = q * y + r} \pland \ssbr{r < y} & \qquad
    & \osp_{9}  && {} \colon \ssbr{x = q * y + x}, \ssbr{x = q * y + r}
    \\
    & \osp_{2}  && {} \colon \pltrue, \ssbr{x = q * y + r} \pland \ssbr{r < y} &
    & \osp_{10} && {} \colon \ssbr{x = q * y + r}, \ssbr{x = q * y + r} \pland \plnot \ssbr{y \leq r}
    \\
    & \osp_{3}  && {}  \colon \pltrue, \ssbr{x = q * y + r} &
    & \osp_{11} && {} \colon \ssbr{x = q * y + r} \pland \ssbr{y \leq r}, \ssbr{x = q * y + r}
    \\
    & \osp_{4}  && {} \colon \ssbr{x = q * y + r}, \ssbr{x = q * y + r} \pland \ssbr{r < y} &
    & \osp_{12} && {} \colon \ssbr{x = \rbr{q + 1} * y + \rbr{r - y}}, \ssbr{x = q * y + r}
    \\
    & \osp_{5}  && {} \colon \pltrue, \ssbr{x = q * y + r} &
    & \osp_{13} && {} \colon \ssbr{x = \rbr{q + 1} * y + \rbr{r - y}}, \ssbr{x = q * y + \rbr{r - y}}
    \\
    & \osp_{6}  && {} \colon \pltrue, \ssbr{x = q * y + x} &
    & \osp_{14} && {} \colon \ssbr{x = q * y + \rbr{r - y}}, \ssbr{x = q * y + r}
    \\
    & \osp_{7}  && {} \colon \ssbr{x = q * y + x}, \ssbr{x = q * y + r} &
    & \osp_{15} && {} \colon \ssbr{x = \rbr{q + 1} * y + \rbr{r - y}}, \ssbr{x = q * y + \rbr{r - y}}
    \\
    & \osp_{8}  && {} \colon \ssbr{x = 0 * y + x}, \ssbr{x = q * y + x} &
    & \osp_{16} && {} \colon \ssbr{x = q * y + \rbr{r - y}}, \ssbr{x = q * y + r}
  \end{alignat*}
\end{sidewaysfigure}
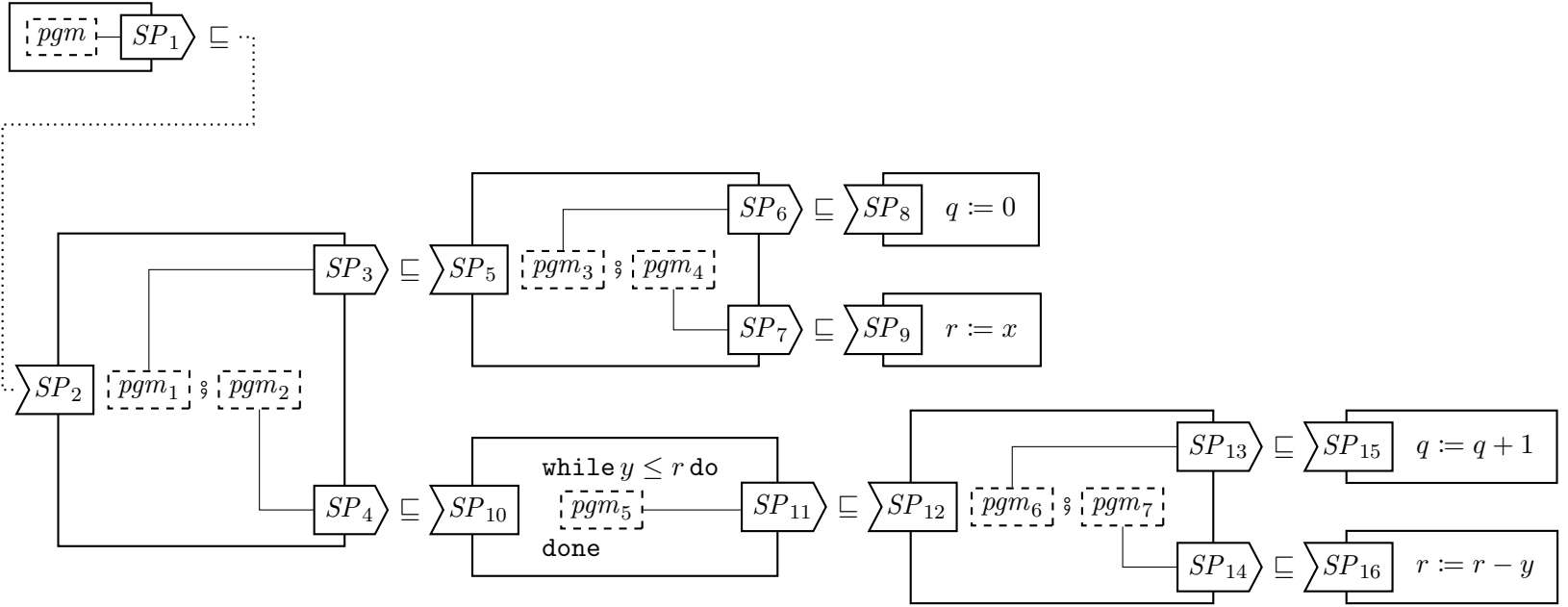

The view that program expressions can be seen as defining `service orchestrations' through which structured programs can be built in a compositional way originates from~\cite{Fiadeiro:The-many-faces-of-complexity-in-software-design-2012}.
Intuitively, we can see the rules of the Hoare calculus~\cite{Hoare:An-axiomatic-basis-for-computer-programming-1969} as defining `clauses' in the sense of logic programming, where unification is controlled through the refinement of pre/post-conditions as specifications of provided/required services, and resolution binds program statements (terms) to variables in program expressions.
In Figure~\ref{figure:program-modules} we depict Hoare rules in a notation that is closer to that of service modules, which also brings out their clausal form: the specification (a pair of a pre- and a post-condition) on the left-hand side corresponds to the consequent of the clause (which relates to a `provides-point' of the service), while those on the right-hand side correspond to the antecedent of the clause (i.e.\ to the `requires-points' of the service) -- the specifications of what remains to be discovered and bound to the program expression (the `service orchestration' inside the box) to produce a program.
In Figure~\ref{figure:program-derivation}, we retrace Hoare's original example of constructing a program that computes the quotient and the remainder resulting from the division of two natural numbers as an instance of the unification and resolution mechanisms particular to logic programming.
We will further discuss these mechanisms in more detail in Subsection~\ref{subsection:resolution-as-service-discovery-and-binding}.

\begin{figure}[h]
  \centering
  
  \begin{minipage}{.44\textwidth}
    \centering
    
    \begin{tikzpicture}
      \node (skip) {\(\opname{skip}\)};
      
      \node (orc) [fit=(skip)] {};

      \node [provides-point] (pspec) [left=.5em of orc.west] {\mathstrut\hspace{.5em}\(\rho, \rho\)};
      
      \coordinate (north) at ($(pspec.north) + (0em, 1ex)$);
      \coordinate (south) at ($(pspec.south) - (0em, 1ex)$);
      \coordinate (west)  at ($(pspec.west)  + (1.5em, 0ex)$);
      \coordinate (east)  at ($(orc.east)    + (.5em, 0ex)$);
      
      \begin{pgfonlayer}{background} 
        \node [orchestration] [fit=(orc) (north) (south) (west) (east)] {};
      \end{pgfonlayer}
    \end{tikzpicture}
    
    \vskip 1ex

    (empty statement)

    \vskip 4ex

    \begin{tikzpicture}
      \node (assign)  {\(x \coloneqq e\)};
      
      \node (orc) [fit=(assign)] {};

      \node [provides-point] (pspec) [left=.5em of orc.west] {\hspace{.5em}\(\rho\rbr{e}, \rho\rbr{x}\)};

      \coordinate (north) at ($(pspec.north) + (0em, 1ex)$);
      \coordinate (south) at ($(pspec.south) - (0em, 1ex)$);
      \coordinate (west)  at ($(pspec.west)  + (1.5em, 0ex)$);
      \coordinate (east)  at ($(orc.east)    + (.5em, 0ex)$);
      
      \begin{pgfonlayer}{background} 
        \node [orchestration] [fit=(orc) (north) (south) (west) (east)] {};
      \end{pgfonlayer}
    \end{tikzpicture}

    \vskip 1ex

    (assignment)

    \vskip 4ex

    \begin{tikzpicture}
      \node        (comp)                     {\(\comp\)};
      \node [pvar] (pgm1) [left=0em of comp]  {\(\ptm_{1}\)};
      \node [pvar] (pgm2) [right=0em of comp] {\(\ptm_{2}\)};
      
      \coordinate (onorth) at ($(comp.north) + (0em,3.5ex)$);
      \coordinate (osouth) at ($(comp.south) - (0em,3.5ex)$);
      \coordinate (owest)  at (pgm1.west);
      \coordinate (oeast)  at (pgm2.east);

      \node (orc) [fit=(onorth) (osouth) (owest) (oeast)] {};

      \node [provides-point] (pspec) [left=.5em of orc.west] {\hspace{.5em}\(\rho, \rho''\)};

      \node [requires-point] (rspec1) [above right=1ex and .5em of orc.east, minimum width=3.25em] {\(\rho, \rho'\)};
      \draw (pgm1) |- (rspec1.west);
      \node [requires-point] (rspec2) [below right=1ex and .5em of orc.east, minimum width=3.25em] {\(\rho', \rho''\)};
      \draw (pgm2) |- (rspec2.west);

      \coordinate (north) at ($(rspec1.north) + (0em, 1ex)$);
      \coordinate (south) at ($(rspec2.south) - (0em, 1ex)$);
      \coordinate (west)  at ($(pspec.west)   + (1.5em, 0ex)$);
      \coordinate (east)  at ($(rspec1.east)  - (1.5em, 0ex)$);

      \begin{pgfonlayer}{background} 
        \node [orchestration] [fit=(orc) (north) (south) (west) (east)] {};
      \end{pgfonlayer}
    \end{tikzpicture}

    \vskip 1ex

    (sequence)
    
  \end{minipage}
  \hfill
  \begin{minipage}{.55\textwidth}
    \centering

    \begin{tikzpicture}
      \node        (if)                                                                    {\(\opname{if}\, C\, \opname{then}\)};
      \node [pvar] (pgm1)  [below right=0ex and 1em of if.south west, anchor=north west]   {\(\ptm_{1}\)};
      \node        (else)  [below left=0ex and 1em of pgm1.south west, anchor=north west]  {\(\opname{else}\)};
      \node [pvar] (pgm2)  [below right=0ex and 1em of else.south west, anchor=north west] {\(\ptm_{2}\)};
      \node        (endif) [below left=0ex and 1em of pgm2.south west, anchor=north west]  {\(\opname{endif}\)};

      \coordinate (onorth) at (if.north);
      \coordinate (osouth) at (endif.south);
      \coordinate (owest)  at (if.west);
      \coordinate (oeast)  at (if.east);

      \node (orc) [fit=(onorth) (osouth) (owest) (oeast)] {};

      \node [provides-point] (pspec) [left=.5em of orc.west] {\hspace{.5em}\(\rho, \rho'\)};

      \node [requires-point] (rspec1) [above right=1.5ex and .5em of orc.east, minimum width=6.75em] {\(\rho \pland \ssbr{C}, \rho'\)};
      \draw (pgm1) -| (rspec1.west);
      \node [requires-point] (rspec2) [below right=1.5ex and .5em of orc.east, minimum width=6.75em] {\(\rho \pland \plnot \ssbr{C}, \rho'\)};
      \draw (pgm2) -| (rspec2.west);

      \coordinate (north) at ($(rspec1.north) + (0em, 1ex)$);
      \coordinate (south) at ($(rspec2.south) - (0em, 1ex)$);
      \coordinate (west)  at ($(pspec.west)   + (1.5em, 0ex)$);
      \coordinate (east)  at ($(rspec1.east)  - (1.5em, 0ex)$);

      \begin{pgfonlayer}{background} 
        \node [orchestration] [fit=(orc) (north) (south) (west) (east)] {};
      \end{pgfonlayer}
    \end{tikzpicture}

    \vskip 1ex

    (selection)

    \vskip 4ex
    
    \begin{tikzpicture}
      \node        (while)                                                                  {\(\opname{while}\, C\, \opname{do}\)};
      \node [pvar] (pgm)   [below right=0ex and 1em of while.south west, anchor=north west] {\(\ptm\)};
      \node        (done)  [below left=0ex and 1em of pgm.south west, anchor=north west]    {\(\opname{done}\)};

      \coordinate (onorth) at (while.north);
      \coordinate (osouth) at (done.south);
      \coordinate (owest)  at (while.west);
      \coordinate (oeast)  at (while.east);

      \node (orc) [fit=(onorth) (osouth) (owest) (oeast)] {};

      \node [provides-point] (pspec) [left=.5em of orc.west] {\hspace{.5em}\(\rho, \rho \pland \plnot \ssbr{C}\)};

      \node [requires-point] (rspec) [right=.5em of orc.east] {\(\rho \pland \ssbr{C}, \rho\)};
      \draw (pgm) -| (rspec.west);

      \coordinate (north) at ($(rspec.north) + (0em, 1ex)$);
      \coordinate (south) at ($(rspec.south) - (0em, 1ex)$);
      \coordinate (west)  at ($(pspec.west)  + (1.5em, 0ex)$);
      \coordinate (east)  at ($(rspec.east)  - (1.5em, 0ex)$);

      \begin{pgfonlayer}{background} 
        \node [orchestration] [fit=(orc) (north) (south) (west) (east)] {};
      \end{pgfonlayer}
    \end{tikzpicture}

    \vskip 1ex

    (iteration)

  \end{minipage}
  
  \caption{Program modules}
  \label{figure:program-modules}
\end{figure}
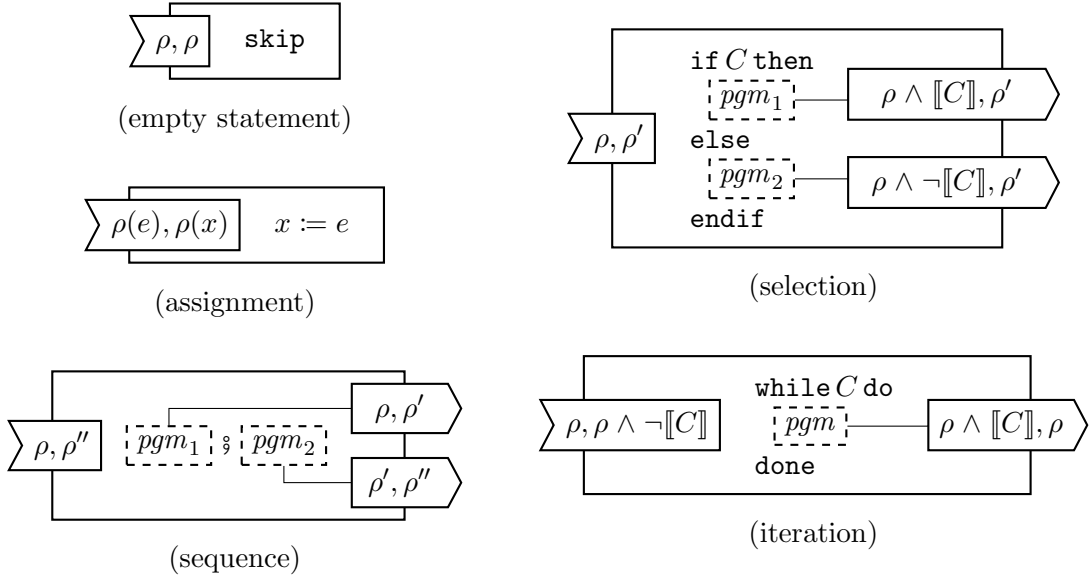

The formal description of program expressions that we consider here follows the presentation given in~\cite{Goguen-Malcolm:Algebraic-Semantics-of-Imperative-Programs-1996} of the algebraic semantics of programs except that, instead of the theory of many-sorted algebra, we rely on the theory of preordered algebra developed in~\cite{Diaconescu-Futatsugi:CafeOBJ-Report-1998}, whose institution we denote by \(\POA\).
In this context, signatures are ordinary algebraic signatures whose denotation is defined over the category of preorders rather than that of sets, with models interpreting the sorts as preordered sets and the operation symbols as monotonic functions.  The sentences are built as in first-order logic based on two kinds of atoms: \emph{equational atoms} \(l = r\) and \emph{preorder atoms} \(l \cto r\), where \(l\) and \(r\) are terms of the same sort; the latter are satisfied by a preordered algebra \(A\) if and only if the interpretations of \(l\) and \(r\) in \(A\) belong to the preorder relation of the carrier of their sort.

In order to fully define the orchestration scheme of program expressions we assume that the programming language we have chosen to analyse is specified through a many-sorted signature \(\abr{S, F}\) equipped with
\begin{itemize}

\item a distinguished set of sorts \(S^{\pgm} \subset S\) corresponding to the types of executable expressions supported by the language, and sorts \(\pstate, \pconfig \in S \setminus S^{\pgm}\) capturing the states of the programs and the various configurations that may arise upon their execution, respectively;

\item operation symbols \(\abr{\_} \colon \pstate \to \pconfig\) and \(\abr{\_, \_} \colon \pexp\, \pstate \to \pconfig\) for sorts \(\pexp \in S^{\pgm}\), which we regard as constructor operators for the sort \(\pconfig\);

\item a (sortwise infinite) \(S^{\pgm}\)\nb-indexed set \(\pVar\) of program variables, and state variables \(\pst, \pst' \colon \pstate\), used to refer to the states that precede or result from executions; and

\item a preordered \(\abr{S, F}\)\nb-algebra \(A\) that describes the semantics of the programming language through the preorder relation associated with the sort \(\pconfig\).\footnote{Alternatively, one could use a theory presentation or a structured specification instead of the algebra \(A\).}

\end{itemize}

\begin{exa}
  \label{example:structured-programs}
  The premises that we consider within this subsection are weak enough to allow the proposed algebraic framework to accommodate a wide variety of programming languages.
  For instance, the program expressions underlying the modules depicted in Figure~\ref{figure:program-modules} are simply terms of sort \(\stname{Pgm}\) that can be formed based on the following five operation symbols (written using the mixfix notation of \CafeOBJ~\cite{Diaconescu-Futatsugi:CafeOBJ-Report-1998} and \CASL~\cite{Mosses:CASL-RM-2004}):
  \begin{iflalign*}
    \text{(empty statement)} \quad & \opname{skip} \colon \to \stname{Pgm}, & \\
    \text{(assignment)}      \quad & \_ \coloneqq \_ \colon \stname{Id}\, \stname{AExp} \to \stname{Pgm}, \\
    \text{(sequence)}        \quad & \_ \comp \_ \colon \stname{Pgm}\, \stname{Pgm} \to \stname{Pgm}, \\
    \text{(selection)}       \quad & \opname{if}\,\_\,\opname{then}\,\_\,\opname{else}\,\_\,\opname{endif} \colon \stname{Cond}\, \stname{Pgm}\, \stname{Pgm} \to \stname{Pgm}, \\
    \text{(iteration)}       \quad & \opname{while}\,\_\,\opname{do}\,\_\,\opname{done} \colon \stname{Cond}\, \stname{Pgm} \to \stname{Pgm}.
  \end{iflalign*}
  
  To simplify our presentation, we omit the details associated with the sorts \(\stname{Id}\) of identifiers, \(\stname{AExp}\) of arithmetic expressions and \(\stname{Cond}\) of conditions;  we also tacitly assume that the signature under consideration  declares the usual operation symbols associated with the names of identifiers, the addition, subtraction and multiplication of arithmetic expressions, and with the atoms and Boolean connectives specific to conditions.
  Moreover, we assume the essential sorts \(\stname{State}\) and \(\stname{Config}\) to be defined, as well as the operation symbols \(\abr{\_}\) and \(\abr{\_, \_}\).
\end{exa}

Algebraic signatures having the aforementioned additional structure induce orchestration schemes in a canonical way, as follows.

\begin{minisection}{Orchestrations.}%
  The \emph{orchestrations} are program expressions, that is \(\abr{S, F \cup \pVar}\)\nb-terms \(\ptm \colon \pexp\), usually denoted simply by \(\ptm\) if there is no danger of confusion, such that \(\pexp\) is a sort in \(S^{\pgm}\).
  The arrows through which they are linked  generalize the subterm relations; in this sense, a \emph{morphism} \(\abr{\psi, \pi}\) between programs \(\ptm_{1} \colon \pexp_{1}\) and \(\ptm_{2} \colon \pexp_{2}\) consists of 
  \begin{itemize}

  \item a substitution \(\psi \colon \pvar\rbr{\ptm_{1}} \to \pvar\rbr{\ptm_{2}}\), mapping the variables that occur in \(\ptm_{1}\) to program expressions defined over the variables of \(\ptm_{2}\), together with

  \item a position \(\pi\) in \(\ptm_{2}\), i.e.\ a sequence of natural numbers that precisely identifies a particular occurrence of a subterm \(\ptm_{2} \reduct_{\pi}\) of \(\ptm_{2}\),

  \end{itemize}
  such that \(\psi^{\sigtm}\rbr{\ptm_{1}} = \ptm_{2} \reduct_{\pi}\).\footnote{Here, we let \(\psi^{\sigtm}\) denote the canonical extension of the substitution \(\psi\) from variables to terms.}
  Their \emph{composition} is defined componentwise, in a way that ensures the commutativity of the following diagram.
  \[
  \xymatrix @C+=5em {
    {\ptm_{1} \colon \pexp_{1}}
    \ar [r] ^-{\abr{\psi_{1}, \pi_{1}}}
    \ar @{->} `d[rr] `[rr]_-{\abr{\psi_{1} \comp \psi_{2}, \pi_{2} \cdot \pi_{1}}} [rr]
    & {\ptm_{2} \colon \pexp_{2}}
    \ar [r] ^-{\abr{\psi_{2}, \pi_{2}}}
    & {\ptm_{3} \colon \pexp_{3}}
  }
  \]
\end{minisection}

\begin{minisection}{Specifications.}%
  For each program expression \(\ptm \colon \pexp\), a \emph{(program) specification} is a triple of the form \(\pspec{\iota}{\rho}{\rho'}\), where \(\iota\) is a position in \(\ptm\) indicating the `subprogram' of \(\ptm\) whose behaviour is being analysed,\footnote{The first component of specifications may be encountered in the literature (e.g.\ in~\cite{Morgan:Programming-from-Specifications-1994}) with a different meaning: the set of identifiers whose values may change during the execution of the program.} and \(\rho\) and \(\rho'\) are \emph{pre-} and \emph{post-conditions} associated with \(\ptm \reduct_{\iota}\), formalized as (quantifier-free) \(\POA\)\nb-sentences over the signature \(\abr{S, F \cup \cbr{\pst \colon \pstate}}\).
  The intuitive interpretation is the usual one: 
  \begin{quotation}
    \noindent Whenever the program \(\ptm \reduct_{\iota}\) is executed in an initial state that satisfies the pre-condition \(\rho\), and the execution terminates, the resulting final state satisfies the post-condition \(\rho'\).
  \end{quotation}
  Note, however, that specifications cannot be evaluated over arbitrary program expressions because, due to the presence of program variables (from \(\pVar\)), some of the programs may not support a well-defined notion of execution.
  We will address this aspect in Section~\ref{section:service-discovery-and-binding} by taking into account translations of specifications along morphisms whose codomains are ground program expressions.
  For now, it suffices to mention that the translation of a program specification \(\pspec{\iota}{\rho}{\rho'}\) of \(\ptm_{1} \colon \pexp_{1}\) along a morphism \(\abr{\psi, \pi} \colon \rbr{\ptm_{1} \colon \pexp_{1}} \to \rbr{\ptm_{2} \colon \pexp_{2}}\) is defined as the specification \(\pspec{\rbr{\pi \cdot \iota}}{\psi\rbr{\rho}}{\psi\rbr{\rho'}}\) of \(\ptm_{2} \colon \pexp_{2}\).
\end{minisection}

\begin{minisection}{Ground orchestrations and properties.}%
  As expected, \emph{ground program expressions} are just program expressions that do not contain variables: \(\abr{S, F}\)\nb-terms \(\ptm \colon \pexp\) whose sort \(\pexp\) belongs to \(S^{\pgm}\).
  Consequently, they have a well-defined operational semantics, which means that we can check whether or not they meet the requirements of a given specification.

  A specification \(\pspec{\iota}{\rho}{\rho'}\) is a \emph{property} of a ground program expression \(\ptm \colon \pexp\) if and only if the following satisfaction condition holds for the preordered algebra \(A\):
  \[
  A \models^{\POA} \univqs{\cbr{\pst, \pst' \colon \pstate}}{\rbr{\rho\rbr{\pst} \pland \abr{\ptm \reduct_{\iota}, \pst} \cto \abr{\pst'}} \plimplies \rho'\rbr{\pst'}}.
  \]
  To keep the notation simple and, at the same time, emphasize the roles of \(\pst\) and \(\pst'\), we used \(\rho\rbr{\pst}\) in the above \(\POA\)\nb-sentence as another name for \(\rho\), while \(\rho'\rbr{\pst'}\) is the sentence derived from \(\rho'\) by replacing the variable \(\pst\) with \(\pst'\).\footnote{Formally, the sentences \(\rho\rbr{\pst}\) and \(\rho'\rbr{\pst'}\) are obtained by translating \(\rho\) and \(\rho'\) along the \(\abr{S, F}\)\nb-substitutions \(\cbr{\pst} \to \cbr{\pst, \pst'}\) given by \(\pst \mapsto \pst\) and \(\pst \mapsto \pst'\), respectively.}
  The same notational convention is used in Figure~\ref{figure:program-modules} to represent the specification attached to the assignment expression.  In that case, \(\rho\) is assumed to be a sentence defined not only over \(\pst \colon \stname{State}\), but also over a variable \(v \colon \stname{AExp}\); the sentences \(\rho\rbr{e}\) and \(\rho\rbr{x}\) are then derived from \(\rho\) by replacing \(v\) with \(e\) and \(x\) (regarded as an atomic arithmetic expression), respectively.
  Another notation used in Figure~\ref{figure:program-modules} (and also in Figure~\ref{figure:program-derivation}) is \(\ssbr{C}\), where \(C\) is a term of sort \(\stname{Cond}\); this follows Iverson's convention (see~\cite{Iverson:A-Programming-Language-1962}, and also~\cite{Graham-Knuth-Patashnik:Concrete-Mathematics-1994}), and corresponds to an atomic \(\POA\)\nb-sentence that captures the semantics of the condition \(C\).
  
  We conclude the presentation of orchestrations as program expressions with Proposition~\ref{proposition:orchestration-scheme-of-program-expressions} below, which guarantees that properties form natural subsets of the sets of specifications; in other words, the morphisms of ground programs preserve properties.

  \begin{prop}
    \label{proposition:orchestration-scheme-of-program-expressions}
    Let \(\abr{\psi, \pi} \colon \rbr{\ptm_{1} \colon \pexp_{1}} \to \rbr{\ptm_{2} \colon \pexp_{2}}\) be a morphism of ground programs.
    For every property \(\pspec{\iota}{\rho}{\rho'}\) of \(\ptm_{1} \colon \pexp_{1}\), the specification \(\SSpec\rbr{\psi, \pi}\rbr{\pspec{\iota}{\rho}{\rho'}}\) is a property of \(\ptm_{2} \colon \pexp_{2}\).
  \end{prop}
  
  \proof
  By the definition of the translation of specifications along morphisms of program expressions, \(\SSpec\rbr{\psi, \pi}\rbr{\pspec{\iota}{\rho}{\rho'}}\) is a property of \(\ptm_{2} \colon \pexp_{2}\) if and only if
  \[
  A \models^{\POA} \univqs{\cbr{\pst, \pst' \colon \pstate}}{
    \rbr{
      \underbracket[.11ex][.5ex]{
        \psi\rbr{\rho}\rbr{\pst}
      }_{\rho\rbr{\pst}}
      \pland
      \abr{
        \underbracket[.11ex][.5ex]{
          \ptm_{2} \reduct_{\pi \cdot \iota}
        }_{\ptm_{1} \reduct_{\iota}}, \pst} \cto \abr{\pst'}
    }
    \plimplies 
    \underbracket[.11ex][.5ex]{
      \psi\rbr{\rho'}\rbr{\pst'}
    }_{\rho'\rbr{\pst'}}
  }.
  \]
  To prove this, notice that all morphisms of ground program expressions share the same underlying substitution: the identity of \(\emptyset\).
  Therefore, \(\psi\rbr{\rho} = \rho\), \(\psi\rbr{\rho'} = \rho'\), and \(\ptm_{2} \reduct_{\pi \cdot \iota} = \ptm_{2} \reduct_{\pi} \reduct_{\iota} = \psi^{\sigtm}\rbr{\ptm_{1}} \reduct_{\iota} = \ptm_{1} \reduct_{\iota}\), from which we immediately deduce that both the evaluation of \(\pspec{\iota}{\rho}{\rho'}\) in \(\ptm_{1} \colon \pexp_{1}\) and that of \(\SSpec\rbr{\psi, \pi}\rbr{\pspec{\iota}{\rho}{\rho'}}\) in \(\ptm_{2} \colon \pexp_{2}\) correspond to the satisfaction by \(A\) of the same \(\POA\)\nb-sentence.
  \qed
\end{minisection}

\subsection{Asynchronous Relational Networks}
\label{subsection:asynchronous-relational-networks}

Asynchronous relational networks as developed in~\cite{Fiadeiro-Lopes:An-interface-theory-for-service-oriented-design-2013} uphold a significantly different perspective on services:  the emphasis is put not on the role of services in addressing design-time organisational aspects of complex, interconnected systems, but rather on their role in managing the run-time interactions that are involved in such systems.
In this paper, we consider a variant of the original theory of asynchronous relational networks that relies on hypergraphs instead of graphs, and uses \(\omega\)\nb-automata~\cite{Thomas:Automata-on-infinite-objects-1990} (see also~\cite{Perrin-Pin:Infinite-Words-2004}) instead of sets of traces as models of behaviour.

The notions discussed within this context depend upon elements of linear temporal logic, and are introduced through dedicated syntactic structures that correspond to specific temporal signatures and signature morphisms.
However, the proposed theory is largely independent of any logical framework of choice -- similarly to the way in which program expressions can be defined over a variety of algebraic signatures -- and can be easily adapted to any institution for which
\begin{enumerate}

\item\label{assumption:ARN-first} the category of signatures is (finitely) cocomplete;

\item there exist cofree models along every signature morphism, meaning that the reduct functors determined by signature morphisms admit right adjoints;

\item the category of models of every signature has (finite) products;

\item\label{assumption:ARN-last} all model homomorphisms reflect the satisfaction of sentences.

\end{enumerate}

In addition to the above requirements, we implicitly assume, as is often done in institutions (see, for example,~\cite{Diaconescu:Institution-Independent-Model-Theory-2008} and~\cite{Sannella-Tarlecki:Foundations-of-Algebraic-Specification-2011} for more details), that the considered logical system is closed under isomorphisms, meaning that the satisfaction of sentences is invariant with respect to isomorphisms of models.  This property holds in most institutions; in particular, it holds in the variant of temporal logic that we use here as a basis for the construction of the orchestration scheme of asynchronous relational networks.

\subsubsection*{Linear Temporal Logic}

In order to capture a more operational notion of service orchestration, we work with an automata-based variant of the institution \(\LTL\) of linear temporal logic~\cite{Fiadeiro-Costa:A-duality-between-specifications-and-models-of-process-behaviour-1996}.  This logical system, denoted \(\aLTL\), has the same syntax as \(\LTL\), which means that signatures are arbitrary sets of \emph{actions}, and that signature morphisms are just functions.  With respect to sentences, for any signature \(A\), the set of \(A\)\nb-sentences is defined as the least set containing the actions in \(A\) that is closed under standard Boolean connectives\footnote{For convenience, we assume that disjunctions, denoted \(\bigplor E\), and conjunctions, denoted \(\bigpland E\), are defined over arbitrary finite sets of sentences \(E\), and we abbreviate \(\bigwedge \cbr{\rho_{1}, \rho_{2}}\) as \(\rho_{1} \pland \rho_{2}\) and \(\bigpland \emptyset\) as \(\pltrue\).} and under the temporal operators \emph{next} (\(\ltlnext \_\)) and \emph{until} (\(\_ \ltluntil \_\)).
As usual, the derived temporal sentences \(\ltleventually \rho\) and \(\ltlalways \rho\) stand for \(\pltrue \ltluntil \rho\) and \(\lnot \rbr{\pltrue \ltluntil \lnot \rho}\), respectively.

{\sloppy The semantics of \(\aLTL\) is defined over (non-deterministic
  finite-state) Muller automata \cite{Muller:Infinite-sequences-and-finite-machines-1963} instead of the more conventional temporal models.  This means that, in the present setting, the models of a signature \(A\) are \emph{Muller automata} \(\Lambda = \abr{Q, \Pset\rbr{A}, \Delta, I, \FS}\), which consist of a (finite) set \(Q\) of \emph{states}, an \emph{alphabet} \(\Pset\rbr{A}\), a \emph{transition relation} \(\Delta \subseteq Q \times \Pset\rbr{A} \times Q\), a subset \(I \subseteq Q\) of \emph{initial states}, and a subset \(\FS \subseteq \Pset\rbr{Q}\) of (non-empty) \emph{final-state sets}.}

The satisfaction relation is based on that of \(\LTL\): an automaton \(\Lambda\) satisfies a sentence \(\rho\) if and only if every trace accepted by \(\Lambda\) satisfies \(\rho\) in the sense of \(\LTL\).
To be more precise, let us first recall that a \emph{trace} over \(A\) is an (infinite) sequence \(\lambda \in \Pset\rbr{A}^{\omega}\), and that a \emph{run} of an automaton \(\Lambda\) defined as above on a trace \(\lambda\) is a state sequence \(\varrho \in Q^{\omega}\) such that \(\varrho\rbr{0} \in I\) and \(\rbr{\varrho\rbr{i}, \lambda\rbr{i}, \varrho\rbr{i + 1}} \in \Delta\) for every \(i \in \omega\).
A run \(\varrho\) is said to be \emph{successful} if its infinity set, i.e.\ the set of states that occur infinitely often in \(\varrho\), denoted \(\Inf\rbr{\varrho}\), is a member of \(\FS\).
Then a trace \(\lambda\) is \emph{accepted} by \(\Lambda\) if and only if there exists a successful run of \(\Lambda\) on \(\lambda\).
Finally, given a trace \(\lambda\) (that can be presumed to be accepted by \(\Lambda\)) and \(i \in \omega\), we use the notation \(\infssuffix{\lambda}{i}\) to indicate the suffix of \(\lambda\) that starts at \(\infsat{\lambda}{i}\).
The satisfaction of temporal sentences by traces can now be defined by structural induction, as follows:
\begin{iflalign*}[\parindent]
  & \lambda \models a\ \text{if and only if}\ a \in \infsat{\lambda}{0}, && \\
  & \lambda \models \plnot \rho\ \text{if and only if}\ \lambda \nmodels \rho, & \\
  & \lambda \models \textstyle \bigplor E\ \text{if and only if}\ \lambda \models \rho\ \text{for some}\ \rho \in E, & \\
  & \lambda \models \ltlnext \rho\ \text{if and only if}\ \infssuffix{\lambda}{1} \models \rho,\ \text{and} & \\
  & \lambda \models \rho_{1} \ltluntil \rho_{2}\ \text{if and only if}\ \infssuffix{\lambda}{i} \models \rho_{2}\ \text{for some}\ i \in \omega,\ \text{and}\ \infssuffix{\lambda}{j} \models \rho_{1}\ \text{for all}\ j < i, &
\end{iflalign*}
where \(a\) is an action in \(A\), \(\rho\), \(\rho_{1}\) and \(\rho_{2}\) are \(A\)\nb-sentences, and \(E\) is a set of \(A\)\nb-sentences.

One can easily see that the first of the hypotheses~\ref{assumption:ARN-first}--\ref{assumption:ARN-last} that form the basis of the present study of asynchronous relational networks is satisfied by \(\aLTL\), as it corresponds to a well-known result about the existence of small colimits in \(\Set\).
In order to check that the remaining three properties hold as well, let us first recall that a \emph{homomorphism} \(h \colon \Lambda_{1} \to \Lambda_{2}\) between Muller automata \(\Lambda_{1} = \abr{Q_{1}, \Pset\rbr{A}, \Delta_{1}, I_{1}, \FS_{1}}\) and \(\Lambda_{2} = \abr{Q_{2}, \Pset\rbr{A}, \Delta_{2}, I_{2}, \FS_{2}}\) (over the same alphabet) is a function \(h \colon Q_{1} \to Q_{2}\) such that \(\rbr{h\rbr{p}, \alpha, h\rbr{q}} \in \Delta_{2}\) whenever \(\rbr{p, \alpha, q} \in \Delta_{1}\), \(h\rbr{I_{1}} \subseteq I_{2}\), and \(h\rbr{\FS_{1}} \subseteq \FS_{2}\).
We also note that for any map \(\sigma \colon A \to A'\), i.e.\ for any signature morphism, and any Muller automaton \(\Lambda' = \abr{Q', \Pset\rbr{A'}, \Delta', I', \FS'}\), the \emph{reduct} \(\Lambda' \reduct_{\sigma}\) is the automaton \(\abr{Q', \Pset\rbr{A}, \Delta' \reduct_{\sigma}, I', \FS'}\) with the same states, initial states and final-state sets as \(\Lambda'\), and with the transition relation given by \(\Delta' \reduct_{\sigma} = \cbr{\rbr{p', \sigma^{-1}\rbr{\alpha'}, q'} \st \rbr{p', \alpha', q'} \in \Delta'}\).

The following results enable us to use the institution \(\aLTL\) as a foundation for the subsequent development of asynchronous relational networks.
In particular, Proposition~\ref{proposition:cofree-models-in-aLTL} ensures the existence of cofree Muller automata along signature morphisms; Proposition~\ref{proposition:products-of-models-in-aLTL} allows us to form products of Muller automata based on a straightforward categorical interpretation of the fact that the sets of traces accepted by Muller automata, i.e.\ regular \(\omega\)\nb-languages, are closed under intersection; and finally, Proposition~\ref{proposition:reflection-of-the-satisfaction-of-sentences-in-aLTL} guarantees that all model homomorphisms reflect the satisfaction of temporal sentences.

\begin{prop}
  \label{proposition:cofree-models-in-aLTL}
  For every morphism of \(\aLTL\)\nb-signatures \(\sigma \colon A \to A'\), the reduct functor \(\_ \reduct_{\sigma} \colon \Mod^{\aLTL}\rbr{A'} \to \Mod^{\aLTL}\rbr{A}\) admits a right adjoint, which we denote by \(\rbr{\_}^{\sigma}\).
\end{prop}

\proof
According to a general result about adjoints, it suffices to show that for any automaton \(\Lambda\) over the alphabet \(\Pset\rbr{A}\) there exists a universal arrow from \(\_ \reduct_{\sigma}\) to \(\Lambda\).

Let us thus consider a Muller automaton \(\Lambda = \abr{Q, \Pset\rbr{A}, \Delta, I, \FS}\) over \(\Pset\rbr{A}\).
We define the automaton \(\Lambda^{\sigma} = \abr{Q, \Pset\rbr{A'}, \Delta^{\sigma}, I, \FS}\) over the alphabet \(\Pset\rbr{A'}\) by
\[
\Delta^{\sigma} = \cbr{\rbr{p, \alpha', q} \st \rbr{p, \sigma^{-1}\rbr{\alpha'}, q} \in \Delta}.
\]
It is straightforward to verify that the identity map \(1_{Q}\) defines a homomorphism of automata \(\Lambda^{\sigma} \reduct_{\sigma} \to \Lambda\): for any transition \(\rbr{p, \alpha, q} \in \Delta^{\sigma} \reduct_{\sigma}\), by the definition of the reduct functor \(\_ \reduct_{\sigma}\), there exists a set \(\alpha' \subseteq A'\) such that \(\sigma^{-1}\rbr{\alpha'} = \alpha\) and \(\rbr{p, \alpha', q} \in \Delta^{\sigma}\); given the definition above of \(\Delta^{\sigma}\), it follows that \(\rbr{p, \sigma^{-1}\rbr{\alpha'}, q} \in \Delta\), and hence \(\rbr{p, \alpha, q} \in \Delta\).
\[
\xymatrix {
  {\Lambda}
  & {\Lambda^{\sigma} \reduct_{\sigma}}
  \ar [l] _{1_{Q}}
  & *+[r]{\Lambda^{\sigma}}
  \\
  & {\Lambda' \reduct_{\sigma}}
  \ar [u] _{h}
  \ar [ul] ^{h}
  & *+[r]{\Lambda'}
  \ar [u] _{h}
}
\]

Let us now assume that \(h \colon \Lambda' \reduct_{\sigma} \to \Lambda\) is another homomorphism of automata, with \(\Lambda' = \abr{Q', \Pset\rbr{A'}, \Delta', I', \FS'}\).
Then for any transition \(\rbr{p', \alpha', q'} \in \Delta'\), by the definition of the functor \(\_ \reduct_{\sigma}\), we have \(\rbr{p', \sigma^{-1}\rbr{\alpha'}, q'} \in \Delta' \reduct_{\sigma}\).
Based on the homomorphism property of \(h\), it follows that \(\rbr{h\rbr{p'}, \sigma^{-1}\rbr{\alpha'}, h\rbr{q'}} \in \Delta\), which further implies, by the definition of \(\Delta^{\sigma}\), that \(\rbr{h\rbr{p'}, \alpha', h\rbr{q'}} \in \Delta^{\sigma}\).
As a result, the map \(h\) is also a homomorphism of automata \(\Lambda' \to \Lambda^{\sigma}\).
Even more, it is obviously the unique homomorphism \(\Lambda' \to \Lambda^{\sigma}\) (in the category of automata over \(\Pset\rbr{A'}\)) such that \(h \comp 1_{Q} = h\) in the category of automata over \(\Pset\rbr{A}\).
\qed

\begin{prop}
  \label{proposition:products-of-models-in-aLTL}
  For any set of actions \(A\), the category \(\Mod^{\aLTL}\rbr{A}\) of Muller automata defined over the alphabet \(\Pset\rbr{A}\) admits (finite) products.
\end{prop}

\proof
Let \(\rbr{\Lambda_{i}}_{i \in J}\) be a (finite) family of Muller automata over the alphabet \(\Pset\rbr{A}\), with \(\Lambda_{i}\) given by \(\abr{Q_{i}, \Pset\rbr{A}, \Delta_{i}, I_{i}, \FS_{i}}\). We define the automaton \(\Lambda = \abr{Q, \Pset\rbr{A}, \Delta, I, \FS}\) by
\begin{iflalign*}
  Q & = \textstyle \prod_{i \in J} Q_{i}, & \\
  \Delta & = \cbr{\rbr{p, \alpha, q} \st \rbr{p\rbr{i}, \alpha, q\rbr{i}} \in \Delta_{i}\ \text{for all}\ i \in J}, \\
  I & = \textstyle \prod_{i \in J} I_{i},\ \text{and} \\
  \FS & = \cbr{S \subseteq Q \st \pi_{i}\rbr{S} \in \FS_{i}\ \text{for all}\ i \in J},
\end{iflalign*}
where the functions \(\pi_{i} \colon Q \to Q_{i}\) are the corresponding projections of the Cartesian product \(\prod_{i \in J} Q_{i}\).
By construction, it immediately follows that for every \(i \in J\), the map \(\pi_{i}\) defines a homomorphism of automata \(\Lambda \to \Lambda_{i}\).
Even more, one can easily see that for any other family of homomorphisms \(\rbr{h_{i} \colon \Lambda' \to \Lambda_{i}}_{i \in J}\), with \(\Lambda' = \abr{Q', \Pset\rbr{A'}, \Delta', I', \FS'}\), the unique map \(h \colon Q' \to Q\) such that \(h \comp \pi_{i} = h_{i}\) for all \(i \in J\) defines a homomorphism of automata as well.
Therefore, the automaton \(\Lambda\) and the projections \(\rbr{\pi_{i}}_{i \in J}\) form the product of \(\rbr{\Lambda_{i}}_{i \in J}\).
\qed

\begin{prop}
  \label{proposition:reflection-of-the-satisfaction-of-sentences-in-aLTL}
  Let \(h \colon \Lambda_{1} \to \Lambda_{2}\) be a homomorphism between automata defined over an alphabet \(\Pset\rbr{A}\).
  Every temporal sentence over \(A\) that is satisfied by \(\Lambda_{2}\) is also satisfied by \(\Lambda_{1}\).
\end{prop}

\proof
Suppose that \(\Lambda_{i} = \abr{Q_{i}, \Pset\rbr{A}, \Delta_{i}, I_{i}, \FS_{i}}\), for \(i \in \cbr{1, 2}\).
Since the map \(h \colon Q_{1} \to Q_{2}\) defines a homomorphism of automata, for every successful run \(\varrho \in Q_{1}^{\omega}\) of \(\Lambda_{1}\) on a trace \(\lambda \in \Pset\rbr{A}^{\omega}\), the composition \(\varrho \comp h\) yields a successful run of \(\Lambda_{2}\) on \(\lambda\).
As a result, \(\Lambda_{2}\) accepts all the traces accepted by \(\Lambda_{1}\), which further implies that \(\Lambda_{1}\) satisfies all temporal sentences that are satisfied by \(\Lambda_{2}\).
\qed

\subsubsection*{Service Components}

Following~\cite{Fiadeiro-Lopes:An-interface-theory-for-service-oriented-design-2013}, we regard service components as networks of processes that interact asynchronously by exchanging messages through communication channels.
Messages are considered to be atomic units of communication.  They can be grouped either into sets of messages that correspond to processes or channels, or into specific structures, called ports, through which processes and channels can be interconnected.

The ports can be viewed as sets of messages with attached polarities.  As in~\cite{Brand-Zafiropulo:Communicating-finite-state-machines-1983,Benatallah-Casati-Toumani:Representing-analysing-and-managing-Web-service-protocols-2006} we distinguish between outgoing or published messages (labelled with a minus sign), and incoming or delivered messages (labelled with a plus sign).

\begin{defi}[Port]
  A \emph{port} \(M\) is a pair \(\abr{M^{-}, M^{+}}\) of disjoint (finite) sets of \emph{published} and \emph{delivered messages}.
  The set of all \emph{messages} of \(M\) is given by \(M^{-} \cup M^{+}\) and is often denoted simply by \(M\).
  Every port \(M\) defines the set of \emph{actions} \(A_{M} = A_{M^{-}} \cup A_{M^{+}}\), where
  \begin{itemize}
    
  \item \(A_{M^{-}}\) is the set \(\cbr{\pact{m} \st m \in M^{-}}\) of \emph{publication actions}, and
    
  \item \(A_{M^{+}}\) is the set \(\cbr{\dact{m} \st m \in M^{+}}\) of \emph{delivery actions}.
    
  \end{itemize}
\end{defi}

Processes are defined by sets of interaction points labelled with ports and by automata that describe their behaviour in terms of observable publication and delivery actions.

\begin{defi}[Process]
  \label{definition:process}
  A \emph{process} is a triple \(\abr{X, \rbr{M_{x}}_{x \in X}, \Lambda}\) that consists of a (finite) set \(X\) of \emph{interaction points}, each point \(x \in X\) being labelled with a port \(M_{x}\), and a Muller automaton \(\Lambda\) over the alphabet \(\Pset\rbr{A_{M}}\), where \(M\) is the port given by
  \[
  M^{\mp} = \biguplus_{x \in X} M_{x}^{\mp} = \cbr{x.m \st x \in X, m \in M_{x}^{\mp}}.
  \]
\end{defi}

\begin{exa}
  \label{example:journey-planner-process}
  In Figure~\ref{figure:journey-planner-process} we depict a process \(\processname{JP}\) (for Journey Planner) that provides directions from a source to a target location. The process interacts with the environment by means of two ports, named \(\portname{JP_{1}}\) and \(\portname{JP_{2}}\).
  The first port is used to communicate with potential client processes -- the request for directions (including the source and the target locations) is encoded into the incoming message \(\msgname{planJourney}\), while the response is represented by the outgoing message \(\msgname{directions}\).
  The second port defines messages that \(\processname{JP}\) exchanges with other processes in order to complete its task -- the outgoing message \(\msgname{getRoutes}\) can be seen as a query for all possible routes between the specified source and target locations, while the incoming messages \(\msgname{routes}\) and \(\msgname{timetables}\) define the result of the query and the timetables of the available transport services for the selected routes.

  \begin{figure}[h]
    \centering
    
    \begin{tikzpicture}
      \node [align=center, minimum height=9ex, minimum width=4em] (JP) {
        \(\processname{JP}\) \\[1ex]
        \(\Lambda_{\processname{JP}}\)
      };
      
      \node [port] (JP1) [left=0em of JP, align=right] {
        \(\dmsgr{planJourney}\) \\
        \(\pmsgr{directions}\)
      };
      \node [port-label] [above=0ex of JP1] {\(\portname{JP_{1}}\)};
      
      \node [port] (JP2) [right=0em of JP, align=left] {
        \(\pmsgl{getRoutes}\) \\
        \(\dmsgl{routes}\) \\
        \(\dmsgl{timetables}\)
      };
      \node [port-label] [above=0ex of JP2] {\(\portname{JP_{2}}\)};

      \begin{pgfonlayer}{background}       
        \node [process] [fit=(JP)] {};
      \end{pgfonlayer}
    \end{tikzpicture}

    \caption{The process \(\processname{JP}\)}
    \label{figure:journey-planner-process}
  \end{figure}
  
  The behaviour of \(\processname{JP}\) is given by the Muller automaton depicted in Figure~\ref{figure:journey-planner-automaton}, whose final-state sets contain \(q_{0}\) whenever they contain \(q_{5}\).
  We can describe it informally as follows: whenever the process \(\processname{JP}\) receives a request \(\msgname{planJourney}\) it immediately initiates the search for the available routes by sending the message \(\msgname{getRoutes}\); it then waits for the delivery of the routes and of the corresponding timetables, and, once it receives both, it compiles the directions and replies to the client.

  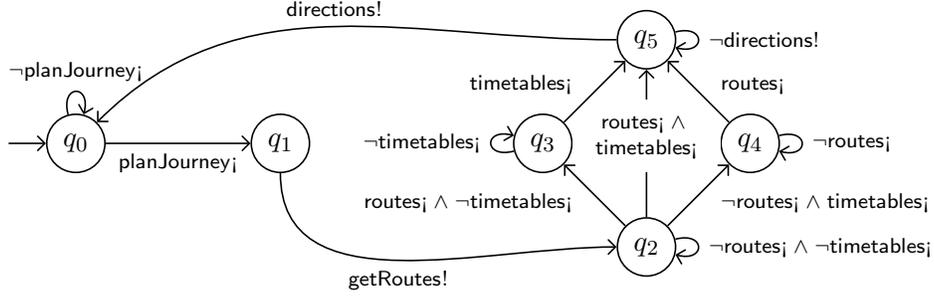
\begin{figure}[h]
    \centering
    
    \begin{tikzpicture}[automaton]
      \node[state, initial, initial text={}] (Qpj)                       {\(q_{0}\)};
      \node[state]                           (Qg)  [right=5em of Qpj]    {\(q_{1}\)};
      \node[state]                           (Qt)  [right=7em of Qg]     {\(q_{3}\)};
      \node[state]                           (Qrt) [below right= of Qt]  {\(q_{2}\)};
      \node[state]                           (Qd)  [above right= of Qt]  {\(q_{5}\)};
      \node[state]                           (Qr)  [above right= of Qrt] {\(q_{4}\)};

      \path [every node/.style={font=\scriptsize}]
      (Qpj) edge [loop above]      node                 {\(\plnot \dact{\msgname{planJourney}}\)}                                      (Qpj)
      (Qpj) edge                   node [swap]          {\(\dact{\msgname{planJourney}}\)}                                             (Qg)
      (Qg)  edge [out=270, in=180] node [swap]          {\(\pact{\msgname{getRoutes}}\)}                                               (Qrt)
      (Qrt) edge [loop right]      node                 {\(\plnot \dact{\msgname{routes}} \pland \plnot \dact{\msgname{timetables}}\)} (Qrt)
      (Qrt) edge                   node [above, pos=.3, fill=white] {
        \(\begin{array}{c}
            \dact{\msgname{routes}} \pland {} \\
            \dact{\msgname{timetables}}
          \end{array}\)
        }                                                                                                                                (Qd)
        (Qrt) edge                   node [swap, pos=.7] {\(\plnot \dact{\msgname{routes}} \pland \dact{\msgname{timetables}}\)}         (Qr)
        (Qrt) edge                   node [pos=.7]       {\(\dact{\msgname{routes}} \pland \plnot \dact{\msgname{timetables}}\)}         (Qt)
        (Qr)  edge                   node [swap, pos=.3] {\(\dact{\msgname{routes}}\)}                                                   (Qd)
        (Qr)  edge [loop right]      node                 {\(\plnot \dact{\msgname{routes}}\)}                                           (Qt)
        (Qt)  edge                   node [pos=.3]       {\(\dact{\msgname{timetables}}\)}                                               (Qd)
        (Qt)  edge [loop left]       node                 {\(\plnot \dact{\msgname{timetables}}\)}                                       (Qt)
        (Qd)  edge [loop right]      node                 {\(\plnot \pact{\msgname{directions}}\)}                                       (Qd)
        (Qd)  edge [out=180, in=45]  node [swap]          {\(\pact{\msgname{directions}}\)}                                              (Qpj);
      \end{tikzpicture}

      \caption{The automaton \(\Lambda_{\processname{JP}}\)\protect\footnotemark}
      \label{figure:journey-planner-automaton}
    \end{figure}
    \footnotetext{In the graphical representation, transitions are labelled with propositional sentences, as in~\cite{Alpern-Schneider:Recognizing-safety-and-liveness-1987}; this means that there exists a transition for any propositional model (i.e.\ set of actions) of the considered sentence.}
  \end{exa}

  \begin{rem}
    \label{remark:abstract-process}
    To generalize Definition~\ref{definition:process} to an arbitrary institution (subject to the four technical assumptions listed at the beginning of the subsection), we first observe that every polarity-preserving map \(\theta\) between ports \(M\) and \(M'\) defines a function \(A_{\theta} \colon A_{M} \to A_{M'}\), i.e.\ a morphism of \(\aLTL\)\nb-signatures, usually denoted simply by \(\theta\), that maps every publication action \(\pact{m}\) to \(\pact{\theta\rbr{m}}\) and every delivery action \(\dact{m}\) to \(\dact{\theta\rbr{m}}\).
    Moreover, for any process \(\abr{X, \rbr{M_{x}}_{x \in X}, \Lambda}\), the injections \(\rbr{x.\_ \colon A_{M_{x}} \to A_{M}}_{x \in X}\) define a coproduct in the category of \(\aLTL\)\nb-signatures.
    This allows us to introduce an abstract notion of process as a triple \(\abr{X, \rbr{\iota_{x} \colon \Sigma_{x} \to \Sigma}_{x \in X}, \Lambda}\) that consists of a set \(X\) of \emph{interaction points}, each point \(x \in X\) being labelled with a \emph{port signature} \(\Sigma_{x}\), a \emph{process signature} \(\Sigma\) together with morphisms \(\iota_{x} \colon \Sigma_{x} \to \Sigma\) for \(x \in X\) (usually defining a coproduct), and a model \(\Lambda\) of \(\Sigma\).
  \end{rem}

  Processes communicate by transmitting messages through channels. As in~\cite{Brand-Zafiropulo:Communicating-finite-state-machines-1983,Fiadeiro-Lopes:An-interface-theory-for-service-oriented-design-2013}, channels are bidirectional: they may transmit both incoming and outgoing messages.

  \begin{defi}[Channel]
    \label{definition:channel}
    A \emph{channel} is a pair \(\abr{M, \Lambda}\) that consists of a (finite) set \(M\) of \emph{messages} and a Muller automaton \(\Lambda\) over the alphabet \(\Pset\rbr{A_{M}}\), where \(A_{M}\) is given by the union \(A_{M}^{-} \cup A_{M}^{+}\) of the sets of actions \(A_{M}^{-} = \cbr{\pact{m} \st m \in M}\) and \(A_{M}^{+} = \cbr{\dact{m} \st m \in M}\).
  \end{defi}

  Note that channels do not provide any information about the communicating entities.  In order to enable given processes to exchange messages, channels need to be attached to their ports, thus forming connections.

  \begin{defi}[Connection]
    \label{definition:connection}
    A \emph{connection} \(\abr{M, \Lambda, \rbr{\mu_{x} \colon M \pto M_{x}}_{x \in X}}\) between the ports \(\rbr{M_{x}}_{x \in X}\) consists of a channel \(\abr{M, \Lambda}\) and a (finite) family of partial \emph{attachment injections} \(\rbr{\mu_{x} \colon M \pto M_{x}}_{x \in X}\) such that \(M = \bigcup_{x \in X} \dom\rbr{\mu_{x}}\) and for any point \(x \in X\),
    \[
    \mu_{x}^{-1}\rbr{M_{x}^{\mp}} \subseteq \bigcup_{y \in X \setminus \cbr{x}} \mu_{y}^{-1}\rbr{M_{y}^{\pm}}.
    \]
  \end{defi}

  \noindent This notion of connection generalizes the one found in~\cite{Fiadeiro-Lopes:An-interface-theory-for-service-oriented-design-2013} so that messages can be transmitted between more than two ports.
  The additional condition ensures in this case that messages are well paired: every published message of \(M_{x}\), for \(x \in X\), is paired with a delivered message of \(M_{y}\), for \(y \in X \setminus \cbr{x}\), and vice versa.
  One can also see that for any binary connection, the attachment injections have to be total functions; therefore, any binary connection is also a connection in the sense of~\cite{Fiadeiro-Lopes:An-interface-theory-for-service-oriented-design-2013}.

  \begin{exa}
    \label{example:journey-planner-connection}
    In order to illustrate how the process \(\processname{JP}\) can send or receive messages, we consider the connection \(\connectionname{C}\) depicted in Figure~\ref{figure:journey-planner-connection} that moderates the flow of messages between the port named \(\portname{JP_{2}}\) and two other ports, named \(\portname{R_{1}}\) and \(\portname{R_{2}}\).

    \begin{figure}[h]
      \centering

      \begin{tikzpicture}
        \node [port] (JP2) [align=left] {
          \(\pmsgl{getRoutes}\) \\
          \(\dmsgl{routes}\) \\
          \(\dmsgl{timetables}\)
        };
        \node [port-label] [above=.5ex of JP2] {\(\portname{JP_{2}}\)};
        
        \node [port] (R1) [above right=-1ex and 3em of JP2, align=right] {
          \(\dmsgr{getRoutes}\) \\
          \(\pmsgr{routes}\)
        };
        \node [port-label] [above=.5ex of R1] {\(\portname{R_{1}}\)};

        \node [port] (R2) [below right=-1ex and 3em of JP2, align=right] {
          \(\dmsgr{routes}\) \\
          \(\pmsgr{timetables}\)
        };
        \node [port-label] [above=.5ex of R2] {\(\portname{R_{2}}\)};

        \draw [rounded corners]
        ($(JP2.north west) + (.5,.1)$)
        to                   ($(JP2.north east) + (-.5,.1)$)
        to [out=0, in=180]   ($(R1.north west)  + (.25,.1)$)
        to                   ($(R1.north east)  + (.1,.1)$)
        to                   ($(R1.south east)  + (.1,-.1)$)
        to                   ($(R1.south west)  + (.25,-.1)$)
        to [out=180, in=180] ($(R2.north west)  + (.25,.1)$)
        to                   ($(R2.north east)  + (.1,.1)$)
        to                   ($(R2.south east)  + (.1,-.1)$)
        to                   ($(R2.south west)  + (.25,-.1)$)
        to [out=180, in=0]   ($(JP2.south east) + (-.5,-.1)$)
        to                   ($(JP2.south west) + (-.1,-.1)$)
        to                   ($(JP2.north west) + (-.1,.1)$)
        to                   ($(JP2.north west) + (.5,.1)$);
        
        \path (JP2) -- node [connection] {
          \(\connectionname{C}\) \\[1ex]
          \(\Lambda_{\connectionname{C}}\)
        } (JP2 -| R1.west);
      \end{tikzpicture}

      \caption{The Journey Planner's connection}
      \label{figure:journey-planner-connection}
    \end{figure}
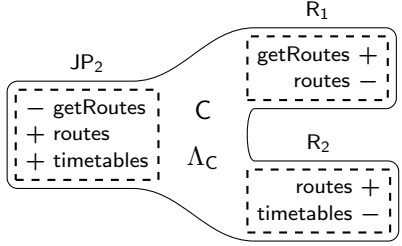
    
    The underlying channel of \(\connectionname{C}\) is given by the set of messages \(M = \cbr{g, r, t}\) together with the automaton \(\Lambda_{\connectionname{C}}\) that specifies the delivery of all published messages without any delay; \(\Lambda_{\connectionname{C}}\) can be built as the product of the automata \(\Lambda_{m}\), for \(m \in M\), whose transition map is depicted in Figure~\ref{figure:connection-LM-automaton}, and whose sets of states are all marked as final.

    \begin{figure}[h]
      \centering
      
      \begin{tikzpicture}[automaton]
        \node[state, initial, initial text={}] (q0)                   {\(q_{0}\)};
        \node[state]                           (q1) [right=5em of q0] {\(q_{1}\)};

        \path [every node/.style={font=\scriptsize}]
        (q0) edge [loop above] node {\(\lnot \pact{m}\)}                (q0)
        (q0) edge [bend left]  node {\(\pact{m}\)}                      (q1)
        (q1) edge [loop above] node {\(\pact{m} \land \dact{m}\)}       (q1)
        (q1) edge [bend left]  node {\(\lnot \pact{m} \land \dact{m}\)} (q0);
      \end{tikzpicture}

      \caption{The automaton \(\Lambda_{m}\)}
      \label{figure:connection-LM-automaton}
    \end{figure}
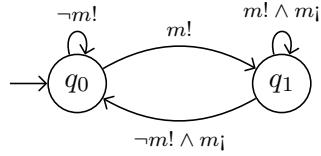

    The channel is attached to the ports \(\portname{JP_{2}}\), \(\portname{R_{1}}\) and \(\portname{R_{2}}\) through the partial injections
    \begin{itemize}
      
    \item \(\mu_{\portname{{JP_{2}}}} \colon M \to M_{\portname{JP_{2}}}\) given by \(g \mapsto \msgname{getRoutes}\), \(r \mapsto \msgname{routes}\) and \(t \mapsto \msgname{timetables}\),
      
    \item \(\mu_{\portname{R_{1}}} \colon M \to M_{\portname{R_{1}}}\) given by \(g \mapsto \msgname{getRoutes}\) and \(r \mapsto \msgname{routes}\), and
      
    \item \(\mu_{\portname{R_{2}}} \colon M \to M_{\portname{R_{2}}}\) given by \(r \mapsto \msgname{routes}\) and \(t \mapsto \msgname{timetables}\).
      
    \end{itemize}
    These injections specify the actual senders and receivers of messages.  For instance, the message \(g\) is delivered only to the port \(\portname{R_{1}}\) (because \(\mu_{\portname{R_{2}}}\) is not defined on \(g\)), whereas \(r\) is simultaneously delivered to both \(\portname{JP_{2}}\) and \(\portname{R_{2}}\).
  \end{exa}

  As already suggested in Examples~\ref{example:journey-planner-process} and~\ref{example:journey-planner-connection}, processes and connections have dual roles, and they interpret the polarities of messages accordingly. In this sense, processes are responsible for publishing messages (i.e.\ they regard delivered messages as inputs and published messages as outputs), while connections are responsible for delivering messages.
  This dual nature of connections can be made explicit by taking into account, for every connection \(\abr{M, \Lambda, \rbr{\mu_{x} \colon M \pto M_{x}}_{x \in X}}\), partial translations \(\rbr{A_{\mu_{x}} \colon A_{M} \pto A_{M_{x}}}_{x \in X}\) of the actions defined by the channel into actions defined by the ports, as follows:
  \begin{iflalign*}
    \dom\rbr{A_{\mu_{x}}} & = \cbr{\pact{m} \st m \in \mu_{x}^{-1}\rbr{M_{x}^{-}}} \cup \cbr{\dact{m} \st m \in \mu_{x}^{-1}\rbr{M_{x}^{+}}}, & \\
    A_{\mu_{x}}\rbr{\pact{m}} & = \pact{\mu_{x}\rbr{m}}\ \text{for all messages}\ m \in \mu_{x}^{-1}\rbr{M_{x}^{-}}, \\
    A_{\mu_{x}}\rbr{\dact{m}} & = \dact{\mu_{x}\rbr{m}}\ \text{for all messages}\ m \in \mu_{x}^{-1}\rbr{M_{x}^{+}}.
  \end{iflalign*}
  We usually designate the partial maps \(A_{\mu_{x}}\) simply by \(\mu_{x}\) if there is no danger of confusion.

  \begin{rem}
    \label{remark:abstract-connection}
    Just as in the case of processes, we can define connections based on an arbitrary logical system, without relying on messages.
    To achieve this goal, note that, in \(\aLTL\), every connection \(\abr{M, \Lambda, \rbr{\mu_{x} \colon M \pto M_{x}}_{x \in X}}\) determines a family of spans 
    \[
    \xymatrix @1 {
      {A_{M}}
      & {\dom\rbr{\mu_{x}}}
      \ar [l] _-{\supseteq}
      \ar [r] ^-{\mu_{x}} 
      & {A_{M_{x}}}
    }
    \]
    indexed by points \(x \in X\).
    Then we can consider connections more generally as triples \(\abr{\Sigma, \Lambda, \rbr{\iota_{x} \colon \Sigma'_{x} \to \Sigma, \mu_{x} \colon \Sigma'_{x} \to \Sigma_{x}}_{x \in X}}\) in which the signature \(\Sigma\) and the model \(\Lambda\) of \(\Sigma\) abstract the \emph{channel} component, and the spans of signature morphisms \(\rbr{\iota_{x}, \mu_{x}}_{x \in X}\) provide the means of attaching port signatures to the channel.
  \end{rem}

  We can now define asynchronous networks of processes as hypergraphs having vertices labelled with ports and hyperedges labelled with processes or connections.

  \begin{defi}[Hypergraph]  
    A \emph{hypergraph} \(\abr{X, E, \gamma}\) consists of a set \(X\) of \emph{vertices} or \emph{nodes}, a set \(E\) of \emph{hyperedges}, disjoint from \(X\), and an \emph{incidence map} \(\gamma \colon E \to \Pset\rbr{X}\), defining for every hyperedge \(e \in E\) a non-empty set \(\gamma_{e} \subseteq X\) of vertices it is incident with.

    A hypergraph \(\abr{X, E, \gamma}\) is said to be \emph{edge-bipartite} if it admits a distinguished partition of \(E\) into subsets \(F\) and \(G\) such that no adjacent hyperedges belong to the same part, i.e.\ for every \(e_{1}, e_{2} \in E\) such that \(\gamma_{e_{1}} \cap \gamma_{e_{2}} \neq \emptyset\), either \(e_{1} \in F\) and \(e_{2} \in G\), or \(e_{1} \in G\) and \(e_{2} \in F\).
  \end{defi}

  Hypergraphs have been used extensively in the context of graph-rewriting-based approaches to concurrency, including service-oriented computing (e.g.~\cite{Bruni-Gadducci-LluchLafuente:A-graph-syntax-for-processes-and-services-2009,Ferrari-Hirsch-Lanese-Montanari-Tuosto:Synchronised-hyperedge-replacement-2005}).  We use them instead of graphs~\cite{Fiadeiro-Lopes:An-interface-theory-for-service-oriented-design-2013} because they offer a more flexible mathematical framework for handling the notions of variable and variable binding required in Section~\ref{section:service-discovery-and-binding}.

  \begin{defi}[Asynchronous relational network -- \newacronym{ARN}{Asynchronous relational network}]
    \label{definition:ARN}
    An \emph{asynchronous relational network} \(\arnet = \abr{X, P, C, \gamma, M, \mu, \Lambda}\) consists of a (finite) edge-bipartite hypergraph \(\abr{X, P, C, \gamma}\) of \emph{points} \(x \in X\), \emph{computation hyperedges} \(p \in P\) and \emph{communication hyperedges} \(c \in C\), and of
    \begin{itemize}
      
    \item a port \(M_{x}\) for every point \(x \in X\),
      
    \item a process \(\abr{\gamma_{p}, \rbr{M_{x}}_{x \in \gamma_{p}}, \Lambda_{p}}\) for every hyperedge \(p \in P\), and
      
    \item a connection \(\abr{M_{c}, \Lambda_{c}, \rbr{\mu^{c}_{x} \colon M_{c} \pto M_{x}}_{x \in \gamma_{c}}}\) for every hyperedge \(c \in C\).
      
    \end{itemize}
  \end{defi}

  \begin{exa}
    \label{example:journey-planner}
    By putting together the process and the connection presented in Examples~\ref{example:journey-planner-process} and~\ref{example:journey-planner-connection}, we obtain the \acn{ARN} \(\arnname{JourneyPlanner}\) depicted in Figure~\ref{figure:journey-planner-ARN}.  Its underlying hypergraph consists of the points \(\portname{JP_{1}}\), \(\portname{JP_{2}}\), \(\portname{R_{1}}\) and \(\portname{R_{2}}\), the computation hyperedge \(\processname{JP}\), the communication hyperedge \(\connectionname{C}\), and the incidence map \(\gamma\) given by \(\gamma_{\processname{JP}} = \cbr{\portname{JP_{1}}, \portname{JP_{2}}}\) and \(\gamma_{\connectionname{C}} = \cbr{\portname{JP_{2}}, \portname{R_{1}}, \portname{R_{2}}}\).

    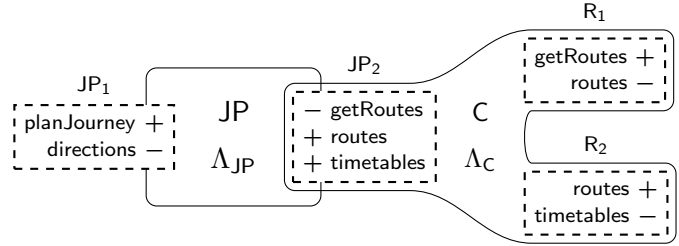
\begin{figure}[h]
      \centering

      \begin{tikzpicture}
        % the process JP
        
        \node [align=center, minimum height=9ex, minimum width=4em] (JP) {
          \(\processname{JP}\) \\[1ex]
          \(\Lambda_{\processname{JP}}\)
        };
        
        \node [port] (JP1) [left=0em of JP, align=right] {
          \(\dmsgr{planJourney}\) \\
          \(\pmsgr{directions}\)
        };
        \node [port-label] [above=0ex of JP1] {\(\portname{JP_{1}}\)};
        
        \node [port] (JP2) [right=0em of JP, align=left] {
          \(\pmsgl{getRoutes}\) \\
          \(\dmsgl{routes}\) \\
          \(\dmsgl{timetables}\)
        };
        \node [port-label] [above=.5ex of JP2] {\(\portname{JP_{2}}\)};

        % the port R1
        
        \node [port] (R1) [above right=-1ex and 3em of JP2, align=right] {
          \(\dmsgr{getRoutes}\) \\
          \(\pmsgr{routes}\)
        };
        \node [port-label] [above=.5ex of R1] {\(\portname{R_{1}}\)};

        % the port R2
        
        \node [port] (R2) [below right=-1ex and 3em of JP2, align=right] {
          \(\dmsgr{routes}\) \\
          \(\pmsgr{timetables}\)
        };
        \node [port-label] [above=.5ex of R2] {\(\portname{R_{2}}\)};

        % the connection
        
        \draw [rounded corners]
        ($(JP2.north west) + (.5,.1)$)
        to                   ($(JP2.north east) + (-.5,.1)$)
        to [out=0, in=180]   ($(R1.north west)  + (.25,.1)$)
        to                   ($(R1.north east)  + (.1,.1)$)
        to                   ($(R1.south east)  + (.1,-.1)$)
        to                   ($(R1.south west)  + (.25,-.1)$)
        to [out=180, in=180] ($(R2.north west)  + (.25,.1)$)
        to                   ($(R2.north east)  + (.1,.1)$)
        to                   ($(R2.south east)  + (.1,-.1)$)
        to                   ($(R2.south west)  + (.25,-.1)$)
        to [out=180, in=0]   ($(JP2.south east) + (-.5,-.1)$)
        to                   ($(JP2.south west) + (-.1,-.1)$)
        to                   ($(JP2.north west) + (-.1,.1)$)
        to                   ($(JP2.north west) + (.5,.1)$);
        
        \path (JP2) -- node [connection] {
          \(\connectionname{C}\) \\[1ex]
          \(\Lambda_{\connectionname{C}}\)
        } (JP2 -| R1.west);

        \begin{pgfonlayer}{background}
          \node [process] [fit=(JP)] {};
        \end{pgfonlayer}
      \end{tikzpicture}

      \caption{The \acn{ARN} \(\arnname{JourneyPlanner}\)}
      \label{figure:journey-planner-ARN}
    \end{figure}
  \end{exa}

  \subsubsection*{The Orchestration Scheme of Asynchronous Relational Networks}

  Let us now focus on the manner in which \acn{ARN}s can be organized to form an orchestration scheme.
  We begin with a brief discussion on the types of points of \acn{ARN}s, which will enable us to introduce notions of morphism of \acn{ARN}s and ground \acn{ARN}.

  An \emph{interaction point} of an \acn{ARN} \(\arnet\) is a point of \(\arnet\) that is not bound to both computation and communication hyperedges.
  We distinguish between two types of interaction points, called requires- and provides-points, as follows.

  \begin{defi}[Requires- and provides-point]
    A \emph{requires-point} of an \acn{ARN} \(\arnet\) is a point of \(\arnet\) that is incident only with a communication hyperedge.
    Similarly, a \emph{provides-point} of \(\arnet\) is a point incident only with a computation hyperedge.
  \end{defi}

  \noindent For the \acn{ARN} \(\arnname{JourneyPlanner}\) depicted in Figure~\ref{figure:journey-planner-ARN}, the points \(\portname{R_{1}}\) and \(\portname{R_{2}}\) are requires-points (incident with the communication hyperedge \(\connectionname{C}\)), whereas \(\portname{JP_{1}}\) is a provides-point (incident with the computation hyperedge \(\processname{JP}\)).

  \begin{minisection}{Orchestrations.}%
    In order to describe \acn{ARN}s as orchestrations we first need to equip them with appropriate notions of morphism and composition of morphisms.
    Morphisms of \acn{ARN}s correspond to injective homomorphisms between their underlying hypergraphs, and are required to preserve all labels, except those associated with points that, like the requires-points, are not incident with computation hyperedges.

    \begin{defi}[Homomorphism of hypergraphs]
      A \emph{homomorphism} \(h\) between hypergraphs \(\abr{X_{1}, E_{1}, \gamma^{1}}\) and \(\abr{X_{2}, E_{2}, \gamma^{2}}\) consists of functions \(h \colon X_{1} \to X_{2}\) and \(h \colon E_{1} \to E_{2}\)\footnote{To simplify the notation, we denote both the translation of vertices and of hyperedges simply by \(h\).} such that for any vertex \(x \in X_{1}\) and hyperedge \(e \in E_{1}\), \(x \in \gamma^{1}_{e}\) if and only if \(h\rbr{x} \in \gamma^{2}_{h\rbr{e}}\).
    \end{defi}

    \begin{defi}[Morphism of \acn{ARN}s]
      Given two \acn{ARN}s \(\arnet_{1} = \abr{X_{1}, P_{1}, C_{1}, \gamma^{1}, M^{1}, \mu^{1}, \Lambda^{1}}\) and \(\arnet_{2} = \abr{X_{2}, P_{2}, C_{2}, \gamma^{2}, M^{2}, \mu^{2}, \Lambda^{2}}\), a \emph{morphism} \(\theta \colon \arnet_{1} \to \arnet_{2}\) consists of
      \begin{itemize}
        
      \item an injective homomorphism \(\theta \colon \abr{X_{1}, P_{1}, C_{1}, \gamma^{1}} \to \abr{X_{2}, P_{2}, C_{2}, \gamma^{2}}\) between the underlying hypergraphs of \(\arnet_{1}\) and \(\arnet_{2}\) such that \(\theta\rbr{P_{1}} \subseteq P_{2}\) and \(\theta\rbr{C_{1}} \subseteq C_{2}\), and
        
      \item a family \(\theta^{\arnpt}\) of polarity-preserving injections \(\theta^{\arnpt}_{x} \colon M^{1}_{x} \to M^{2}_{\theta\rbr{x}}\), for \({x \in X_{1}}\),
        
      \end{itemize}
      such that
      \begin{itemize}

      \item for every point \(x \in X_{1}\) incident with a computation hyperedge, \(\theta^{\arnpt}_{x} = 1_{M^{1}_{x}}\),
        
      \item for every computation hyperedge \(p \in P_{1}\), \(\Lambda^{1}_{p} = \Lambda^{2}_{\theta\rbr{p}}\), and
        
      \item for every communication hyperedge \(c \in C_{1}\), \(M^{1}_{c} = M^{2}_{\theta\rbr{c}}\), \(\Lambda^{1}_{c} = \Lambda^{2}_{\theta\rbr{c}}\) and the following diagram commutes, for every point \(x \in \gamma^{1}_{c}\).
        \[
        \xymatrix{
          {\mathllap{M^{1}_{c} = {}}M^{2}_{\theta\rbr{c}}}
          \ar @{-^{>}} [r] ^-{\mu^{1, c}_{x}}
          \ar @{-^{>}} [dr] _{\mu^{2, \theta\rbr{c}}_{\theta\rbr{x}}}
          & *+[r]{M^{1}_{x}}
          \ar [d] ^{\theta^{\arnpt}_{x}}
          \\
          & *+[r]{M^{2}_{\theta\rbr{x}}}
        }\]

      \end{itemize}
    \end{defi}

\noindent    It is straightforward to verify that the morphisms of \acn{ARN}s can be composed in terms of their components.
    Their composition is associative and has left and right identities given by morphisms that consists solely of set-theoretic identities.
    We obtain in this way the first result supporting the construction of an orchestration scheme of \acn{ARN}s.

    \begin{prop}
      The morphisms of \acn{ARN}s form a category, denoted \(\ARN\).
      \qed
    \end{prop}
    \vspace{-\topsep}
  \end{minisection}

  \begin{minisection}{Specifications.}%
    To define specifications over given \acn{ARN}s, we label their points with linear temporal sentences, much in the way we used pre- and post-conditions as labels for positions in terms when defining specifications of program expressions.

    \begin{defi}[Specification over an \acn{ARN}]
      For any \acn{ARN} \(\arnet\), the set \(\SSpec\rbr{\arnet}\) of \emph{\(\arnet\)\nb-specifications} is the set of pairs \(\abr{x, \rho}\), usually denoted \(\lsen{x}{\rho}\), where \(x\) is a point of \(\arnet\) and \(\rho\) is an \(\aLTL\)\nb-sentence over \(A_{M_{x}}\), i.e.\ over the set of actions defined by the port that labels \(x\).
    \end{defi}
    
    The \emph{translation} of specifications along morphisms of \acn{ARN}s presents no difficulties: for every morphism  \(\theta \colon \arnet \to \arnet'\), the map \(\SSpec\rbr{\theta} \colon \SSpec\rbr{\arnet} \to \SSpec\rbr{\arnet'}\) is given by
    \[
    \SSpec\rbr{\theta}\rbr{\lsen{x}{\rho}} = \lsen{\theta\rbr{x}}{\Sen^{\aLTL}\rbr{\theta^{\arnpt}_{x}}\rbr{\rho}}
    \]
    for each point \(x\) of \(\arnet\) and each \(\aLTL\)\nb-sentence \(\rho\) over the actions of \(x\).
    Furthermore, it can be easily seen that it inherits the functoriality of the translation of sentences in \(\aLTL\), thus giving rise to the functor \(\SSpec \colon \ARN \to \Set\) that we are looking for.
  \end{minisection}

  \begin{minisection}{Ground orchestrations.}%
    Morphisms of \acn{ARN}s can also be regarded as refinements, as they formalize the embedding of networks with an intuitively simpler behaviour into networks that are more complex.  This is achieved essentially by mapping each of the requires-points of the source \acn{ARN} to a potentially non-requires-point of the target \acn{ARN}, a point which can be looked at as the `root' of a particular subnetwork of the target \acn{ARN}.
    To explain this aspect in more detail we introduce the notions of dependency and \acn{ARN} defined by a point.

    \begin{defi}[Dependency]
      Let \(x\) and \(y\) be points of an \acn{ARN} \(\arnet\).
      The point \(x\) is said to be \emph{dependent} on \(y\) if there exists a path from \(x\) to \(y\) that begins with a computation hyperedge, i.e.\ if there exists an alternating sequence \(x\, e_{1}\, x_{1}\, \dotso\, e_{n}\, y\) of (distinct) points and hyperedges of the underlying hypergraph \(\abr{X, P, C, \gamma}\) of \(\arnet\) such that \(x \in \gamma_{e_{1}}\), \(y \in \gamma_{e_{n}}\), \(x_{i} \in \gamma_{e_{i}} \cap \gamma_{e_{i+1}}\) for every \(1 \leq i < n\), and \(e_{1} \in P\).
    \end{defi}

    \begin{defi}[Network defined by a point]
      The \emph{\acn{ARN} defined by a point} \(x\) of an \acn{ARN} \(\arnet\) is the full sub-\acn{ARN} \(\arnet_{x}\) of \(\arnet\) determined by \(x\) and the points on which \(x\) is dependent.
    \end{defi}

    One can now see that any morphism of \acn{ARN}s \(\theta \colon \arnet_{1} \to \arnet_{2}\) assigns to each requires-point \(x\) of the source network \(\arnet_{1}\) the sub-\acn{ARN} \(\arnet_{2, \theta\rbr{x}}\) of \(\arnet_{2}\) defined by \(\theta\rbr{x}\).

    \begin{exa}
      \label{example:journey-planner-net}
      In Figure~\ref{figure:journey-planner-net-ARN} we outline an extension of the \acn{ARN} \(\arnname{JourneyPlanner}\) discussed in Example~\ref{example:journey-planner} that is obtained by attaching the processes \(\processname{MS}\) (for Map Services) and \(\processname{TS}\) (for Transport System) to the requires-points \(\portname{R_{1}}\) and \(\portname{R_{2}}\) of \(\arnname{JourneyPlanner}\).
      Formally, the link between \(\arnname{JourneyPlanner}\) and the resulting \acn{ARN} \(\arnname{JourneyPlannerNet}\) is given by a morphism \(\theta \colon \arnname{JourneyPlanner} \to \arnname{JourneyPlannerNet}\) that preserves all the labels, points and hyperedges of \(\arnname{JourneyPlanner}\), with the exception of the requires-points \(\portname{R_{1}}\) and \(\portname{R_{2}}\), which are mapped to \(\portname{MS_{1}}\) and \(\portname{TS_{1}}\), respectively.

      In this case, \(\portname{MS_{1}}\) only depends on itself, hence the sub-\acn{ARN} of \(\arnname{JourneyPlannerNet}\) defined by \(\portname{MS_{1}}\), i.e.\ the \acn{ARN} assigned to the requires-point \(\portname{R_{1}}\) of \(\arnname{JourneyPlanner}\), is given by the process \(\processname{MS}\) and its port \(\portname{MS_{1}}\).
      In contrast, the point \(\portname{JP_{1}}\) depends on all the other points of \(\arnname{JourneyPlannerNet}\), and thus it defines the entire \acn{ARN} \(\arnname{JourneyPlannerNet}\).

      \begin{figure}[h]
        \centering

        \begin{tikzpicture}
          % the process JP
          
          \node [align=center, minimum height=9ex, minimum width=4em] (JP) {
            \(\processname{JP}\) \\[1ex]
            \(\Lambda_{\processname{JP}}\)
          };
          
          \node [port] (JP1) [left=0em of JP, align=right] {
            \(\dmsgr{planJourney}\) \\
            \(\pmsgr{directions}\)
          };
          \node [port-label] [above=0ex of JP1] {\(\portname{JP_{1}}\)};
          
          \node [port] (JP2) [right=0em of JP, align=left] {
            \(\pmsgl{getRoutes}\) \\
            \(\dmsgl{routes}\) \\
            \(\dmsgl{timetables}\)
          };
          \node [port-label] [above=.5ex of JP2] {\(\portname{JP_{2}}\)};

          % the process MS
          
          \node [port] (MS1) [above right=0ex and 3em of JP2, align=right] {
            \(\dmsgr{getRoutes}\) \\
            \(\pmsgr{routes}\)
          };
          \node [port-label] [above=.5ex of MS1] {\(\portname{MS_{1}}\)};

          \node [align=center, minimum height=7ex, minimum width=4em] (MS) [right=0em of MS1] {
            \(\processname{MS}\) \\[1ex]
            \(\Lambda_{\processname{MS}}\)
          };

          % the process TS
          
          \node [port] (TS1) [below right=0ex and 3em of JP2, align=right] {
            \(\dmsgr{routes}\) \\
            \(\pmsgr{timetables}\)
          };
          \node [port-label] [above=.5ex of TS1] {\(\portname{TS_{1}}\)};

          \node [align=center, minimum height=7ex, minimum width=4em] (TS) [right=0em of TS1] {
            \(\processname{TS}\) \\[1ex]
            \(\Lambda_{\processname{TS}}\)
          };
          
          % the connection
          
          \draw [rounded corners]
          ($(JP2.north west) + (.5,.1)$)
          to                   ($(JP2.north east) + (-.5,.1)$)
          to [out=0, in=180]   ($(MS1.north west) + (.25,.1)$)
          to                   ($(MS1.north east) + (.1,.1)$)
          to                   ($(MS1.south east) + (.1,-.1)$)
          to                   ($(MS1.south west) + (.3,-.1)$)
          to [out=180, in=180] ($(TS1.north west) + (.3,.1)$)
          to                   ($(TS1.north east) + (.1,.1)$)
          to                   ($(TS1.south east) + (.1,-.1)$)
          to                   ($(TS1.south west) + (.25,-.1)$)
          to [out=180, in=0]   ($(JP2.south east) + (-.5,-.1)$)
          to                   ($(JP2.south west) + (-.1,-.1)$)
          to                   ($(JP2.north west) + (-.1,.1)$)
          to                   ($(JP2.north west) + (.5,.1)$);
          
          \path (JP2) -- node [connection] {
            \(\connectionname{C}\) \\[1ex]
            \(\Lambda_{\connectionname{C}}\)
          } (JP2 -| MS1.west);

          \begin{pgfonlayer}{background}       
            \node [process] [fit=(JP)] {};
            \node [process] [fit=(MS)] {};
            \node [process] [fit=(TS)] {};
          \end{pgfonlayer}
        \end{tikzpicture}

        \caption{The \acn{ARN} \(\arnname{JourneyPlannerNet}\)}
        \label{figure:journey-planner-net-ARN}
      \end{figure}
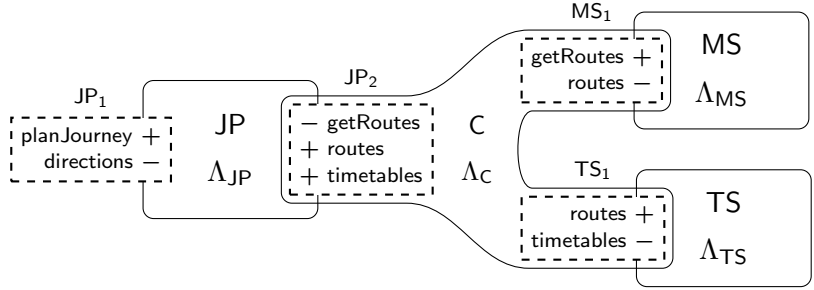
    \end{exa}

    In view of the above observation, we may consider the requires-points of networks as counterparts of the variables used in program expressions, and their morphisms as substitutions. 
    This leads us to the following definition of ground \acn{ARN}s.

    \begin{defi}[Ground \acn{ARN}]
      An \acn{ARN} is said to be \emph{ground} if it has no requires-points.
    \end{defi}
    \vspace{-\topsep}
  \end{minisection}

  \begin{minisection}{Properties.}%
    The evaluation of specifications with respect to ground \acn{ARN}s relies on the concepts of diagram of a network and automaton (i.e.\ \(\aLTL\)\nb-model) defined by a point, whose purpose is to describe the observable behaviour of a ground \acn{ARN} through one of its points.
    We start by extending Remarks~\ref{remark:abstract-process} and~\ref{remark:abstract-connection} to \acn{ARN}s.

    \begin{fact}[Diagram of an \acn{ARN}]
      Every \acn{ARN} \(\arnet = \abr{X, P, C, \gamma, M, \mu, \Lambda}\) defines a diagram \(D_{\arnet} \colon \catnamecap{J}_{\arnet} \to \Sig^{\aLTL}\) as follows:
      \begin{itemize}
        
      \item \(\mathbb{J}_{\arnet}\) is the free preordered category given by the set of objects
        \[
        X \cup P \cup C \cup \cbr{\abr{c, x}_{\arnet} \st c \in C, x \in \gamma_{c}}
        \]
        and the arrows
        \begin{itemize}
          
        \item \(\cbr{x \to p \st p \in P, x \in \gamma_{p}}\) for computation hyperedges, and
          
        \item \(\cbr{c \leftarrow \abr{c, x}_{\arnet} \rightarrow x \st c \in C, x \in \gamma_{c}}\) for communication hyperedges;
          
        \end{itemize}
        
      \item \(D_{\arnet}\) is the functor that provides the sets of actions of ports, processes and channels, together with the appropriate mappings between them.
        For example, given a communication hyperedge \(c \in C\) and a point \(x \in \gamma_{c}\),
        \begin{itemize}
          
        \item \(D_{\arnet}\rbr{c} = A_{M_{c}}\), \(D_{\arnet}\rbr{\abr{c, x}_{\arnet}} = \dom\rbr{\mu^{c}_{x}}\), \(D_{\arnet}\rbr{x} = A_{M_{x}}\),
          
        \item \(D_{\arnet}\rbr{\abr{c, x}_{\arnet} \to c} = \rbr{\dom\rbr{\mu^{ c}_{x}} \subseteq A_{M_{c}}}\), and
          
        \item \(D_{\arnet}\rbr{\abr{c, x}_{\arnet} \to x} = \mu^{c}_{x}\).
          
        \end{itemize}
        
      \end{itemize}
    \end{fact}

\noindent     We define the signature of an \acn{ARN} by taking the colimit of its diagram, which is guaranteed to exist because the category \(\Sig^{\aLTL}\), i.e.\ \(\Set\), is finitely cocomplete.

    \begin{defi}[Signature of an \acn{ARN}]
      The signature of an \acn{ARN} \(\arnet\) is the colimiting cocone \(\xi \colon D_{\arnet} \To A_{\arnet}\) of the diagram \(D_{\arnet}\).
    \end{defi}

    The most important construction that allows us to define properties of ground \acn{ARN}s is the one that defines the observed behaviour of a (ground) network at one of its points.

    \begin{defi}[Automaton defined by a point]
      \label{definition:observed-automaton}
      Let \(x\) be a point of a ground \acn{ARN} \(\grnet\).
      The \emph{observed automaton} \(\Lambda_{x}\) at \(x\) is given by the reduct \(\Lambda_{\grnet_{x}} \reduct_{\xi_{x}}\), where
      \begin{itemize}
        
      \item \(\grnet_{x} = \abr{X, P, C, \gamma, M, \mu, \Lambda}\) is the sub-\acn{ARN} of \(\grnet\) defined by \(x\),
        
      \item \(\xi \colon D_{\grnet_{x}} \To A_{\grnet_{x}}\) is the signature of \(\grnet_{x}\),
        
      \item \(\Lambda_{\grnet_{x}}\) is the product automaton \(\prod_{e \in P \cup C} \Lambda^{\grnet_{x}}_{e}\), and
        
      \item \(\Lambda^{\grnet_{x}}_{e}\) is the cofree expansion of \(\Lambda_{e}\) along \(\xi_{e}\), for any hyperedge \(e \in P \cup C\).
        
      \end{itemize}
    \end{defi}

    \begin{exa}
      Consider once again the (ground) \acn{ARN} represented in Figure~\ref{figure:journey-planner-net-ARN}.
      The automaton defined by the point \(\portname{MS_{1}}\) is just \(\Lambda_{\processname{MS}} \reduct_{A_{\portname{MS}_{1}}}\); this follows from the observation that the \acn{ARN} defined by \(\portname{MS_{1}}\) consists exclusively of the process \(\processname{MS}\) and the port \(\portname{MS_{1}}\).
      On the other hand, in order to obtain the automaton defined by the provides-point \(\portname{JP_{1}}\) one needs to compute the product of the cofree expansions of all four automata \(\Lambda_{\processname{JP}}\), \(\Lambda_{\connectionname{C}}\), \(\Lambda_{\processname{MS}}\) and \(\Lambda_{\processname{TS}}\).
      Based on Propositions~\ref{proposition:cofree-models-in-aLTL} and~\ref{proposition:products-of-models-in-aLTL}, the resulting automaton has to accept precisely the projections to \(A_{M_{\portname{JP_{1}}}}\) of those traces accepted by \(\Lambda_{\processname{JP}}\) that are compatible with traces accepted by \(\Lambda_{\connectionname{C}}\), \(\Lambda_{\processname{MS}}\) and \(\Lambda_{\processname{TS}}\), in the sense that together they give rise, by amalgamation, to traces over the alphabet of the network.
    \end{exa}

    We now have all the necessary concepts for defining properties of ground \acn{ARN}s.

    \begin{defi}[Property of an \acn{ARN}]
      Let \(\lsen{x}{\rho}\) be a specification over a ground \acn{ARN} \(\grnet\).
      Then \(\lsen{x}{\rho}\) is a \emph{property} of \(\grnet\) if and only if the automaton \(\Lambda_{x}\) observed at the point \(x\) in \(\grnet\) satisfies (according to the definition of satisfaction in \(\aLTL\)) the temporal sentence \(\rho\).
      \[
      \Lambda_{x} \models^{\aLTL} \rho
      \]
    \end{defi}

    \begin{rem}
      It is important to notice that not only the signature of an \acn{ARN}, but also the various cofree expansions and products considered in Definition~\ref{definition:observed-automaton} are unique only up to an isomorphism.  Consequently, the automaton defined by a point of a ground \acn{ARN} is also unique only up to an isomorphism, which means that the closure of \(\aLTL\) under isomorphisms plays a crucial role in ensuring that the evaluation of specifications with respect to ground \acn{ARN}s is well defined.
    \end{rem}

    All we need now in order to complete the construction of the orchestration scheme of \acn{ARN}s is to show that the morphisms of ground \acn{ARN}s preserve properties.
    This result depends upon the last of the four hypotheses we introduced at the beginning of the subsection: the reflection of the satisfaction of sentences by the model homomorphisms of the institution used as foundation for the construction of \acn{ARN}s.

    \begin{prop}
      \label{orchestration-scheme-of-ARNs}
      For every morphism of ground \acn{ARN}s \(\theta \colon \grnet_{1} \to \grnet_{2}\) and every property \(\lsen{x}{\rho}\) of \(\grnet_{1}\), the specification \(\SSpec\rbr{\theta}\rbr{\lsen{x}{\rho}}\) is a property of \(\grnet_{2}\).
    \end{prop}

    \proof
    Let \(\grnet_{1}^{x}\) and \(\grnet_{2}^{x}\) be the sub-\acn{ARN}s of \(\grnet_{1}\) and \(\grnet_{2}\) determined by \(x\) and \(\theta\rbr{x}\) respectively, and let us also assume that \(\grnet_{i}^{x} = \abr{X_{i}, P_{i}, C_{i}, \gamma^{i}, M^{i}, \mu^{i}, \Lambda^{i}}\) and that \(\xi^{i} \colon D_{\grnet_{i}^{x}} \To A_{\grnet_{i}^{x}}\) is the signature of \(\grnet_{i}^{x}\), for \(i \in \cbr{1, 2}\).
    Since \(\lsen{x}{\rho}\) is a property of \(\grnet_{1}\), we know that the automaton \(\Lambda^{1}_{x}\) observed at the point \(x\) in \(\grnet_{1}\) satisfies \(\rho\).
    We also know that \(\theta \colon \grnet_{1} \to \grnet_{2}\) defines the \(\aLTL\)\nb-signature morphism \(\theta^{\arnpt}_{x} \colon A_{M_{1, x}} \to A_{M_{2, \theta\rbr{x}}}\) as the identity of \(A_{M_{1, x}}\) (because \(\grnet_{1}\) is ground); hence, the automaton \(\Lambda^{2}_{\theta\rbr{x}}\) observed at \(\theta\rbr{x}\) in \(\grnet_{2}\) is also a model of \(A_{M_{1, x}}\).

    By Proposition~\ref{proposition:reflection-of-the-satisfaction-of-sentences-in-aLTL}, \(\aLTL\)\nb-model homomorphisms reflect the satisfaction of sentences; therefore, in order to prove that \(\Lambda^{2}_{\theta\rbr{x}}\) satisfies \(\rho\) -- and in this way, that \(\SSpec\rbr{\theta}\rbr{\lsen{x}{\rho}}\) is a property of \(\grnet_{2}\) -- it suffices to determine the existence of a homomorphism \(\Lambda^{2}_{\theta\rbr{x}} \to \Lambda^{1}_{x}\).

    Recall that \(\Lambda^{1}_{x}\) and \(\Lambda^{2}_{\theta\rbr{x}}\) are the reducts \(\Lambda_{\grnet_{1}^{x}} \reduct_{\xi^{1}_{x}}\) and \(\Lambda_{\grnet_{2}^{x}} \reduct_{\xi^{2}_{\theta\rbr{x}}}\), where, for \(i \in \cbr{1, 2}\),
    \begin{itemize}
      
    \item \(\Lambda_{\grnet_{i}^{x}}\) is the product \(\prod_{e \in P_{i} \cup C_{i}} \Lambda^{\grnet_{i}^{x}}_{e}\), equipped with projections \(\pi^{i}_{e} \colon \Lambda_{\grnet_{i}^{x}} \to \Lambda^{\grnet_{i}^{x}}_{e}\), and

    \item \(\Lambda^{\grnet_{i}^{x}}_{e}\), for \(e \in P_{i} \cup C_{i}\), is the cofree expansion of \(\Lambda^{i}_{e}\) along \(\xi^{i}_{e}\), for which we denote the universal morphism from \(\_ \reduct_{\xi^{i}_{e}}\) to \(\Lambda^{i}_{e}\) by \(\varepsilon^{i}_{e} \colon \Lambda^{\grnet_{i}^{x}}_{e} \reduct_{\xi^{i}_{e}} \to \Lambda^{i}_{e}\).
      
    \end{itemize}
    According to the description of the \acn{ARN}s defined by given points, we can restrict \(\theta\) to a morphism of \acn{ARN}s from \(\grnet_{1}^{x}\) to \(\grnet_{2}^{x}\).
    Since \(\grnet_{1}^{x}\) is ground, we further obtain, based on this restriction, a functor \(F \colon \catnamecap{J}_{\grnet_{1}^{x}} \to \catnamecap{J}_{\grnet_{2}^{x}}\) that makes the following diagram commutative.
    \[
    \xymatrix @R-4ex {
      {\catnamecap{J}_{\grnet_{1}^{x}}}
      \ar [dr] ^-{D_{\grnet_{1}^{x}}}
      \ar [dd] _{F}
      \\
      & {\Sig^{\aLTL}}
      \\
      {\catnamecap{J}_{\grnet_{2}^{x}}}
      \ar [ur] _-{D_{\grnet_{2}^{x}}}
    }
    \]
    This allows us to define the derived cocone \(F \cdot \xi^{2} \colon D_{\grnet_{1}^{x}} \To A_{\grnet_{2}^{x}}\), whose components are given, for example, by \(\rbr{F \cdot \xi^{2}}_{x} = \xi^{2}_{\theta\rbr{x}}\).
    Since \(\xi^{1}\) is the colimit of \(D_{\grnet_{1}^{x}}\) it follows that there exists a (unique) morphism of cocones \(\sigma \colon \xi^{1} \to F \cdot \xi^{2}\), i.e.\ an \(\aLTL\)\nb-signature morphism \(\sigma \colon A_{\grnet_{1}^{x}} \to A_{\grnet_{2}^{x}}\) that satisfies, in particular, \(\xi^{1}_{e} \comp \sigma = \xi^{2}_{\theta\rbr{e}}\) for every hyperedge \(e \in P_{1} \cup C_{1}\).

    We obtain in this way, for every hyperedge \(e \in P_{1} \cup C_{1}\), the composite morphism \(\pi^{2}_{\theta\rbr{e}} \reduct_{\xi^{2}_{\theta\rbr{e}}} \comp \varepsilon^{2}_{\theta\rbr{e}}\) from \(\Lambda_{\grnet_{2}^{x}} \reduct_{\xi^{2}_{\theta\rbr{e}}} = \Lambda_{\grnet_{2}^{x}} \reduct_{\sigma} \reduct_{\xi^{1}_{e}}\) to \(\Lambda^{1}_{e} = \Lambda^{2}_{\theta\rbr{e}}\).
    \[
    \xymatrix @C+4em {
      {\mathllap{\Lambda^{1}_{e} = {}}\Lambda^{2}_{\theta\rbr{e}}}
      & {\Lambda^{\grnet_{1}^{x}}_{e} \reduct_{\xi^{1}_{e}}}
      \ar [l] _-{\varepsilon^{1}_{e}}
      & {\Lambda^{\grnet_{1}^{x}}_{e}}
      \\
      {\Lambda^{\grnet_{2}^{x}}_{\theta\rbr{e}} \reduct_{\xi^{2}_{\theta\rbr{e}}}}
      \ar [u] ^{\varepsilon^{2}_{\theta\rbr{e}}} 
      & {\Lambda_{\grnet_{2}^{x}} \reduct_{\xi^{2}_{\theta\rbr{e}}}\mathrlap{{} = \Lambda_{\grnet_{2}^{x}} \reduct_{\sigma} \reduct_{\xi^{1}_{e}}}}
      \ar [u] _{h_{e} \reduct_{\xi^{1}_{e}}}
      \ar [l] ^-{\pi^{2}_{\theta\rbr{e}} \reduct_{\xi^{2}_{\theta\rbr{e}}}}
      & {\Lambda_{\grnet_{2}^{x}} \reduct_{\sigma}}
      \ar [u] _{h_{e}}
    }
    \]
    Given that \(\Lambda^{\grnet_{1}^{x}}_{e}\) is the cofree expansion of \(\Lambda^{1}_{e}\) along \(\xi^{1}_{e}\), we deduce that there exists a (unique) morphism \(h_{e} \colon \Lambda_{\grnet_{2}^{x}} \reduct_{\sigma} \to \Lambda^{\grnet_{1}^{x}}_{e}\) such that the above diagram is commutative.
    This implies, by the universal property of the product \(\Lambda_{\grnet_{1}^{x}}\), the existence of a (unique) morphism \(h \colon \Lambda_{\grnet_{2}^{x}} \reduct_{\sigma} \to \Lambda_{\grnet_{1}^{x}}\) such that \(h \comp \pi^{1}_{e} = h_{e}\) for every \(e \in P_{1} \cup C_{1}\).
    \[
    \xymatrix {
      {\Lambda^{1}_{e}}
      & {\Lambda_{\grnet_{1}^{x}}}
      \ar [l] _-{\pi^{1}_{e}}
      \\
      & 
      {\Lambda_{\grnet_{2}^{x}} \reduct_{\sigma}}
      \ar [u] _{h}
      \ar [ul] ^-{h_{e}}
    }
    \]
    It follows that the reduct \(h \reduct_{\xi^{1}_{x}}\) is a morphism from \(\Lambda_{\grnet_{2}^{x}} \reduct_{\sigma} \reduct_{\xi^{1}_{x}}\) to \(\Lambda_{\grnet_{1}^{x}} \reduct_{\xi^{1}_{x}}\).  Then, to complete the proof, we only need to notice that \(\Lambda_{\grnet_{2}^{x}} \reduct_{\sigma} \reduct_{\xi^{1}_{x}} = \Lambda_{\grnet_{2}^{x}} \reduct_{\xi^{2}_{\theta\rbr{x}}} = \Lambda^{2}_{\theta\rbr{x}}\) and \(\Lambda_{\grnet_{1}^{x}} \reduct_{\xi^{1}_{x}} = \Lambda^{1}_{x}\).
    \qed
  \end{minisection}

  \section{A Logical View on Service Discovery and Binding}
  \label{section:service-discovery-and-binding}

  Building on the results of Section~\ref{section:orchestration-schemes}, let us now investigate how the semantics of the service overlay can be characterized using fundamental computational aspects of the logic-programming paradigm such as unification and resolution.
  Our approach is founded upon a simple and intuitive analogy between concepts of service-oriented computing like service module and client application~\cite{Fiadeiro-Lopes-Bocchi:An-abstract-model-for-service-discovery-and-binding-2011}, and concepts such as clause and query that are specific to (the relational variant of) logic programming~\cite{Lloyd:Foundations-of-Logic-Programming-1987}.
  In order to clarify this analogy we rely on the institutional framework that we put forward in~\cite{Tutu-Fiadeiro:Institution-independent-logic-programming-2015} to address the model-theoretic foundations of logic programming.

  We begin by briefly recalling the most basic structure that underlies both the denotational and the operational semantics of relational logic programming: the substitution system of (sets of) variables and substitutions over a given (single-sorted) first-order signature.
  Its definition relies technically on the category \(\Room\) of \emph{institution rooms} and \emph{corridors} (see e.g.~\cite{Mossakowski:Comorphism-based-Grothendieck-logics-2002}). The objects of \(\Room\) are triples \(\abr{S, \catnamecap{M}, \models}\) consisting of a set \(S\) of \emph{sentences}, a category \(\catnamecap{M}\) of \emph{models}, and a \emph{satisfaction relation} \({\models} \subseteq \obj{\catnamecap{M}} \times S\). They are related through corridors \(\abr{\alpha, \beta} \colon \abr{S, \catnamecap{M}, \models} \to \abr{S', \catnamecap{M}', \models'}\) that abstract the change of notation within or between logics by defining a \emph{sentence-translation function} \(\alpha \colon S \to S'\) and a \emph{model-reduction functor} \(\beta \colon \catnamecap{M}' \to \catnamecap{M}\) such that the following condition holds for all \(M' \in \obj{\catnamecap{M}'}\) and \(\rho \in S\):
  \[
  M' \models' \alpha\rbr{\rho}
  \qquad \text{if and only if} \qquad 
  \beta\rbr{M'} \models \rho.
  \]

  \begin{defi}[Substitution system]
    A \emph{substitution system} is a triple \(\abr{\Subst, G, \SubSys}\), often denoted simply by \(\SubSys\), that consists of
    \begin{itemize}
      
    \item a category \(\Subst\) of \emph{signatures of variables} and \emph{substitutions}, 
      
    \item a room \(G\) of \emph{ground sentences} and \emph{models}, and 
      
    \item a functor \(\SubSys \colon \Subst \to \commacat{G}{\Room}\), defining for every signature of variables \(X\) the corridor \(\SubSys\rbr{X} \colon G \to G\rbr{X}\) from \(G\) to the room \(G\rbr{X}\) of \emph{\(X\)\nb-sentences} and \emph{\(X\)\nb-models}.

    \end{itemize}
  \end{defi}

  \begin{exa}
    \label{example:FP-substitution-system}
    In the case of conventional logic programming, every single-sorted first-order signature \(\abr{F, P}\) determines a substitution system 
    \[
    \rbr[\big]{\aFOLssneqGS}_{\abr{F, P}} \colon \Subst_{\abr{F, P}} \to \commacat{\aFOLssneq\rbr{F, P}}{\Room}\footnote{Through \(\aFOLssneq\) we refer to the institution that corresponds to the atomic fragment of the single-sorted variant of first-order logic without equality.}
    \]
    %in which
    where \(\Subst_{\abr{F, P}}\) is simply the category whose objects are sets of variables (defined over the signature \(\abr{F, P}\)), and whose arrows are first-order substitutions.  The room \(\aFOLssneq\rbr{F, P}\) accounts for the (ground) atomic sentences given by \(\abr{F, P}\), the models of \(\abr{F, P}\), as well as the standard satisfaction relation between them.  And finally, the functor \(\rbr[\big]{\aFOLssneqGS}_{\abr{F, P}}\) maps every signature (i.e.\ set) of variables \(X\) to the corridor \(\abr{\alpha_{\abr{F, P}, X}, \beta_{\abr{F, P}, X}}\),
    \[
    \xymatrix @C=0em @M=1pt @H+2.5ex {
      {\langle \Sen\rbr{F, P},}
      \ar @{->} `u[r] `[rrrr] ^{\alpha_{\abr{F, P}, X}} [rrrr]
      & {\Mod\rbr{F, P},}
      & {\models_{\abr{F, P}} \rangle}
      & {\hspace{3em}}
      & {\langle \Sen\rbr{F \cup X, P},}
      & {\Mod\rbr{F \cup X, P},}
      \ar @{->} `d[l] `[llll] ^{\beta_{\abr{F, P}, X}} [llll]
      & {\models_{\abr{F \cup X, P}} \rangle}
    }
    \]
    where \(\alpha_{\abr{F, P}, X}\) and \(\beta_{\abr{F, P}, X}\) are the translation of sentences and the reduction of models that correspond to the inclusion of signatures \(\abr{F, P} \subseteq \abr{F \cup X, P}\).
  \end{exa}

  Substitution systems are particularly useful when reasoning about the semantics of clauses and queries.
  For instance, the above substitution system can be used to define (definite) clauses over \(\abr{F, P}\) as syntactic structures \(\dclause{X}{C}{H}\), also written
  \[
  \arrclause{X}{C}{H}
  \]
  such that \(X\) is a signature of variables, \(C\) is sentence over \(\abr{F \cup X, P}\), and \(H\) is a (finite) set of sentences over \(\abr{F \cup X, P}\).\footnote{Note that, in relational logic programming, the variables are often distinguished from other symbols through notational conventions; for this reason, the set \(X\) of variables is at times omitted.}
  The semantics of such a construction is given by the class of models of \(\abr{F, P}\), i.e.\ of ground models of the substitution system, whose expansions to \(\abr{F \cup X, P}\) satisfy \(C\) whenever they satisfy all sentences in \(H\) -- this reflects the usual interpretation of logic-programming clauses as universally quantified sentences \(\univqs{X}{\bigpland H \plimplies C}\).

  Similarly to institutions, the axiomatic approach to logic programming on which we rely in this paper is parameterized by the signature used.
  In categorical terms, this means that the morphisms of signatures induce appropriate morphisms between their corresponding substitution systems, and moreover, that this mapping is functorial.
  As regards our inquiry on the semantics of the service overlay, it suffices to recall that the category \(\SubstSys\) of substitution systems results from the Grothendieck construction~\cite{Tarlecki-Burstall-Goguen:Fundamental-algebraic-tools-III-1991} for the functor \(\FCat{\_}{\commacat{\_}{\Room}} \colon \rbr{\Cat \times \Room}^{\op} \to \Cat\) that maps
  \begin{itemize}

  \item every category \(\Subst\) and room \(G\) to the category of functors \(\FCat{\Subst}{\commacat{G}{\Room}}\),

  \item every functor \(\Psi \colon \Subst \to \Subst'\) and corridor \(\kappa \colon G \to G'\) to the canonical composition functor 
    \(\Psi \_ \rbr{\commacat{\kappa}{\Room}} \colon \FCat{\Subst'}{\commacat{G'}{\Room}} \to \FCat{\Subst}{\commacat{G}{\Room}}\).

  \end{itemize}
  This allows us to introduce the next notion of generalized substitution system.

  \begin{defi}[Generalized substitution system]
    A \emph{generalized substitution system} is a pair \(\abr{\Sig, \GS}\) given by a category \(\Sig\) of \emph{signatures}, and a functor \(\GS \colon \Sig \to \SubstSys\).
  \end{defi}

  \noindent In order to provide a better understanding of the complex structure of generalized substitution systems, we consider the following notational conventions and terminology:
  \begin{itemize}[label=$-$]

  \item For every signature \(\Sigma\) of a generalized substitution system \(\GS\), we denote the (local) substitution system \(\GS\rbr{\Sigma}\) by \(\GS_{\Sigma} \colon \Subst_{\Sigma} \to \commacat{G_{\Sigma}}{\Room}\), and we refer to the objects and morphisms of \(\Subst_{\Sigma}\) as \emph{signatures of \(\Sigma\)\nb-variables} and \emph{\(\Sigma\)\nb-substitutions}.
    The room \(G_{\Sigma}\) is assumed to comprise the set \(\Sen\rbr{\Sigma}\) of \emph{ground \(\Sigma\)\nb-sentences}, the category \(\Mod\rbr{\Sigma}\) of \emph{\(\Sigma\)\nb-models}, and the \emph{\(\Sigma\)\nb-satisfaction relation} \({\models_{\Sigma}} \subseteq \obj{\Mod\rbr{\Sigma}} \times \Sen\rbr{\Sigma}\).

  \item On objects, \(\GS_{\Sigma}\) maps every signature of \(\Sigma\)\nb-variables \(X\) to the corridor \(\GS_{\Sigma}\rbr{X} = \abr{\alpha_{\Sigma, X}, \beta_{\Sigma, X}}\) from \(G_{\Sigma}\) to the room \(G_{\Sigma}\rbr{X} = \abr{\Sen_{\Sigma}\rbr{X}, \Mod_{\Sigma}\rbr{X}, \models_{\Sigma, X}}\) of \emph{\(X\)\nb-sentences} and \emph{\(X\)\nb-models}.
    \[
    \alpha_{\Sigma, X} \colon \Sen\rbr{\Sigma} \to \Sen_{\Sigma}\rbr{X}
    \qquad
    \beta_{\Sigma, X} \colon \Mod_{\Sigma}\rbr{X} \to \Mod\rbr{\Sigma}
    \]

  \item On arrows, \(\GS_{\Sigma}\) maps every \(\Sigma\)\nb-substitution \(\psi \colon X \to Y\) to the corridor \(\GS_{\Sigma}\rbr{\psi} = \abr{\Sen_{\Sigma}\rbr{\psi},\linebreak[0] \Mod_{\Sigma}\rbr{\psi}}\) from \(G_{\Sigma}\rbr{X}\) to \(G_{\Sigma}\rbr{Y}\), which satisfies, by definition, \(\GS_{\Sigma}\rbr{X} \comp \GS_{\Sigma}\rbr{\psi} = \GS_{\Sigma}\rbr{Y}\).
    \[
    \xymatrix @C=0em {
      & {\Sen\rbr{\Sigma}}
      \ar [dl] _{\alpha_{\Sigma, X}}
      \ar [dr] ^{\alpha_{\Sigma, Y}}
      \\
      {\Sen_{\Sigma}\rbr{X}}
      \ar [rr] _{\Sen_{\Sigma}\rbr{\psi}}
      &
      & {\Sen_{\Sigma}\rbr{Y}}
    }
    \qquad
    \xymatrix @C=0em {
      & {\Mod\rbr{\Sigma}}
      \\
      {\Mod_{\Sigma}\rbr{X}}
      \ar [ur] ^{\beta_{\Sigma, X}}
      &
      & {\Mod_{\Sigma}\rbr{Y}}
      \ar [ul] _{\beta_{\Sigma, Y}}
      \ar [ll] ^{\Mod_{\Sigma}\rbr{\psi}}
    }
    \]

  \item With respect to signature morphisms, every \(\varphi \colon \Sigma \to \Sigma'\) determines a morphism of substitution systems \(\GS_{\varphi} \colon \GS_{\Sigma} \to \GS_{\Sigma'}\) in the form of a triple \(\abr{\Psi_{\varphi}, \kappa_{\varphi}, \tau_{\varphi}}\), where \(\Psi_{\varphi}\) is a functor \(\Subst_{\Sigma} \to \Subst_{\Sigma'}\), \(\kappa_{\varphi}\) is a corridor \(\abr{\Sen\rbr{\varphi}, \Mod\rbr{\varphi}} \colon G_{\Sigma} \to G_{\Sigma'}\), and for every signature of \(\Sigma\)\nb-variables \(X\), \(\tau_{\varphi, X}\) is a (natural) corridor \(\abr{\alpha_{\varphi, X}, \beta_{\varphi, X}} \colon G_{\Sigma}\rbr{X} \to G_{\Sigma'}\rbr{\Psi_{\varphi}\rbr{X}}\).
    \[
    \xymatrix {
      {\Sen\rbr{\Sigma}}
      \ar [r] ^-{\Sen\rbr{\varphi}}
      \ar [d] _{\alpha_{\Sigma, X}}
      & {\Sen\rbr{\Sigma'}}
      \ar [d] ^{\alpha_{\Sigma', \Psi_{\varphi}\rbr{X}}}
      \\
      {\Sen_{\Sigma}\rbr{X}}
      \ar [r] _-{\alpha_{\varphi, X}}
      & {\Sen_{\Sigma'}\rbr{\Psi_{\varphi}\rbr{X}}}
    }
    \qquad
    \xymatrix {
      {\Mod\rbr{\Sigma}}
      & {\Mod\rbr{\Sigma'}}
      \ar [l] _-{\Mod\rbr{\varphi}}
      \\
      {\Mod_{\Sigma}\rbr{X}}
      \ar [u] ^{\beta_{\Sigma, X}}
      & {\Mod_{\Sigma'}\rbr{\Psi_{\varphi}\rbr{X}}}
      \ar [u] _{\beta_{\Sigma', \Psi_{\varphi}\rbr{X}}}
      \ar [l] ^-{\beta_{\varphi, X}}
    }
    \]

  \end{itemize}
 \noindent In addition, we adopt notational conventions that are similar to those used for institutions.
  For example, we may use superscripts as in \(\Subst^{\GS}_{\Sigma}\) is order to avoid potential ambiguities; or we may drop the subscripts of \(\models_{\Sigma, X}\) when there is no danger of confusion.
  Also, we will often denote the functions \(\Sen\rbr{\varphi}\), \(\alpha_{\Sigma, X}\) and \(\Sen_{\Sigma}\rbr{\psi}\) by \(\varphi\rbr{\_}\), \(X\rbr{\_}\) and \(\psi\rbr{\_}\), respectively, and the functors \(\Mod\rbr{\varphi}\), \(\beta_{\Sigma, X}\) and \(\Mod_{\Sigma}\rbr{\psi}\) by \(\_ \reduct_{\varphi}\), \(\_ \reduct_{\Sigma}\) and \(\_ \reduct_{\psi}\).

  \begin{exa}
    Relational logic programming is based upon the generalized substitution system \(\aFOLssneqGS\) of the atomic fragment of single-sorted first-order logic  without equality.
    \[
    \aFOLssneqGS \colon \Sig^{\aFOLssneq} \to \SubstSys
    \]
    In this case, the category \(\Sig^{\aFOLssneq}\) is just the category of single-sorted first-order signatures.
    Every signature \(\abr{F, P}\) is  mapped to a substitution system \(\rbr[\big]{\aFOLssneqGS}_{\abr{F, P}}\) as described in Example~\ref{example:FP-substitution-system}, while every signature morphism \(\varphi \colon \abr{F, P} \to \abr{F', P'}\) resolves to a morphism of substitution systems for which \(\Psi_{\varphi}\) is the obvious translation of \(\abr{F, P}\)\nb-substitutions along \(\varphi\), and \(\kappa_{\varphi}\) is the corridor \(\aFOLssneq\rbr{\varphi}\).
    A more detailed presentation of first-order generalized substitution systems can be found in~\cite{Tutu-Fiadeiro:Institution-independent-logic-programming-2015}.
  \end{exa}

  \subsection{A Generalized Substitution System of Orchestration Schemes}

  What is essential about orchestration schemes with respect to the development of the service-oriented variant of logic programming is that they can be organized as a category \(\OS\) from which there exists a functor \(\OrcScheme\) into \(\SubstSys\) that allows us to capture some of the most basic aspects of service-oriented computing by means of logic-programming constructs.
  More precisely, orchestration schemes form the signatures of a generalized substitution system
  \[
  \OrcScheme \colon \OS \to \SubstSys
  \]
  through which the notions of service module, application, discovery and binding emerge as particular instances of the abstract notions of clause, query, unification and resolution.
  In this sense, \(\OrcScheme\) and \(\aFOLssneqGS\) can be regarded as structures having the same role in the description of service-oriented and relational logic programming, respectively.

  Morphisms of orchestration schemes are, intuitively, a way of encoding orchestrations.
  In order to understand how they arise in practice, let us consider a morphism \(\varphi\) between two algebraic signatures \(\Sigma\) and \(\Sigma'\) used in defining program expressions.  For instance, we may assume \(\Sigma\) to be the signature of structured programs discussed in Example~\ref{example:structured-programs}, and \(\varphi \colon \Sigma \to \Sigma'\) its extension with a new operation symbol \(\opname{repeat}\,\_\,\opname{until}\,\_ \colon \stname{Pgm}\, \stname{Cond} \to \stname{Pgm}\).
  Then, it is easy to notice that the translation of \(\Sigma\)\nb-terms (over a given set of program variables) along \(\varphi\) generalizes to a functor \(F\) between the categories of program expressions defined over \(\Sigma\) and \(\Sigma'\).
  Moreover, the choice of \(\varphi\) enables us to define a second functor \(U\), from program expressions over \(\Sigma'\) to program expression over \(\Sigma\), based on the derived signature morphism (see e.g.~\cite{Sannella-Tarlecki:Foundations-of-Algebraic-Specification-2011}) \(\Sigma' \to \Sigma\) that encodes the \(\opname{repeat}\,\_\,\opname{until}\,\_\) operation as the term \(\underline{1}\, \comp\, \opname{while}\, \opname{not}\, \underline{2}\, \opname{do}\, \underline{1}\, \opname{done}\).\footnote{In this context, \(\underline{1} \colon \stname{Pgm}\) and \(\underline{2} \colon \stname{Cond}\) are variables corresponding to the arguments of the derived operation.}
  The functor \(U\) is clearly a right inverse of \(F\) with respect to ground program expressions, whereas in general, for every program expression \(\ptm\) over \(\Sigma\) we actually obtain a morphism \(\eta_{\ptm} \colon \ptm \to U\rbr{F\rbr{\ptm}}\) as a result of the potential renaming of program variables; thus, the morphism \(\eta_{\ptm}\) accounts for translation of the program variables of \(\ptm\) along \(F \comp U\).
  Furthermore, for every program expression \(\ptm'\) over \(\Sigma'\), the translation of \(\Sigma\)\nb-sentences determined by \(\varphi\) extends to a map between the specifications over \(U\rbr{\ptm'}\) and the specifications over \(\ptm'\), which, as we will see, can be used to define a translation of the specifications over a program expression \(\ptm\) (given by \(\Sigma\)) to specifications over \(F\rbr{\ptm}\).
  With respect to the semantics, it is natural to expect that every program expression \(\ptm\) over \(\Sigma\) has the same behaviour as \(F\rbr{\ptm}\) and, even more, that every program expression \(\ptm'\) over \(\Sigma'\) (that may be built using \(\opname{repeat}\,\_\,\opname{until}\,\_\)), behaves in the same way as \(U\rbr{\ptm'}\).
  These observations lead us to the following formalization of the notion of morphism of orchestration schemes.

  \begin{defi}[Morphism of orchestration schemes]
    \label{definition:morphism-of-orchestration-schemes}
    A \emph{morphism} between orchestration schemes \(\abr{\Orc, \SSpec, \Grc, \Prop}\) and \(\abr{\Orc', \SSpec', \Grc', \Prop'}\) is a tuple \(\abr{F, U, \eta, \sigma}\), where
    \[
    \xymatrix {
      {\Orc}
      \ar @/^/ [r] ^{F}
      & {\Orc'}
      \ar @/^/ [l] ^{U}
    }
    \]
    \begin{itemize}

    \item \(F\) and \(U\) are functors as depicted above such that \(F\rbr{\Grc} \subseteq \Grc'\) and \(U\rbr{\Grc'} \subseteq \Grc\),

    \item \(\eta\) is a natural transformation \(1_{\Orc} \To F \comp U\) such that \(\eta_{\grc} = 1_{\grc}\) for every \(\grc \in \obj{\Grc}\), and
      
    \item \(\sigma\) is a natural transformation \(U \comp \SSpec \To \SSpec'\) such that for every ground orchestration \(\grc' \in \obj{\Grc'}\) and specification \(\rho \in \SSpec\rbr{U\rbr{\grc'}}\),
      \[
      \sigma_{\grc'}\rbr{\rho} \in \Prop'\rbr{\grc'}
      \qquad \text{if and only if} \qquad
      \rho \in \Prop\rbr{U\rbr{\grc'}}.
      \]

    \end{itemize}
  \end{defi}

  \begin{exa}
    \label{example:morphisms-of-orchestration-schemes-of-ARNs}
    Let \(\I = \abr{\Sig, \Sen, \Mod, \models}\) and \(\I' = \abr{\Sig', \Sen', \Mod', \models'}\) be two institutions suitable for defining orchestration schemes of \acn{ARN}s (according to the hypotheses introduced in Subsection~\ref{subsection:asynchronous-relational-networks}), and let \(\abr{\Upsilon, \alpha, \beta}\) be a morphism of institutions \(\I' \to \I\) such that \(\Upsilon \colon \Sig' \to \Sig\) is cocontinuous and \(\beta \colon \Mod' \To \Upsilon^{\op} \comp \Mod\) preserves cofree expansions and products.
    If \(\Upsilon\) and \(\beta\) admit sections, that is if there exist a functor \(\Phi \colon \Sig \to \Sig'\) such that \(\Phi \comp \Upsilon = 1_{\Sig}\) and a natural transformation \(\tau \colon \Mod \To \Phi^{\op} \comp \Mod'\) such that \(\tau \comp \rbr{\Phi^{\op} \cdot \beta} = 1_{\Mod}\), then \(\abr{\Upsilon, \alpha, \beta}\) gives rise to a morphism \(\abr{F, U, \eta, \sigma}\) between the orchestration schemes of \acn{ARN}s defined over \(\I\) and \(\I'\).
    In particular, the functor \(F\) maps the diagram and the models that label an \acn{ARN} defined over \(\I\) to their images under \(\Phi\) and \(\tau\); similarly, \(U\) maps \acn{ARN}s defined over \(\I'\) according to \(\Upsilon\) and \(\beta\); the natural transformation \(\eta\) is just an identity, and \(\sigma\) extends the \(\alpha\)\nb-translation of sentences to specifications. The additional properties of \(\Upsilon\) and \(\beta\) are essential for ensuring that the observable behaviour of ground networks is preserved.

    One may consider, for instance, the extension of \(\aLTL\) (in the role of \(\I\)) with new temporal modalities such as \emph{previous} and \emph{since}, as in~\cite{Knapp-Marczynski-Wirsing-Zawlocki:A-heterogeneous-approach-to-service-oriented-systems-specification-2010}; this naturally leads to a morphism of orchestration schemes for which both \(\Upsilon\) and \(\beta\) would be identities.
    Alternatively, one may explore the correspondence between deterministic weak \(\omega\)\nb-automata -- which form a subclass of Muller automata -- and sets of traces that are both B\"{u}chi and co\nb-B\"{u}chi deterministically recognizable -- for which a minimal automaton can be shown to exist (see e.g.~\cite{Maler-Staiger:Syntactic-congruences-for-omega-languages-1997,Loding:Efficient-minimization-of-deterministic-weak-omega-automata-2001}). In this case, in the roles of \(\I\) and \(\I'\) we could consider variants of \(\aLTL\) with models given by sets of traces and deterministic weak automata, respectively;\footnote{Note that, to ensure that model reducts are well defined for deterministic automata, one may need to restrict signature morphisms to injective maps.} \(\Upsilon\) and \(\alpha\) would be identities, \(\beta\) would define the language recognized by a given automaton, and \(\tau\) would capture the construction of minimal automata.
  \end{exa}

  It it easy to see that the morphisms of orchestration schemes compose in a natural way in terms of their components, thus giving rise to a category of orchestration schemes.

  \begin{prop}
    The morphisms of orchestration schemes can be composed as follows:
    \[
    \abr{F, U, \eta, \sigma} \comp {\abr{F', U', \eta', \sigma'}} = \abr{F \comp F', U' \comp U, \eta \comp {\rbr{F \cdot \eta' \cdot U}}, \rbr{U' \cdot \sigma} \comp \sigma'}.
    \]
    Under this composition, orchestration schemes and their morphisms form a category \(\OS\).
    \qed
  \end{prop}

  The definition of the functor \(\OrcScheme\) is grounded on two simple ideas:
  \begin{enumerate}

  \item Orchestrations can be regarded as signatures of variables; they provide sentences in the form of specifications, and models as morphisms into ground orchestrations -- which can also be seen, in the case of \acn{ARN}s, for example, as collections of ground networks assigned to the `variables' of the considered orchestration.  In addition, we can define a satisfaction relation between the models and the sentences of an orchestration based of the evaluation of specifications with respect to ground orchestrations.  In this way, every orchestration scheme yields an institution whose composition resembles that of the so-called institutions of extended models~\cite{Schroder-Mossakowski-Luth:Type-class-polymorphism-2004}.

  \item There is a one-to-one correspondence between institutions and substitution systems defined over the initial room \(\abr{\emptyset, \tcat, \emptyset}\) -- the room given by the empty set of sentences, the terminal category \(\tcat\), and the empty satisfaction relation.  The effect of this is that a clause can be described as `correct' whenever it is satisfied by the sole model of \(\abr{\emptyset, \tcat, \emptyset}\); therefore, we obtain precisely the notion of correctness of a service module~\cite{Fiadeiro-Lopes-Bocchi:An-abstract-model-for-service-discovery-and-binding-2011}: all models of the underlying signature of variables, i.e.\ of the orchestration, that satisfy the antecedent of the clause satisfy its consequent as well.

  \end{enumerate}
  Formally, \(\OrcScheme\) results from the composition of two functors, \(\osIns \colon \OS \to \coIns\) and \(\osSS \colon \coIns \to \SubstSys\), that implement the general constructions outlined above.
  \[
  \xymatrix @C+=7em @!0 {
    {\OS}
    \ar [r] ^-{\osIns}
    \ar @{-<} `u[rr] `[rr] ^-{\OrcScheme} [rr]
    & {\coIns}
    \ar [r] ^-{\osSS}
    & {\SubstSys}
  }
  \]
  The functor \(\osIns\) carries most of the complexity of \(\OrcScheme\), and is discussed in detail in Theorem~\ref{theorem:osIns}.
  Concerning \(\osSS\), we recall from~\cite{Tutu-Fiadeiro:Institution-independent-logic-programming-2015} that the category \(\coIns\) of institution comorphisms can also be described as the category \(\GrCat{\FCat{\_}{\Room}}\) of functors into \(\Room\), and that any functor \(G \colon \catnamecap{K} \to \catnamecap{K}'\) can be extended to a functor \(\GrCat{\FCat{\_}{\catnamecap{K}}} \to \GrCat{\FCat{\_}{\catnamecap{K}'}}\) that is given essentially by the right-composition with \(G\).
  In particular, the isomorphism \(\Room \to \commacat{\abr{\emptyset, \tcat, \emptyset}}{\Room}\) that maps every room \(\abr{S, \catnamecap{M}, \models}\) to the unique corridor \(\abr{\emptyset, \tcat, \emptyset} \to \abr{S, \catnamecap{M}, \models}\) generates an isomorphism of categories between \(\GrCat{\FCat{\_}{\Room}}\), i.e.\ \(\coIns\), and \(\GrCat{\FCat{\_}{\commacat{\abr{\emptyset, \tcat, \emptyset}}{\Room}}}\).  The latter is further embedded into \(\SubstSys\), defining in this way, by composition, the required functor \(\osSS\).
  To sum up, \(\osSS\) maps every institution \(\I \colon \Sig \to \Room\) to the substitution system \(\SubSys \colon \Sig \to \commacat{\abr{\emptyset, \tcat, \emptyset}}{\Room}\) for which \(\SubSys\rbr{\Sigma}\), for every signature \(\Sigma \in \obj{\Sig}\), is the unique corridor between \(\abr{\emptyset, \tcat, \emptyset}\) and \(\I\rbr{\Sigma}\).

  \begin{thm}
    \label{theorem:osIns}
    The following map defines a functor \(\osIns \colon \OS \to \coIns\).
    \begin{itemize}

    \item For any orchestration scheme \(\Oscheme = \abr{\Orc, \SSpec, \Grc, \Prop}\), \(\osIns\rbr{\Oscheme}\) is the institution whose category of signatures is \(\Orc\), sentence functor is \(\SSpec\), model functor is \(\commacat{\_}{\Grc}\), and whose family of satisfaction relations is given by
      \[
      \rbr{\delta \colon \orc \to \grc} \models_{\orc} \osp
      \qquad \text{if and only if} \qquad
      \SSpec\rbr{\delta}\rbr{\osp} \in \Prop\rbr{\grc}
      \]
      for every orchestration \(\orc\), every \(\orc\)\nb-model \(\delta\), i.e.\ every morphism of orchestrations \(\delta \colon \orc \to \grc\) such that \(\grc\) is ground, and every specification \(\osp\) over \(\orc\).\footnote{Moreover, \(\osIns\rbr{\Oscheme}\) is exact, because the functor \(\commacat{\_}{\Grc} \colon \Orc^{\op} \to \Cat\) is continuous (see e.g.~\cite{Meseguer:General-logics-1989}).}

    \item For any morphism of orchestration schemes \(\abr{F, U, \eta, \sigma} \colon \Oscheme \to \Oscheme'\), with \(\Oscheme\) as above and \(\Oscheme'\) given by \(\abr{\Orc', \SSpec', \Grc', \Prop'}\), \(\osIns\rbr{F, U, \eta, \sigma}\) is the comorphism of institutions \(\abr{F, \alpha, \beta} \colon \osIns\rbr{\Oscheme} \to \osIns\rbr{\Oscheme'}\) defined by
      \begin{align*}
        \alpha_{\orc} & = \SSpec\rbr{\eta_{\orc}} \comp \sigma_{F\rbr{\orc}} \\
        \beta_{\orc} & = \upsilon_{F\rbr{\orc}} \comp {\rbr{\commacat{\eta_{\orc}}{\Grc}}}
      \end{align*}
      for every orchestration \(\orc \in \obj{\Orc}\), where \(\upsilon \colon \rbr{\commacat{\_}{\Grc'}} \To U^{\op} \comp {\rbr{\commacat{\_}{\Grc}}}\) is the natural transformation given by \(\upsilon_{\orc'}\rbr{x} = U\rbr{x}\) for every orchestration \(\orc' \in \obj{\Orc'}\) and every object or arrow \(x\) of the comma category \(\commacat{\orc'}{\Grc'}\).

    \end{itemize}
  \end{thm}

  \proof
  For the first part, all we need to show is that the satisfaction condition holds; but this follows easily since for every morphism of orchestrations \(\theta \colon \orc_{1} \to \orc_{2}\), every \(\orc_{1}\)\nb-specification \(\osp\) and every \(\orc_{2}\)\nb-model \(\delta \colon \orc_{2} \to \grc\),
  \begin{align*}
    \delta \models_{\orc_{2}} \SSpec\rbr{\theta}\rbr{\osp}
    & \qquad \text{if and only if} \qquad \SSpec\rbr{\theta \comp \delta}\rbr{\osp} \in \Prop\rbr{\grc} \\
    & \qquad \text{if and only if} \qquad \rbr{\commacat{\theta}{\Grc}}\rbr{\delta} = \theta \comp \delta \models_{\orc_{2}} \osp.
  \end{align*}

  As regards the second part of the statement, let us begin by noticing that \(\alpha\) and \(\beta\) are the natural transformations \(\rbr{\eta \cdot \SSpec} \comp {\rbr{F \cdot \sigma}}\) and \(\rbr{\eta^{\op} \cdot \rbr{\commacat{\_}{\Grc}}} \comp {\rbr{F^{\op} \cdot \upsilon}}\), respectively.
  Then, in order to verify that \(\abr{F, \alpha, \beta}\) is indeed a comorphism \(\osIns\rbr{\Oscheme} \to \osIns\rbr{\Oscheme'}\), consider an orchestration \(\orc\) in \(\Orc\), a model \(\delta' \colon F\rbr{\orc} \to \grc'\) of \(F\rbr{\orc}\), and a specification \(\osp\) over \(\orc\).
  Assuming that \(\models'\) is the family of satisfaction relations of \(\osIns\rbr{\Oscheme'}\), we deduce that
  \begin{align*}
    \MoveEqLeft \delta' \models'_{F\rbr{\orc}} \alpha_{\orc}\rbr{\osp} \\
 & \text{iff} \quad \SSpec'\rbr{\delta'}\rbr{\alpha_{\orc}\rbr{\osp}} \in \Prop'\rbr{\grc'}
 && \text{by the definition of \(\models'_{F\rbr{\orc}}\)} \\
 & \text{iff} \quad \SSpec'\rbr{\delta'}\rbr{\sigma_{F\rbr{\orc}}\rbr{\SSpec\rbr{\eta_{\orc}}\rbr{\osp}}} \in \Prop'\rbr{\grc'}
 && \text{by the definition of \(\alpha_{\orc}\)} \\
 & \text{iff} \quad \sigma_{\grc'}\rbr{\SSpec\rbr{\eta_{\orc} \comp U\rbr{\delta'}}\rbr{\osp}} \in \Prop'\rbr{\grc'}
 && \text{by the naturality of \(\sigma\)} \\
 & \text{iff} \quad \SSpec\rbr{\eta_{\orc} \comp U\rbr{\delta'}}\rbr{\osp} \in \Prop\rbr{U\rbr{\grc'}}
 && \text{since \(\Prop\rbr{U\rbr{\grc'}} = \sigma_{\grc'}^{-1}\rbr{\Prop'\rbr{\grc'}}\)} \\
 & \text{iff} \quad \eta_{\orc} \comp U\rbr{\delta'} \models_{\orc} \osp
 && \text{by the definition of \(\models_{\orc}\)} \\
 & \text{iff} \quad \beta_{\orc}\rbr{\delta'} \models_{\orc} \osp
 && \text{by the definition of \(\beta_{\orc}\)}.
  \end{align*}

\noindent  Finally, it is easy to see that \(\osIns\) preserves identities.
  To prove that it also preserves composition, let \(\abr{F, U, \eta, \sigma}\) and \(\abr{F', U', \eta', \sigma'}\) be  morphisms of orchestration schemes as below, and suppose that \(\osIns\rbr{F, U, \eta, \sigma} = \abr{F, \alpha, \beta}\) and \(\osIns\rbr{F', U', \eta', \sigma'} = \abr{F', \alpha', \beta'}\).
  \[
  \xymatrix @C=0em @L+.5ex {
    {\abr{\Orc, \SSpec, \Grc, \Prop}}
    \ar [rr] ^-{\abr{F, U, \eta, \sigma}}
    \ar `d[rrrr] `[rrrr] _-{\abr{F \comp F', U' \comp U, \eta \comp {\rbr{F \cdot \eta' \cdot U}}, \rbr{U' \cdot \sigma} \comp \sigma'}} [rrrr]
    & \hspace{2em}
    & {\abr{\Orc', \SSpec', \Grc', \Prop'}}
    \ar [rr] ^-{\abr{F', U', \eta', \sigma'}}
    & \hspace{3.25em}
    & {\abr{\Orc'', \SSpec'', \Grc'', \Prop''}}
  }
  \]
  In addition, let \(\upsilon \colon \rbr{\commacat{\_}{\Grc'}} \To U^{\op} \comp {\rbr{\commacat{\_}{\Grc}}}\) and \(\upsilon' \colon \rbr{\commacat{\_}{\Grc''}} \To U^{\prime\op} \comp {\rbr{\commacat{\_}{\Grc'}}}\) be the natural transformations involved in the definitions of \(\beta\) and \(\beta'\), respectively.
  Based on the composition of morphisms of orchestration schemes and on the definition of \(\osIns\), it follows that \(\osIns\rbr{\abr{F, U, \eta, \sigma} \comp {\abr{F', U', \eta', \sigma'}}}\) is a comorphism of institutions of the form \(\abr{F \comp F', \alpha'', \beta''}\), where \(\alpha''\) and \(\beta''\) are given by
  \begin{align*}
    \alpha''_{\orc} & = \SSpec\rbr{\rbr{\eta \comp {\rbr{F \cdot \eta' \cdot U}}}_{\orc}} \comp {\rbr{\rbr{U' \cdot \sigma} \comp \sigma'}_{\rbr{F \comp F'}\rbr{\orc}}} \\
    \beta''_{\orc} & = \rbr{\upsilon' \comp {\rbr{U^{\prime\op} \cdot \upsilon}}}_{\rbr{F \comp F'}\rbr{\orc}} \comp {\rbr{\commacat{\rbr{\eta \comp {\rbr{F \cdot \eta' \cdot U}}}_{\orc}}{\Grc}}}.
  \end{align*}
  In order to complete the proof we need to show that \(\alpha'' = \alpha \comp {\rbr{F \cdot \alpha'}}\) and \(\beta'' = \rbr{F \cdot \beta'} \comp \beta\).
  Each of these equalities follows from a sequence of straightforward calculations that relies on the naturality of \(\sigma\) (in the case of \(\alpha''\)), or on the naturality of \(\upsilon\) (in the case of \(\beta''\)).
  \begin{align*}
    \alpha''_{\orc} & = \SSpec\rbr{\eta_{\orc}} \comp 
                      \underbracket[.11ex][.5ex]{
                      \SSpec\rbr{U\rbr{\eta'_{F\rbr{\orc}}}} \comp \sigma_{\rbr{F \comp F' \comp U'}\rbr{\orc}}
                      } \comp \sigma'_{\rbr{F \comp F'}\rbr{\orc}} \\
                    & = \SSpec\rbr{\eta_{\orc}} \comp 
                      \sigma_{F\rbr{\orc}} \comp \SSpec'\rbr{\eta'_{F\rbr{\orc}}}
                      \comp \sigma'_{\rbr{F \comp F'}\rbr{\orc}} \\
                    & = \alpha_{\orc} \comp \alpha'_{F\rbr{\orc}} \\
    \beta''_{\orc} & = \upsilon'_{\rbr{F \comp F'}\rbr{\orc}} \comp 
                     \underbracket[.11ex][.5ex]{
                     \upsilon_{\rbr{F \comp F' \comp U'}\rbr{\orc}} \comp {\rbr{\commacat{U\rbr{\eta'_{F\rbr{\orc}}}}{\Grc}}}
                     } \comp {\rbr{\commacat{\eta_{\orc}}{\Grc}}} \\
                    & = \upsilon'_{\rbr{F \comp F'}\rbr{\orc}} \comp 
                      {\rbr{\commacat{\eta'_{F\rbr{\orc}}}{\Grc'}}} \comp \upsilon_{F\rbr{\orc}}
                      \comp {\rbr{\commacat{\eta_{\orc}}{\Grc}}} \\
                    & = \beta'_{F\rbr{\orc}} \comp \beta_{\orc}
                    \rlap{\hbox to281 pt{\hfill\qEd}}
  \end{align*}

  \begin{cor}
    The pair \(\abr{\OS, \OrcScheme}\) defines a generalized substitution system.
    \qed
  \end{cor}

  We recall from~\cite{Tutu-Fiadeiro:Institution-independent-logic-programming-2015} that, in order to be used as semantic frameworks for logic programming, generalized substitution systems need to ensure a weak model-amalgamation property between the models that are ground and those that are defined by signatures of variables.
  This property entails that the satisfaction of quantified sentences (and in particular, of clauses and queries) is invariant under change of notation.
  In the case of \(\OrcScheme\), this means, for example, that the correctness property of service modules does not depend on the actual orchestration scheme over which the modules are defined.

  \begin{defi}[Model amalgamation]
    A generalized substitution system \(\GS \colon \Sig \to \SubstSys\) has \emph{weak model amalgamation} when for every signature morphism \(\varphi \colon \Sigma \to \Sigma'\) and every signature of \(\Sigma\)\nb-variables \(X\), the diagram depicted below is a weak pullback.
    \[
    \xymatrix {
      {\obj{\Mod\rbr{\Sigma}}}
      & {\obj{\Mod\rbr{\Sigma'}}}
      \ar [l] _-{\_ \reduct_{\varphi}}
      \\
      {\obj{\Mod_{\Sigma}\rbr{X}}}
      \ar [u] ^{\_ \reduct_{\Sigma}}
      & {\obj{\Mod_{\Sigma'}\rbr{\Psi_{\varphi}\rbr{X}}}}
      \ar [u] _{\_ \reduct_{\Sigma'}}
      \ar [l] ^-{\beta_{\varphi, X}}
    }
    \]
  \end{defi}

  \noindent This means that for every model \(\Sigma'\)\nb-model \(M'\) and every \(X\)\nb-model \(N\) such that \(M' \reduct_{\varphi} = N \reduct_{\Sigma}\) there exists a \(\Psi_{\varphi}\rbr{X}\)\nb-model \(N'\) that satisfies \(N' \reduct_{\Sigma'} = M'\) and \(\beta_{\varphi, X}\rbr{N'} = N\).

  \begin{prop}
    \label{proposition:weak-model-amalgamation-in-OrcScheme}
    The generalized substitution system \(\OrcScheme \colon \OS \to \SubstSys\) has weak model amalgamation.
  \end{prop}

  \proof
  Let \(\varphi\) be a morphism \(\abr{F, U, \eta, \sigma}\) between orchestration schemes \(\Oscheme\) and \(\Oscheme'\) as in Definition~\ref{definition:morphism-of-orchestration-schemes}, and let \(\orc\) be an orchestration of \(\Oscheme\).
  Since orchestrations define substitution systems over the initial room \(\abr{\emptyset, \tcat, \emptyset}\), we can redraw the diagram of interest as follows:
  \[
  \xymatrix {
    {\obj{\tcat}}
    & {\obj{\tcat}}
    \ar [l] _-{\_ \reduct_{\varphi}}
    \\
    {\obj{\commacat{\orc}{\Grc}}}
    \ar [u] ^{\_ \reduct_{\Oscheme}}
    & {\obj{\commacat{F\rbr{\orc}}{\Grc'}}}
    \ar [u] _{\_ \reduct_{\Oscheme'}}
    \ar [l] ^-{\beta_{\varphi, \orc}}
  }
  \]
  It is easy to see that the above diagram depicts a weak pullback if and only if \(\beta_{\varphi, \orc}\) is surjective on objects.
  By Theorem~\ref{theorem:osIns}, we know that \(\beta_{\varphi, \orc}\rbr{\delta'} = \eta_{\orc} \comp U\rbr{\delta'}\) for every object \(\delta' \colon F\rbr{\orc} \to \grc'\) in \(\obj{\commacat{F\rbr{\orc}}{\Grc'}}\).
  Therefore, for every \(\delta \colon \orc \to \grc\) in \(\obj{\commacat{\orc}{\Grc}}\) we obtain 
  \begin{align*}
    \beta_{\varphi, \orc}\rbr{F\rbr{\delta}} & = \eta_{\orc} \comp U\rbr{F\rbr{\delta}} \\
                                             & = \delta \comp \eta_{\grc}
                                             && \text{by the naturality of \(\eta\)} \\
                                             & = \delta
                                             && \text{because, by definition, \(\eta_{\grc}\) is an identity}.
                                             % \qedhere
  \end{align*}
  \qed

  \begin{rem}
    \label{remark:preservation-of-satisfaction-in-OrcScheme}
    In addition to model amalgamation, it is important to notice that, similarly to \(\aFOLssneqGS\), in \(\OrcScheme\) the satisfaction of sentences is preserved by model homomorphisms.
    This is an immediate consequence of the fact that, in every orchestration scheme, the morphisms of ground orchestrations preserve properties: given an orchestration \(\orc\), a specification \(\osp\) over \(\orc\), and a homomorphism \(\zeta\) between \(\orc\)\nb-models \(\delta_{1}\) and \(\delta_{2}\) as depicted below, if \(\SSpec\rbr{\delta_{1}}\rbr{\osp}\) is a property of \(\grc_{1}\) then \(\SSpec\rbr{\delta_{2}}\rbr{\osp} = \SSpec\rbr{\zeta}\rbr{\SSpec\rbr{\delta_{1}}\rbr{\osp}}\) is a property of \(\grc_{2}\); therefore, \(\delta_{1} \models^{\OrcScheme} \osp\) implies \(\delta_{2} \models^{\OrcScheme} \osp\).
    \[
    \xymatrix @C=1em {
      & {\orc}
      \ar [dl] _{\delta_{1}}
      \ar [dr] ^{\delta_{2}}
      \\
      {\grc_{1}}
      \ar [rr] _{\zeta}
      && {\grc_{2}}
    }
    \]
  \end{rem}

  \subsection{The Clausal Structure of Services}

  Given the above constructions, we can now consider a service-oriented notion of clause, defined over the generalized substitution system \(\OrcScheme\) rather than \(\aFOLssneqGS\).
  Intuitively, this means that we replace first-order signatures with orchestration schemes, sets of variables with orchestrations, and first-order sentences (over given sets of variables) with specifications.  Furthermore, certain orchestration schemes allow us to identify structures that correspond to finer-grained notions like variable and term: in the case of program expressions, variables and terms have their usual meaning (although we only take into account executable expressions), whereas in the case of \acn{ARN}s, variables and terms materialize as requires-points and sub-\acn{ARN}s defined by provides-points.

  The following notion of service clause corresponds to the concept of service module presented in~\cite{Fiadeiro-Lopes-Bocchi:An-abstract-model-for-service-discovery-and-binding-2011}, and also to the concept of orchestrated interface discussed in~\cite{Fiadeiro-Lopes:An-interface-theory-for-service-oriented-design-2013}.

  \begin{defi}[Service clause]
    A \emph{(definite) service-oriented clause} over a given orchestration scheme \(\Oscheme = \abr{\Orc, \SSpec, \Grc, \Prop}\) is a structure \(\dclause{\orc}{P}{R}\), also denoted
    \[
    \arrclause{\orc}{P}{R}
    \]
    where \(\orc\) is an orchestration of \(\Oscheme\), \(P\) is a specification over \(\orc\) -- called the \emph{provides-interface} of the clause -- and \(R\) is a finite set of specifications over \(\orc\) -- the \emph{requires-interface} of the clause.
  \end{defi}

  The semantics of service-oriented clauses is defined just as the semantics of first-order clauses, except they are evaluated within the generalized substitution system \(\OrcScheme\) instead of \(\aFOLssneqGS\).
  As mentioned before, this means that we can only distinguish whether or not a clause is correct.

  \begin{defi}[Correct clause]
    A service-oriented clause \(\dclause{\orc}{P}{R}\) is \emph{correct} if for every morphism \(\delta \colon \orc \to \grc\) such that \(\grc\) is a ground orchestration and \(\SSpec\rbr{\delta}\rbr{R}\) consists only of properties of \(\grc\), the specification \(\SSpec\rbr{\delta}\rbr{P}\) is also a property of \(\grc\).
  \end{defi}

  \noindent In other words, a service clause is correct if the specification given by its provides-interface is ensured by its orchestration and the specifications of its requires-interface.

  \begin{exa}
    \label{example:journey-planner-module}
    We have already encountered several instances of service clauses in the form of the program modules depicted in Figure~\ref{figure:program-modules}.  Their provides- and requires-interfaces are placed on the left- and right-hand side of their orchestrations, and are represented using symbolic forms that are traditionally associated with services.

    To illustrate how service modules can be defined as clauses over \acn{ARN}s, notice that the network \(\arnname{JourneyPlanner}\) introduced in Example~\ref{example:journey-planner} can orchestrate a module named Journey Planner that consistently delivers the requested directions, provided that the routes and the timetables can be obtained whenever they are needed.
    This can be described in logical terms through the following (correct) service-oriented clause:
    \[
    \arrclause{
      \arnname{JourneyPlanner}
    }{
      \lsen{\portname{JP_{1}}}{\rho^{\arnname{JP}}}
    }{
      \cbr[\big]{\lsen{\portname{R_{1}}}{\rho^{\arnname{JP}}_{1}}, \lsen{\portname{R_{2}}}{\rho^{\arnname{JP}}_{2}}}
    }
    \]
    where \(\rho^{\arnname{JP}}\), \(\rho^{\arnname{JP}}_{1}\) and \(\rho^{\arnname{JP}}_{2}\) correspond to the \(\aLTL\)\nb-sentences
    \(\ltlalways \rbr{\dact{\msgname{planJourney}} \plimplies \ltleventually \pact{\msgname{directions}}}\),
    \(\ltlalways \rbr{\dact{\msgname{getRoutes}} \plimplies \ltleventually \pact{\msgname{routes}}}\) and
    \(\ltlalways \rbr{\dact{\msgname{routes}} \plimplies \ltleventually \pact{\msgname{timetables}}}\), respectively.
  \end{exa}

  Client applications are captured in the present setting by service-oriented queries.
  The way they are defined is similar to that of service clauses, but their semantics is based on an existential quantification, not on a universal one.

  \begin{defi}[Service query]
    A \emph{service-oriented query} over an orchestration scheme \(\Oscheme = \abr{\Orc, \SSpec, \Grc, \Prop}\) is a structure \(\query{\orc}{Q}\), also written
    \[
    \arrquery{\orc}{Q}
    \]
    such that \(\orc\) is an orchestration of \(\Oscheme\), and \(Q\) is a finite set of specifications over \(\orc\) that defines the \emph{requires-interface} of the query.
  \end{defi}

  \begin{defi}[Satisfiable query]
    A service-oriented query \(\query{\orc}{Q}\) is said to be \emph{satisfiable} if there exists a morphism of orchestrations \(\delta \colon \orc \to \grc\) such that \(\grc\) is ground and all specifications in \(\SSpec\rbr{\delta}\rbr{Q}\) are properties of \(\grc\).
  \end{defi}

  \begin{exa}
    \label{example:traveller-application}
    Figure~\ref{figure:client-ARN} outlines the \acn{ARN} of a possible client application for the service module Journey Planner discussed in Example~\ref{example:journey-planner-module}.  We specify the actual application, called Traveller, through the service query
    \[
    \arrquery{
      \arnname{Traveller}
    }{
      \cbr[\big]{\lsen{\portname{R_{1}}}{\rho^{\arnname{T}}_{1}}}
    }
    \]
    given by the \(\aLTL\)\nb-sentence \(\ltlalways \rbr{\dact{\msgname{getRoute}} \plimplies \ltleventually \pact{\msgname{route}}}\).

    \begin{figure}[h]
      \centering

      \begin{tikzpicture}
        % the process T
        
        \node [align=center, minimum height=7ex, minimum width=4em] (T) {
          \(\processname{T}\) \\[1ex]
          \(\Lambda_{\processname{T}}\)
        };
        
        \node [port] (T1) [right=0em of T, align=left] {
          \(\pmsgl{getRoute}\) \\
          \(\dmsgl{route}\)
        };
        \node [port-label] [above=.5ex of T1] {\(\portname{T_{1}}\)};

        % the port R1
        
        \node [port] (R1) [right=3em of T1, align=right] {
          \(\dmsgr{getRoute}\) \\
          \(\pmsgr{route}\)
        };
        \node [port-label] [above=.5ex of R1] {\(\portname{R_{1}}\)};

        % the connection

        \draw [rounded corners]
        ($(T1.north west) + (0.5,0.1)$)
        to ($(R1.north east)  + (0.1,0.1)$)
        to ($(R1.south east)  + (0.1,-0.1)$)
        to ($(T1.south west) + (-0.1,-0.1)$)
        to ($(T1.north west) + (-0.1,0.1)$)
        to ($(T1.north west) + (0.5,0.1)$);
        
        \path (T1) -- node [connection] {
          \(\connectionname{C}\) \\
          \(\Lambda_{\connectionname{C}}\)
        } (T1 -| R1.west);

        \begin{pgfonlayer}{background}       
          \node [process] [fit=(T)] {};
        \end{pgfonlayer}
      \end{tikzpicture}

      \caption{The \acn{ARN} \(\arnname{Traveller}\)}
      \label{figure:client-ARN}
    \end{figure}
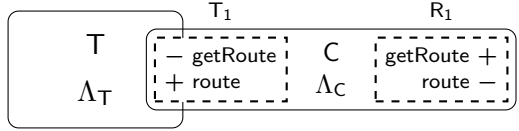

  \end{exa}

  \subsection{Resolution as Service Discovery and Binding}
  \label{subsection:resolution-as-service-discovery-and-binding}

  Let us now turn our attention to the dynamic aspects of service-oriented computing that result from the process of service discovery and binding~\cite{Fiadeiro-Lopes-Bocchi:An-abstract-model-for-service-discovery-and-binding-2011}.
  \emph{Service discovery} represents, as in conventional logic programming, the search for a module (service clause) that can be bound to a given application (service query) in order to take it one step closer to a possible solution, i.e.\ to a `complete' application capable of fulfilling its goal.
  From a technical point of view, both discovery and binding are subject to \emph{matching} the requires-interface of the application, or more precisely, one of its specifications, with the provides-interface of the module under consideration.
  This is usually achieved through a suitable notion of \emph{refinement} of specifications.
  For instance, in the case of program expressions, given specifications \(\pspec{\iota_{1}}{\rho_{1}}{\rho_{1}'}\) and \(\pspec{\iota_{2}}{\rho_{2}}{\rho_{2}'}\) over programs \(\ptm_{1} \colon \pexp_{1}\) and \(\ptm_{2} \colon \pexp_{2}\), respectively, \(\pspec{\iota_{2}}{\rho_{2}}{\rho_{2}'}\) refines \(\pspec{\iota_{1}}{\rho_{1}}{\rho_{1}'}\) up to a cospan
  \[
  \xymatrix @C+2em {
    {\ptm_{1} \colon \pexp_{1}}
    \ar [r] ^-{\abr{\psi_{1}, \pi_{1}}}
    & {\ptm \colon \pexp}
    & {\ptm_{2} \colon \pexp_{2}}
    \ar [l] _-{\abr{\psi_{2}, \pi_{2}}}
  }
  \]
  if by translation we obtain specifications that refer to the same position of \(\ptm \colon \pexp\), i.e.\ \(\pi_{1} \cdot \iota_{1} = \pi_{2} \cdot \iota_{2}\), such that the pre-condition \(\psi_{2}\rbr{\rho_{2}}\) is weaker that \(\psi_{1}\rbr{\rho_{1}}\), and the post-condition \(\psi_{2}\rbr{\rho_{2}'}\) is stronger than \(\psi_{1}\rbr{\rho_{1}'}\), meaning that
  \[
  \psi_{1}\rbr{\rho_{1}} \models^{\POA} \psi_{2}\rbr{\rho_{2}}
  \qquad \text{and} \qquad
  \psi_{2}\rbr{\rho_{2}'} \models^{\POA} \psi_{1}\rbr{\rho_{1}'}.
  \]
  This notion of refinement reflects the rules of consequence introduced in~\cite{Hoare:An-axiomatic-basis-for-computer-programming-1969} (see also~\cite{Morgan:Programming-from-Specifications-1994}, whence we also adopt the notation \(\pspec{\iota_{1}}{\rho_{1}}{\rho_{1}'} \sqsubseteq \pspec{\iota_{2}}{\rho_{2}}{\rho_{2}'}\) used in Figure~\ref{figure:program-derivation}).

  In a similar manner, in the case of \acn{ARN}s, a specification \(\lsen{x_{1}}{\rho_{1}}\) over a network \(\arnet_{1}\) is refined by another specification \(\lsen{x_{2}}{\rho_{2}}\) over a network \(\arnet_{2}\) up to a cospan of morphisms of \acn{ARN}s \(\abr{\theta_{1} \colon \arnet_{1} \to \arnet, \theta_{2} \colon \arnet_{2} \to \arnet}\) when \(\theta_{1}\rbr{x_{1}} = \theta_{2}\rbr{x_{2}}\) and \(\theta^{\arnpt}_{2, x_{2}}\rbr{\rho_{2}} \models^{\aLTL} \theta^{\arnpt}_{1, x_{1}}\rbr{\rho_{1}}\)~\cite{Tutu-Fiadeiro:A-logic-programming-semantics-of-services-2013}.

  Both of these notions of refinement generalize to the following concept of unification.

  \begin{defi}[Unification]
    \label{definition:service-oriented-unification}
    Let \(\osp_{1}\) and \(\osp_{2}\) be specifications defined over orchestrations \(\orc_{1}\) and \(\orc_{2}\), respectively, of an arbitrary but fixed orchestration scheme.
    We say that the ordered pair \(\abr{\osp_{1}, \osp_{2}}\) is \emph{unifiable} if there exists a cospan of morphisms of orchestrations
    \[
    \xymatrix {
      {\orc_{1}}
      \ar [r] ^-{\theta_{1}}
      & {\orc}
      & {\orc_{2}}
      \ar [l] _-{\theta_{2}}
    }
    \]
    called the \emph{unifier} of \(\osp_{1}\) and \(\osp_{2}\), such that \(\theta_{2}\rbr{\osp_{2}} \models^{\OrcScheme} \theta_{1}\rbr{\osp_{1}}\).
  \end{defi}

  \noindent Therefore, \(\abr{\theta_{1}, \theta_{2}}\) is a unifier of \(\osp_{1}\) and \(\osp_{2}\) if and only if, for every morphism of orchestrations \(\delta \colon \orc \to \grc\) such that \(\grc\) is a ground orchestration, if \(\SSpec\rbr{\theta_{2} \comp \delta}\rbr{\osp_{2}}\) is a property of \(g\) then so is \(\SSpec\rbr{\theta_{1} \comp \delta}\rbr{\osp_{1}}\).

  In conventional logic programming, the resolution inference rule simplifies the current goal and at the same time, through unification, yields computed substitutions that could eventually deliver a solution to the initial query.  This process is accurately reflected in the case of service-oriented computing by \emph{service binding}.
  However, unlike relational logic programming, in the case of services the emphasis is put not on the computed morphisms of orchestrations (i.e.\ on substitutions), but on the dynamic reconfiguration of the orchestrations (i.e.\ of the signatures of variables) that underlie the considered applications.

  \begin{defi}[Resolution]
    Let \(\query{\orc_{1}}{Q_{1}}\) be a query and \(\dclause{\orc_{2}}{P_{2}}{R_{2}}\) a clause defined over an arbitrary but fixed orchestration scheme.
    A query \(\query{\orc}{Q}\) is said to be \emph{derived by resolution} from \(\query{\orc_{1}}{Q_{1}}\) and \(\dclause{\orc_{2}}{P_{2}}{R_{2}}\) using the \emph{computed morphism} \(\theta_{1} \colon \orc_{1} \to \orc\) when
    \[
    \xymatrix @H=4ex @R=0ex @C-2em {
      {\mathrlap{\arrquery{\orc_{1}}{
            Q_{1}
          }
        }}
      && {\mathllap{\arrclause{\orc_{2}}{
            P_{2}
          }{
            R_{2}
          }
        }} \\
      & {\arrquery{\orc}{
          \theta_{1}\rbr{Q_{1} \setminus \cbr{\osp_{1}}} \cup \theta_{2}\rbr{R_{2}}
        }}
      \ar @{-} [ul]!DL-<1ex,0ex>;[ur]!DR+<1ex,0ex> |>/1em/*[r]{\theta_{1}}
    }
    \]
    \begin{itemize}
      
    \item \(\theta_{1}\) can be extended to a unifier \(\abr{\theta_{1}, \theta_{2}}\) of a specification \(\osp_{1} \in Q_{1}\) and \(P_{2}\), and
      
    \item \(Q\) is the set of specifications given by the translation along \(\theta_{1}\) and \(\theta_{2}\) of the specifications in \(Q_{1} \setminus \cbr{\osp_{1}}\) and \(R_{2}\).
      
    \end{itemize}
  \end{defi}

  \begin{exa}
    \label{example:service-oriented-resolution}
    Consider the service query and the clause detailed in Examples~\ref{example:traveller-application} and~\ref{example:journey-planner-module}.  One can easily see that the single specification \(\lsen{\portname{R_{1}}}{\rho^{\arnname{T}}_{1}}\) of the requires-interface of the application Traveller and the provides-interface \(\lsen{\portname{JP_{1}}}{\rho^{\arnname{JP}}}\) of the module Journey Planner form a unifiable pair: they admit, for instance, the unifier \(\abr{\theta_{1}, \theta_{2}}\) given by
    \[
    \xymatrix {
      {\arnname{Traveller}}
      \ar [r] ^-{\theta_{1}} 
      & {\arnname{JourneyPlannerApp}}
      & {\arnname{JourneyPlanner}}
      \ar [l] _-{\theta_{2}}
    }
    \]
    \begin{itemize}
      
    \item the \acn{ARN} \(\arnname{JourneyPlannerApp}\) depicted in Figure~\ref{figure:journey-planner-app-ARN},
      
    \item the morphism \(\theta_{1}\) that maps the point \(\portname{R_{1}}\) to \(\portname{JP_{1}}\), the communication hyperedge \(\connectionname{C}\) to \(\connectionname{CJP}\) and the messages \(\msgname{getRoute}\) and \(\msgname{route}\) of \(M_{\portname{R_{1}}}\) to \(\msgname{planJourney}\) and \(\msgname{directions}\), respectively, while preserving all the remaining elements of \(\arnname{Traveller}\), and

    \item the inclusion \(\theta_{2}\) of \(\arnname{JourneyPanner}\) into \(\arnname{JourneyPlannerApp}\).
      
    \end{itemize}
    
    \begin{figure}[h]
      \centering

      \begin{tikzpicture}
        % the process T
        
        \node [align=center, minimum height=7ex, minimum width=4em] (T) {
          \(\processname{T}\) \\[1ex]
          \(\Lambda_{\processname{T}}\)
        };
        
        \node [port] (T1) [right=0em of T, align=left] {
          \(\pmsgl{getRoute}\) \\
          \(\dmsgl{route}\)
        };
        \node [port-label] [above=.5ex of T1] {\(\portname{T_{1}}\)};

        % the process JP
        
        \node [port] (JP1) [right=3em of T1, align=right] {
          \(\dmsgr{planJourney}\) \\
          \(\pmsgr{directions}\)
        };
        \node [port-label] [above=.5ex of JP1] {\(\portname{JP_{1}}\)};

        \node [align=center, minimum height=9ex, minimum width=4em] (JP) [right=0em of JP1] {
          \(\processname{JP}\) \\[1ex]
          \(\Lambda_{\processname{JP}}\)
        };      
        
        \node [port] (JP2) [right=0em of JP, align=left] {
          \(\pmsgl{getRoutes}\) \\
          \(\dmsgl{routes}\) \\
          \(\dmsgl{timetables}\)
        };
        \node [port-label] [above=.5ex of JP2] {\(\portname{JP_{2}}\)};

        % the port R1
        
        \node [port] (R1) [above right=-1ex and 3em of JP2, align=right] {
          \(\dmsgr{getRoutes}\) \\
          \(\pmsgr{routes}\)
        };
        \node [port-label] [above=.5ex of R1] {\(\portname{R_{1}}\)};

        % the port R2
        
        \node [port] (R2) [below right=-1ex and 3em of JP2, align=right] {
          \(\dmsgr{routes}\) \\
          \(\pmsgr{timetables}\)
        };
        \node [port-label] [above=.5ex of R2] {\(\portname{R_{2}}\)};

        % the connections

        \draw [rounded corners]
        ($(T1.north west) + (0.5,0.1)$)
        to ($(JP1.north east) + (0.1,0.1)$)
        to ($(JP1.south east) + (0.1,-0.1)$)
        to ($(T1.south west) + (-0.1,-0.1)$)
        to ($(T1.north west) + (-0.1,0.1)$)
        to ($(T1.north west) + (0.5,0.1)$);
        
        \path (T1) -- node [connection] {
          \(\connectionname{CJP}\) \\
          \(\Lambda_{\connectionname{CJP}}\)
        } (T1 -| JP1.west);

        \draw [rounded corners]
        ($(JP2.north west) + (.5,.1)$)
        to                   ($(JP2.north east) + (-.5,.1)$)
        to [out=0, in=180]   ($(R1.north west)  + (.25,.1)$)
        to                   ($(R1.north east)  + (.1,.1)$)
        to                   ($(R1.south east)  + (.1,-.1)$)
        to                   ($(R1.south west)  + (.25,-.1)$)
        to [out=180, in=180] ($(R2.north west)  + (.25,.1)$)
        to                   ($(R2.north east)  + (.1,.1)$)
        to                   ($(R2.south east)  + (.1,-.1)$)
        to                   ($(R2.south west)  + (.25,-.1)$)
        to [out=180, in=0]   ($(JP2.south east) + (-.5,-.1)$)
        to                   ($(JP2.south west) + (-.1,-.1)$)
        to                   ($(JP2.north west) + (-.1,.1)$)
        to                   ($(JP2.north west) + (.5,.1)$);
        
        \path (JP2) -- node [connection] {
          \(\connectionname{C}\) \\[1ex]
          \(\Lambda_{\connectionname{C}}\)
        } (JP2 -| R1.west);

        \begin{pgfonlayer}{background}       
          \node [process] [fit=(T)] {};
          \node [process] [fit=(JP)] {};
        \end{pgfonlayer}
      \end{tikzpicture}
      
      \caption{The \acn{ARN} \(\arnname{JourneyPlannerApp}\)}
      \label{figure:journey-planner-app-ARN}
    \end{figure}
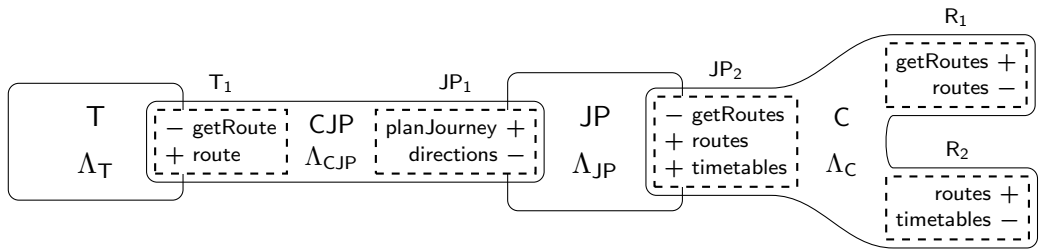

    \noindent It follows that we can derive by resolution a new service query defined by the network \(\arnname{JourneyPlannerApp}\) and the requires-specifications \(\lsen{\portname{R_{1}}}{\rho^{\arnname{JP}}_{1}}\) and \(\lsen{\portname{R_{2}}}{\rho^{\arnname{JP}}_{2}}\).

    \[
    \xymatrix @H=4ex @R=0ex @C+4em {
      {\mathrlap{\arrquery{
            \arnname{Traveller}
          }{
            \cbr[\big]{
              \lsen{\portname{R_{1}}}{\rho^{\arnname{T}}_{1}}
            }
          }
        }}
      && {\mathllap{\arrclause{
            \arnname{JourneyPlanner}
          }{
            \lsen{\portname{JP_{1}}}{\rho^{\arnname{JP}}}
          }{
            \cbr[\big]{
              \lsen{\portname{R_{1}}}{\rho^{\arnname{JP}}_{1}}, \lsen{\portname{R_{2}}}{\rho^{\arnname{JP}}_{2}}
            }
          }
        }} \\
      & {\arrquery{
          \arnname{JourneyPlannerApp}
        }{
          \cbr[\big]{
            \lsen{\portname{R_{1}}}{\rho^{\arnname{JP}}_{1}}, \lsen{\portname{R_{2}}}{\rho^{\arnname{JP}}_{2}}
          }
        }}
      \ar @{-} [ul]!DL-<1ex,0ex>;[ur]!DR+<1ex,0ex> |>/1em/*[r]{\theta_{1}}
    }
    \]
  \end{exa}

  \begin{minisection}{The logic-programming framework of services.}%
    The crucial property of the above notions of service clause, query, and resolution is that, together with the generalized substitution system \(\OrcScheme\) used to define them, they give rise to a logic-programming framework~\cite{Tutu-Fiadeiro:Institution-independent-logic-programming-2015}.
    The construction is to a great extent self-evident, and it requires little additional consideration apart from the fact that, from a technical point of view, in order to define clauses and queries as quantified sentences, we need to extend \(\OrcScheme\) by closing the sets of sentences that it defines under propositional connectives such as implication and conjunction.
    It should be noted, however, that the properties that guarantee the well-definedness of the resulting logic-programming framework such as the fact that its underlying generalized substitution system has weak model amalgamation (ensured by Proposition~\ref{proposition:weak-model-amalgamation-in-OrcScheme}), and also the fact that the satisfaction of specifications is preserved by model homomorphisms (detailed in Remark~\ref{remark:preservation-of-satisfaction-in-OrcScheme}), are far from trivial, especially when taking into account particular orchestration schemes (see e.g.\ Proposition~\ref{orchestration-scheme-of-ARNs}).
  \end{minisection}

  By describing service discovery and binding as instances of unification and resolution (specific to the logic-programming framework of services) we obtain not only a rigorously defined analogy between service-oriented computing and relational logic programming, but also a way to apply the general theory of logic programming to the particular case of services.
  For example, we gain a concept of solution to a service query that reflects the rather intuitive service-oriented notion of solution and, moreover, through Herbrand's theorem, a characterization of satisfiable queries as queries that admit solutions.

  \begin{defi}[Solution]
    A \emph{solution}, or \emph{correct answer}, to a service-oriented query \(\query{\orc}{Q}\) consists of a morphism of orchestrations \(\psi \colon \orc \to \orc'\) such that \(\orc'\) has models, and every one of them satisfies the \(\psi\)\nb-translations of the specifications in \(Q\).
  \end{defi}

  \begin{prop}
    \label{proposition:characterization-of-satisfiable-service-queries}
    A service query is satisfiable if and only if it admits a solution.
    \qed
  \end{prop}

\begin{sidewaysfigure}
    \centering

    \begin{tikzpicture}
      % the Traveller application
      
      % the process T
      
      \node [align=center, minimum height=7ex, minimum width=4em] (T) {
        \(\processname{T}\) \\[1ex]
        \(\Lambda_{\processname{T}}\)
      };
      
      \node [port] (T1) [right=0em of T, align=left] {
        \(\pmsgl{getRoute}\) \\
        \(\dmsgl{route}\)
      };
      \node [port-label] [above=.5ex of T1] {\(\portname{T_{1}}\)};

      % the port TR1
      
      \node [port] (TR1) [right=3em of T1, align=right] {
        \(\dmsgr{getRoute}\) \\
        \(\pmsgr{route}\)
      };
      \node [port-label] [above=.5ex of TR1] {\(\portname{R_{1}}\)};

      % the connection of Traveller

      \draw [rounded corners]
      ($(T1.north west) + (0.5,0.1)$)
      to ($(TR1.north east) + (0.1,0.1)$)
      to ($(TR1.south east) + (0.1,-0.1)$)
      to ($(T1.south west)  + (-0.1,-0.1)$)
      to ($(T1.north west)  + (-0.1,0.1)$)
      to ($(T1.north west)  + (0.5,0.1)$);
      
      \path (T1) -- node [connection] {
        \(\connectionname{C}\) \\
        \(\Lambda_{\connectionname{C}}\)
      } (T1 -| TR1.west);
      
      % the orchestration of Traveller
      
      \node (Tapp-orc) [fit=(T) (T1) (TR1), inner sep=1.05em] {};

      % the requires-point of Traveller
      
      \node [requires-point] (Tapp-rspec) [right=.5em of TR1.east] {%
        \(\lsen{\portname{R_{1}}}{\rho^{\arnname{T}}_{1}}\)% 
      };

      \coordinate (Tapp-north) at ($(Tapp-rspec.north) + (0em, 1ex)$);
      \coordinate (Tapp-south) at ($(Tapp-rspec.south) - (0em, 1ex)$);
      \coordinate (Tapp-west)  at ($(Tapp-orc.west)    - (.5em, 0ex)$);
      \coordinate (Tapp-east)  at ($(Tapp-rspec.east)  - (1.5em, 0ex)$);

      \node (Tapp-rspecr) [right=0em of Tapp-rspec] {\hspace{.6em}\reflectbox{\(\models\)}};

      % the Journey Planner service

      % the provides-point of Journey Planner

      \node [provides-point] (JPmod-pspec) [below=25ex of Tapp-orc.south west, anchor=north west] {%
        \hspace{.25em}\(\lsen{\portname{JP_{1}}}{\rho^{\arnname{JP}}}\)% 
      };
      \draw [dotted, thick] (Tapp-rspecr)
      -- ($(Tapp-rspecr.east) + (.5em, 0ex)$)
      -- ($(Tapp-rspecr.east) + (.5em, -13.25ex)$)
      -| ($(JPmod-pspec.west) - (.5em, -15ex)$)
      -- ($(JPmod-pspec.west) - (.5em, 0ex)$)
      -- (JPmod-pspec.west);

      % the process JP
      
      \node [port] (JP1) [right=.25em of JPmod-pspec, align=right] {
        \(\dmsgr{planJourney}\) \\
        \(\pmsgr{directions}\)
      };
      \node [port-label] [above=0ex of JP1] {\(\portname{JP_{1}}\)};

      \node [align=center, minimum height=9ex, minimum width=4em] (JP) [right=0em of JP1] {
        \(\processname{JP}\) \\[1ex]
        \(\Lambda_{\processname{JP}}\)
      };      
      
      \node [port] (JP2) [right=0em of JP, align=left] {
        \(\pmsgl{getRoutes}\) \\
        \(\dmsgl{routes}\) \\
        \(\dmsgl{timetables}\)
      };
      \node [port-label] [above=.5ex of JP2] {\(\portname{JP_{2}}\)};

      % the port JPR1
      
      \node [port] (JPR1) [above right=3ex and 3em of JP2, align=right] {
        \(\dmsgr{getRoutes}\) \\
        \(\pmsgr{routes}\)
      };
      \node [port-label] [above=.5ex of JPR1] {\(\portname{R_{1}}\)};

      % the port JPR2
      
      \node [port] (JPR2) [below right=3ex and 3em of JP2, align=right] {
        \(\dmsgr{routes}\) \\
        \(\pmsgr{timetables}\)
      };
      \node [port-label] [above=.5ex of JPR2] {\(\portname{R_{2}}\)};

      % the connection
      
      \draw [rounded corners]
      ($(JP2.north west) + (.5,.1)$)
      to                   ($(JP2.north east)  + (-.5,.1)$)
      to [out=0, in=180]   ($(JPR1.north west) + (.25,.1)$)
      to                   ($(JPR1.north east) + (.1,.1)$)
      to                   ($(JPR1.south east) + (.1,-.1)$)
      to                   ($(JPR1.south west) + (.6,-.1)$)
      to [out=180, in=180] ($(JPR2.north west) + (.6,.1)$)
      to                   ($(JPR2.north east) + (.1,.1)$)
      to                   ($(JPR2.south east) + (.1,-.1)$)
      to                   ($(JPR2.south west) + (.25,-.1)$)
      to [out=180, in=0]   ($(JP2.south east)  + (-.5,-.1)$)
      to                   ($(JP2.south west)  + (-.1,-.1)$)
      to                   ($(JP2.north west)  + (-.1,.1)$)
      to                   ($(JP2.north west)  + (.5,.1)$);
      
      \path (JP2) -- node [connection] {
        \(\connectionname{C}\) \\[1ex]
        \(\Lambda_{\connectionname{C}}\)
      } (JP2 -| JPR1.west);

      % the orchestration of Journey Planner
      
      \node (JPmod-orc) [fit=(JP) (JP1) (JP2) (JPR1) (JPR2), inner sep=1.5em] {};
      
      % the requires-points of Journey Planner
      
      \node [requires-point] (JPmod-rspec1) [right=.5em of JPR1.east] {%
        \(\lsen{\portname{R_{1}}}{\rho^{\arnname{JP}}_{1}}\)% 
      };
      \node [requires-point] (JPmod-rspec2) [right=.5em of JPR2.east] {%
        \(\lsen{\portname{R_{2}}}{\rho^{\arnname{JP}}_{2}}\)% 
      };

      \coordinate (JPmod-north) at ($(JPmod-rspec1.north) + (0em, 1ex)$);
      \coordinate (JPmod-south) at ($(JPmod-rspec2.south) - (0em, 1ex)$);
      \coordinate (JPmod-west)  at ($(JPmod-pspec.west)   + (1.5em, 0ex)$);
      \coordinate (JPmod-east)  at ($(JPmod-rspec1.east)  - (1.5em, 0ex)$);

      % the Map Services service

      % the provides-point of Map Services
      
      \node [provides-point] (MSmod-pspec) [right=2em of JPmod-rspec1.east] {%
        \hspace{.25em}\(\lsen{\portname{MS_{1}}}{\rho^{\arnname{MS}}}\)% 
      };
      \path (JPmod-rspec1) -- node {\hspace{.5em}\reflectbox{\(\models\)}} (MSmod-pspec);
      
      % the process MS
      
      \node [port] (MS1) [right=.25em of MSmod-pspec, align=right] {
        \(\dmsgr{getRoutes}\) \\
        \(\pmsgr{routes}\)
      };
      \node [port-label] [above=0ex of MS1] {\(\portname{MS_{1}}\)};

      \node [align=center, minimum height=7ex, minimum width=4em] (MS) [right=0em of MS1] {
        \(\processname{MS}\) \\[1ex]
        \(\Lambda_{\processname{MS}}\)
      };

      % the orchestration of Map Services
      
      \node (MSmod-orc) [fit=(MS) (MS1), inner sep=1.05em] {};

      \coordinate (MSmod-north) at ($(MSmod-pspec.north) + (0em, 1ex)$);
      \coordinate (MSmod-south) at ($(MSmod-pspec.south) - (0em, 1ex)$);
      \coordinate (MSmod-west)  at ($(MSmod-pspec.west)  + (1.5em, 0ex)$);
      \coordinate (MSmod-east)  at ($(MSmod-orc.east)    + (.5em, 0ex)$);

      % the Transport System service

      % the provides-point of Transport System
      
      \node [provides-point] (TSmod-pspec) [right=2em of JPmod-rspec2.east] {%
        \hspace{.25em}\(\lsen{\portname{TS_{1}}}{\rho^{\arnname{TS}}}\)% 
      };
      \path (JPmod-rspec2) -- node {\hspace{.5em}\reflectbox{\(\models\)}} (TSmod-pspec);
      
      % the process TS
      
      \node [port] (TS1) [right=.25em of TSmod-pspec, align=right] {
        \(\dmsgr{routes}\) \\
        \(\pmsgr{timetables}\)
      };
      \node [port-label] [above=0ex of TS1] {\(\portname{TS_{1}}\)};

      \node [align=center, minimum height=7ex, minimum width=4em] (TS) [right=0em of TS1] {
        \(\processname{TS}\) \\[1ex]
        \(\Lambda_{\processname{TS}}\)
      };

      % the orchestration of Transport System
      
      \node (TSmod-orc) [fit=(TS) (TS1), inner sep=1.05em] {};

      \coordinate (TSmod-north) at ($(TSmod-pspec.north) + (0em, 1ex)$);
      \coordinate (TSmod-south) at ($(TSmod-pspec.south) - (0em, 1ex)$);
      \coordinate (TSmod-west)  at ($(TSmod-pspec.west)  + (1.5em, 0ex)$);
      \coordinate (TSmod-east)  at ($(TSmod-orc.east)    + (.5em, 0ex)$);

      \begin{pgfonlayer}{background}
        \node [orchestration] (Tapp-border) [fit=(Tapp-orc) (Tapp-north) (Tapp-south) (Tapp-west) (Tapp-east)] {};
        \node [below=1ex of Tapp-border.south] {Traveller};
        \node [process] [fit=(T)] {};
        \node [orchestration] (JPmod-border) [fit=(JPmod-orc) (JPmod-north) (JPmod-south) (JPmod-west) (JPmod-east)] {};
        \node [below=1ex of JPmod-border.south] {Journey Planner};
        \node [process] [fit=(JP)] {};
        \node [orchestration] (MSmod-border) [fit=(MSmod-orc) (MSmod-north) (MSmod-south) (MSmod-west) (MSmod-east)] {};
        \node [below=1ex of MSmod-border.south] {Map Services};
        \node [process] [fit=(MS)] {};
        \node [orchestration] (TSmod-border) [fit=(TSmod-orc) (TSmod-north) (TSmod-south) (TSmod-west) (TSmod-east)] {};
        \node [below=1ex of TSmod-border.south] {Transport System};
        \node [process] [fit=(TS)] {};
      \end{pgfonlayer}
    \end{tikzpicture}
    
    \caption{The derivation of an answer to the Traveller application}
    \label{figure:service-oriented-derivation}

    \begin{alignat*}{4}
      & \rho^{\arnname{T}}_{1}  && {} \colon \ltlalways \rbr{\dact{\msgname{getRoute}} \plimplies \ltleventually \pact{\msgname{route}}} & \qquad
      & \rho^{\arnname{JP}}_{2} && {} \colon \ltlalways \rbr{\dact{\msgname{routes}} \plimplies \ltleventually \pact{\msgname{timetables}}} 
      \\
      & \rho^{\arnname{JP}}    && {} \colon \ltlalways \rbr{\dact{\msgname{planJourney}} \plimplies \ltleventually \pact{\msgname{directions}}} &
      & \rho^{\arnname{MS}}    && {} \colon \ltlalways \rbr{\dact{\msgname{getRoutes}} \plimplies \ltleventually \pact{\msgname{routes}}}
      \\
      & \rho^{\arnname{JP}}_{1} && {} \colon \ltlalways \rbr{\dact{\msgname{getRoutes}} \plimplies \ltleventually \pact{\msgname{routes}}} &
      & \rho^{\arnname{TS}}    && {} \colon \ltlalways \rbr{\dact{\msgname{routes}} \plimplies \ltleventually \pact{\msgname{timetables}}}
    \end{alignat*}
  \end{sidewaysfigure}
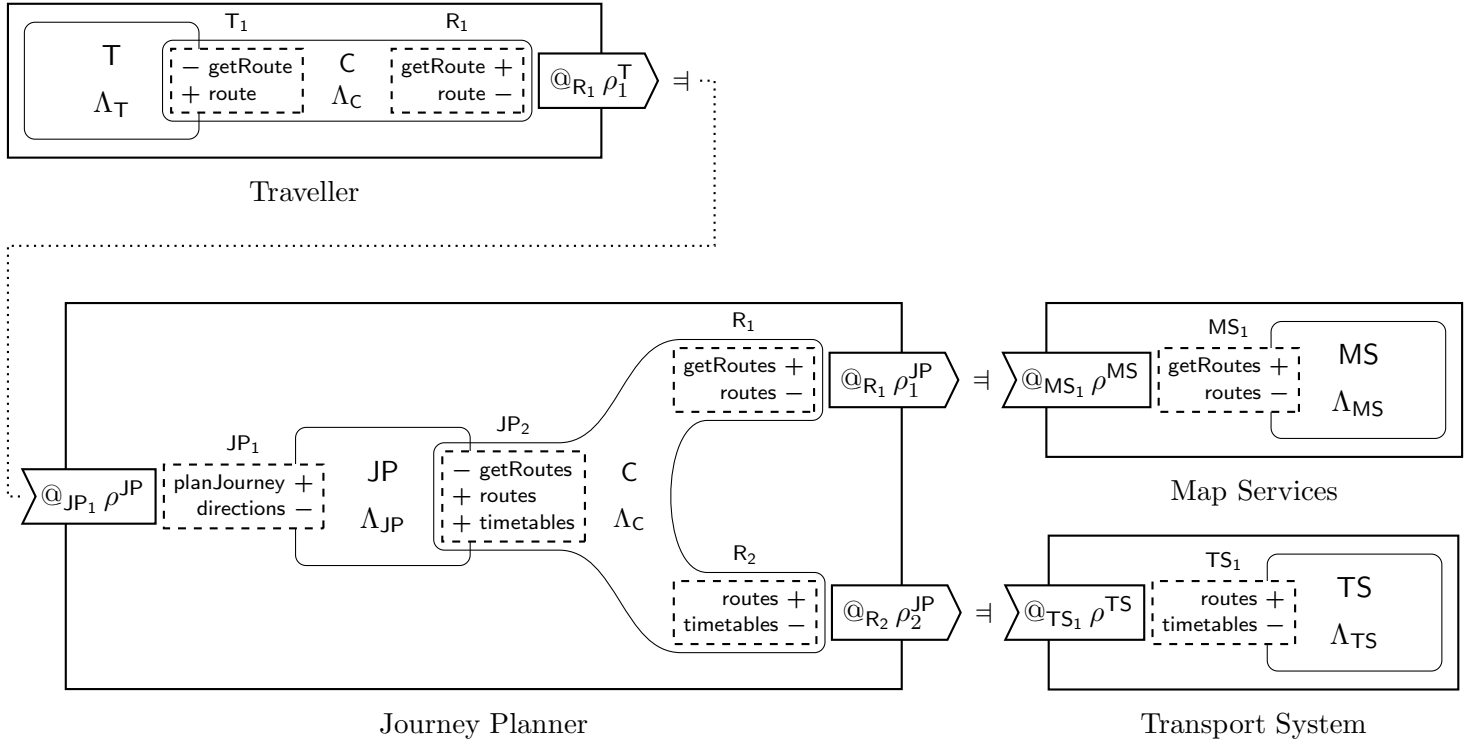

  Even more significant is the fact that logic programming provides us with a general search procedure that can be used to compute solutions to queries.
  The search is triggered by a query \(\query{\orc}{Q}\) and consists in the iterated application of resolution, that is of service discovery and binding, until the requires-interface of the derived service query consists solely of trivial specifications (tautologies); these are specifications whose translation along morphisms into ground orchestrations always gives rise to properties.
  Thus, whenever the search procedure successfully terminates we obtain a \emph{computed answer} to the original query by sequentially composing the resulting computed morphisms.
  This is the process that led, for example, to the derivation of the program that calculates the quotient and the remainder obtained on dividing two natural numbers illustrated in Figure~\ref{figure:program-derivation}.
  The computed answer is given in this case by the sequence of substitutions
  \[
  \setlength{\arraycolsep}{0pt}
  \begin{array}{rl}
    \ptm & {}
           \mapsto \ptm_{1} \comp \ptm_{2}
           \mapsto \rbr{\ptm_{3} \comp \ptm_{4}} \comp \ptm_{2}
           \mapsto \dotsb \\[1ex]
         & {\begin{array}{rl}
              {} \mapsto \rbr{q \coloneqq 0 \comp r \coloneqq x} \comp {} & \opname{while}\, y \leq r\, \opname{do} \\
                                                                          & \hspace{1em} q \coloneqq q + 1 \comp r \coloneqq r - y \\
                                                                          & \opname{done}.
            \end{array}}
  \end{array}
  \]
  In a similar manner, we can continue Example~\ref{example:service-oriented-resolution} towards the derivation of an answer to the Traveller application.
  To this purpose, we assume that Map Services and Transport System are two additional service modules that correspond to the processes \(\processname{MS}\) and \(\processname{TS}\) used in Example~\ref{example:journey-planner-net}, and whose provides-interfaces meet the requires-specifications of the module Journey Planner.
  We obtain in this way the construction outlined in Figure~\ref{figure:service-oriented-derivation}.

  The soundness of resolution, detailed in Proposition~\ref{proposition:soundness-of-service-oriented-resolution} below, entails that the search for solutions is sound as well, in the sense that every computed answer to \(\query{\orc}{Q}\) is also a solution to \(\query{\orc}{Q}\).  This fundamental result, originally discussed in~\cite{Tutu-Fiadeiro:Institution-independent-logic-programming-2015} in the context of abstract logic programming, ensures, in combination with Proposition~\ref{proposition:characterization-of-satisfiable-service-queries}, that the operational semantics of the service overlay given by discovery and binding is sound with respect to the notion of satisfiability of a service query.

  \begin{prop}
    \label{proposition:soundness-of-service-oriented-resolution}
    Let \(\query{\orc}{Q}\) be a service query derived by resolution from \(\query{\orc_{1}}{Q_{1}}\) and \(\dclause{\orc_{2}}{P_{2}}{R_{2}}\) using the computed morphism \(\theta_{1} \colon \orc_{1} \to \orc\).
    If \(\dclause{\orc_{2}}{P_{2}}{R_{2}}\) is correct then, for any solution \(\psi\) to \(\query{\orc}{Q}\), the composed morphism \(\theta_{1} \comp \psi\) is a solution to \(\query{\orc_{1}}{Q_{1}}\).
    \qed
  \end{prop}

  \section{Conclusions}

  We have shown how the integration of the declarative and the operational semantics of conventional logic programming can be generalized to service-oriented computing, thus offering a unified semantics for the static and the dynamic aspects of this paradigm.  That is, we have provided, for the first time, an algebraic framework that accounts for the mechanisms through which service interfaces can be orchestrated, as well as for those mechanisms that allow applications to discover and bind to services.

  The analogy that we have established is summarized in Table~\ref{table:correspondence-between-relational-and-service-oriented-logic-programming}.
  Our approach to the logic-programming semantics of services is based on the identification of the binding of terms to variables in logic programming with the binding of orchestrations of services to those of software applications in service-oriented computing; the answer to a service query -- the request for external services -- is obtained through resolution using service clauses -- orchestrated service interfaces -- that are available from a repository.
  This departs from other works on the logic-programming semantics of service-oriented computing such as~\cite{Kona-Bansal-Gupta:Automatic-composition-of-semantic-Web-services-2007} that actually considered implementations of the service discovery and binding mechanisms based on constraint logic programming.

  \begin{sidewaystable}
    \centering
    \caption{Correspondence between concepts of relational and service-oriented logic programming}
    \label{table:correspondence-between-relational-and-service-oriented-logic-programming}

    \vskip \abovecaptionskip

    \renewcommand{\arraystretch}{1.4}
    \overfullrule=0 pt
    \begin{tabular}{@{}
      >{\raggedright}p{.11\textwidth}
      >{\centering\arraybackslash}p{.23\textwidth}
      >{\centering\arraybackslash}p{.28\textwidth}
      >{\centering\arraybackslash}p{.33\textwidth}
      @{}}
      \toprule
      & Relational logic programming
      & \multicolumn{2}{c}{Service-oriented logic programming}
      \\
      \cmidrule(lr){2-2} \cmidrule(l){3-4}
      Concept
      & over a signature \(\abr{F, P}\)
      & over program expressions
      & over asynchronous relational networks
      \\
      \midrule
      Variable
      & pair \(\rbr{x, F_{0}}\)
      & program variable \(\ptm \colon \pexp\)
      & requires-point \(x \in X\)
      \\
      Term
      & structure \(\sigma\rbr{t_{1}, \dotsc, t_{n}}\)
      & program statement

        \vskip .5\baselineskip
        
        \begin{tikzpicture}
          \node        (while)                                                                  {\(\opname{while}\, C\, \opname{do}\)};
          \node [pvar] (pgm)   [below right=0ex and 1em of while.south west, anchor=north west] {\(\ptm\)};
          \node        (done)  [below left=0ex and 1em of pgm.south west, anchor=north west]    {\(\opname{done}\)};
        \end{tikzpicture}
      & subnetwork determined by a point
        
        \vskip .5\baselineskip
        
        \begin{tikzpicture}
          % the process MS
          
          \node [port] (MS1) [align=right] {
            \(\dmsgr{getRoutes}\) \\
            \(\pmsgr{routes}\)
          };
          \node [port-label] [above=0ex of MS1] {\(\portname{MS_{1}}\)};

          \node [align=center, minimum height=7ex, minimum width=4em] (MS) [right=0em of MS1] {
            \(\processname{MS}\) \\[1ex]
            \(\Lambda_{\processname{MS}}\)
          };

          \begin{pgfonlayer}{background}  
            \node [process] [fit=(MS)] {};
          \end{pgfonlayer}
        \end{tikzpicture}
      \\
      Clause
      & universally quantified implication
        
        \vskip .5\baselineskip
        
        \(\arrclause{X}{C}{H}\)
      & program module

        \vskip .5\baselineskip

        \begin{tikzpicture}
          \node        (comp)                     {\(\comp\)};
          \node [pvar] (pgm1) [left=0em of comp]  {\phantom{\(\bullet\)}};
          \node [pvar] (pgm2) [right=0em of comp] {\phantom{\(\bullet\)}};
          
          \coordinate (onorth) at ($(comp.north) + (0em,3.5ex)$);
          \coordinate (osouth) at ($(comp.south) - (0em,3.5ex)$);
          \coordinate (owest)  at (pgm1.west);
          \coordinate (oeast)  at (pgm2.east);

          \node (orc) [fit=(onorth) (osouth) (owest) (oeast)] {};

          \node [provides-point] (pspec) [left=.25em of orc.west] {\hspace{.5em}\(\rho, \rho''\)};

          \node [requires-point] (rspec1) [above right=1ex and .25em of orc.east, minimum width=3.25em] {\(\rho, \rho'\)};
          \draw (pgm1) |- (rspec1.west);
          \node [requires-point] (rspec2) [below right=1ex and .25em of orc.east, minimum width=3.25em] {\(\rho', \rho''\)};
          \draw (pgm2) |- (rspec2.west);

          \coordinate (north) at ($(rspec1.north) + (0em, 1ex)$);
          \coordinate (south) at ($(rspec2.south) - (0em, 1ex)$);
          \coordinate (west)  at ($(pspec.west)   + (1.5em, 0ex)$);
          \coordinate (east)  at ($(rspec1.east)  - (1.5em, 0ex)$);

          \begin{pgfonlayer}{background} 
            \node [orchestration] [fit=(orc) (north) (south) (west) (east)] {};
          \end{pgfonlayer}
        \end{tikzpicture}
      & service module

        \vskip .5\baselineskip

        \begin{tikzpicture}
          % the process JP
          
          \node [port] (JP1) [align=right] {
            \phantom{\(\bullet\)}
          };

          \node [align=center, minimum height=5ex, minimum width=2em] (JP) [right=0em of JP1] {
            \(\processname{JP}\)
          };      
          
          \node [port] (JP2) [right=0em of JP, align=left] {
            \phantom{\(\bullet\)}
          };

          % the port R1
          
          \node [port] (R1) [above right=.75ex and 1.5em of JP2, align=right] {
            \phantom{\(\bullet\)}
          };

          % the port JPR2
          
          \node [port] (R2) [below right=.75ex and 1.5em of JP2, align=right] {
            \phantom{\(\bullet\)}
          };

          % the connection
          
          \draw [rounded corners]
          ($(JP2.north west) + (.1,.1)$)
          to                   ($(JP2.north east)  + (-.35,.1)$)
          to [out=0, in=180]   ($(R1.north west) + (.25,.1)$)
          to                   ($(R1.north east) + (.1,.1)$)
          to                   ($(R1.south east) + (.1,-.1)$)
          to                   ($(R1.south west) + (.3,-.1)$)
          to [out=180, in=180] ($(R2.north west) + (.3,.1)$)
          to                   ($(R2.north east) + (.1,.1)$)
          to                   ($(R2.south east) + (.1,-.1)$)
          to                   ($(R2.south west) + (.25,-.1)$)
          to [out=180, in=0]   ($(JP2.south east)  + (-.35,-.1)$)
          to                   ($(JP2.south west)  + (-.1,-.1)$)
          to                   ($(JP2.north west)  + (-.1,.1)$)
          to                   ($(JP2.north west)  + (.1,.1)$);
          
          \path (JP2) -- node [connection] {
            \(\connectionname{C}\)
          } (JP2 -| R1);

          % the provides- and requires-points

          \node [provides-point] (pspec) [left=.25em of JP1] {%
            \hspace{.25em}\(\lsen{\portname{JP_{1}}}{\rho^{\arnname{JP}}}\)% 
          };

          \node [requires-point] (rspec1) [right=.5em of R1.east] {%
            \(\lsen{\portname{R_{1}}}{\rho^{\arnname{JP}}_{1}}\)% 
          };
          \node [requires-point] (rspec2) [right=.5em of R2.east] {%
            \(\lsen{\portname{R_{2}}}{\rho^{\arnname{JP}}_{2}}\)% 
          };

          \coordinate (north) at ($(rspec1.north) + (0em, 1ex)$);
          \coordinate (south) at ($(rspec2.south) - (0em, 1ex)$);
          \coordinate (west)  at ($(pspec.west)   + (1.5em, 0ex)$);
          \coordinate (east)  at ($(rspec1.east)  - (1.5em, 0ex)$);

          \begin{pgfonlayer}{background}
            \node [orchestration] [fit=(orc) (north) (south) (west) (east)] {};
            \node [process] [fit=(JP)] {};
          \end{pgfonlayer}
        \end{tikzpicture}
      \\
      Query
      & existentially quantified conjunction

        \vskip .5\baselineskip
        
        \(\arrquery{X}{Q}\)
      & program query

        \vskip .5\baselineskip

        \begin{tikzpicture}
          \node [pvar] (app) {\phantom{\(\bullet\)}};
          
          \node (orc) [fit=(app)] {};

          \node [requires-point] (rspec) [right=.5em of orc.east] {%
            \(\rho, \rho'\)% 
          };
          \draw (app) |- (rspec.west);
          
          \coordinate (north) at ($(rspec.north) + (0em, 1ex)$);
          \coordinate (south) at ($(rspec.south) - (0em, 1ex)$);
          \coordinate (west)  at ($(orc.west)    - (.5em, 0ex)$);
          \coordinate (east)  at ($(rspec.east)  - (1.5em, 0ex)$);
          
          \begin{pgfonlayer}{background} 
            \node [orchestration] [fit=(orc) (north) (south) (west) (east)] {};
          \end{pgfonlayer}
        \end{tikzpicture}
      & client application
        
        \vskip .5\baselineskip
        
        \begin{tikzpicture}
          % the process T
          
          \node [align=center, minimum height=5ex, minimum width=2em] (T) {
            \(\processname{T}\)
          };
          
          \node [port] (T1) [right=0em of T, align=left] {
            \phantom{\(\bullet\)}
          };

          % the port R1
          
          \node [port] (R1) [right=2em of T1, align=right] {
            \phantom{\(\bullet\)}
          };

          % the connection

          \draw [rounded corners]
          ($(T1.north west) + (0.5,0.1)$)
          to ($(R1.north east) + (0.1,0.1)$)
          to ($(R1.south east) + (0.1,-0.1)$)
          to ($(T1.south west)  + (-0.1,-0.1)$)
          to ($(T1.north west)  + (-0.1,0.1)$)
          to ($(T1.north west)  + (0.5,0.1)$);
          
          \path (T1) -- node [connection] {
            \(\connectionname{C}\)
          } (T1 -| R1.west);
          
          % the orchestration
          
          \node (orc) [fit=(T) (T1) (R1), inner sep=.9em] {};

          % the requires-point of Traveller
          
          \node [requires-point] (rspec) [right=.5em of R1.east] {%
            \(\lsen{\portname{R_{1}}}{\rho^{\arnname{T}}_{1}}\)% 
          };

          \coordinate (north) at ($(rspec.north) + (0em, 1ex)$);
          \coordinate (south) at ($(rspec.south) - (0em, 1ex)$);
          \coordinate (west)  at ($(orc.west)    - (.5em, 0ex)$);
          \coordinate (east)  at ($(rspec.east)  - (1.5em, 0ex)$);
          
          \begin{pgfonlayer}{background}
            \node [orchestration] [fit=(orc) (north) (south) (west) (east)] {};
            \node [process] [fit=(T)] {};
          \end{pgfonlayer}
        \end{tikzpicture}
      \\
      Unification and resolution
      & term unification and first-order~resolution
      & program discovery and binding

        (see Figure~\ref{figure:program-derivation})
      & service discovery and binding
        
        (see Figure~\ref{figure:service-oriented-derivation})
      \\
      \bottomrule
    \end{tabular}
  \end{sidewaystable}

  The theory of services that we have developed here is grounded on a declarative semantics of service clauses defined over a novel logical system of orchestration schemes.
  The structure of the sentences and of the models of this logical system varies according to the orchestration scheme under consideration.
  For example, when orchestrations are defined as asynchronous relational networks over the institution \(\aLTL\), we obtain sentences as linear-temporal-logic sentences expressing properties observed at given interaction points of a network, and models in the form of ground orchestrations of Muller automata.  Other logics (with corresponding model theory) could have been used instead of the automata-based variant of linear temporal logic, more specifically any institution such that
  \begin{inlinenum}

  \item the category of signatures is (finitely) cocomplete;

  \item there exist cofree models along every signature morphism;

  \item the category of models of every signature has (finite) products; and

  \item model homomorphisms reflect the satisfaction of sentences.

  \end{inlinenum}
  Moreover, the formalism used in defining orchestrations can change by means of morphisms of orchestration schemes.  We could consider, for instance, an encoding of the hypergraphs of processes and connections discussed in this paper into graph-based structures similar to those of~\cite{Fiadeiro-Lopes:Dynamic-reconfiguration-in-service-oriented-architectures-2013}; or we could change their underlying institution by adding new temporal modalities (along the lines of Example~\ref{example:morphisms-of-orchestration-schemes-of-ARNs}) or by considering other classes of automata, like the closed reduced B\"{u}chi automata used in~\cite{Alpern-Schneider:Recognizing-safety-and-liveness-1987,Fiadeiro-Lopes:An-interface-theory-for-service-oriented-design-2013}.
  This encourages us to further investigate aspects related to the heterogeneous foundations of service-oriented computing based on the proposed logical system of orchestration schemes.

  \section*{Acknowledgements}

  The work of the first author has been supported by a grant of the Romanian National Authority for Scientific Research, CNCS-UEFISCDI, project number PN-II-ID-PCE-2011-3-0439.
  The authors also wish to thank Fernando Orejas for suggesting the use of hypergraphs, Ant\'{o}nia Lopes for many useful discussions that led to the present form of this paper, and the anonymous referees for their careful study of the original manuscript.

  \bibliographystyle{alpha}
  \bibliography{references}
\vspace{-30 pt}
\end{document}